\documentclass[oldversion]{aa}
\usepackage{graphicx}
\usepackage{txfonts}
\usepackage{lscape}
\begin{document}
   \title{What ignites on the neutron star of 4U\,0614+091?}

   \author{E. Kuulkers\inst{1}
	  \and
          J.J.M. in 't Zand\inst{2}
          \and
          J.-L. Atteia\inst{3}
          \and
	  A.M. Levine\inst{4}
          \and
	  S. Brandt\inst{5}
          \and
          D.A. Smith\inst{6}
	  \and
	  M. Linares\inst{7}
	  \and
	  M. Falanga\inst{8}
	  \and
	  C.~S\'anchez-Fern\'andez\inst{1}
	  \and
	  C.B. Markwardt\inst{9}
	  \and
	  T.E. Strohmayer\inst{10}
	  \and
          A. Cumming\inst{11}
          \and 
	  M. Suzuki\inst{12}
	  }

   \authorrunning{E.~Kuulkers et al.}
   \titlerunning{Thermonuclear X-ray bursts from 4U\,0614+091}


   \institute{ISOC, ESA, European Space Astronomy Centre (ESAC), P.O.~Box 78, 28691, Villanueva de la Ca\~nada (Madrid), Spain\\
              \email{Erik.Kuulkers@esa.int}
          \and
              SRON Netherlands Institute for Space Research, Sorbonnelaan 2, 3584 CA Utrecht, The Netherlands
          \and
              Laboratoire d'Astrophysique de Toulouse-Tarbes, Observatoire Midi-Pyr\'en\'ees (CNRS-UMR5572/Universit\'e 
              Paul Sabatier Toulouse III), 14 avenue \'Edouard Belin, 31400 Toulouse, France 
          \and
              Kavli Institute for Astrophysics and Space Research,
              Massachusetts Institute of Technology, Cambridge, MA 02139, USA
          \and
	      National Space Institute, DTU, Juliane Maries Vej 30, DK-2100 Copenhagen, Denmark
          \and
              Guilford College, Physics Department, 5800 West Friendly Ave., Greensboro, NC 27410, USA
	  \and
	     Astronomical Institute ``Anton Pannekoek'', University of Amsterdam and Center for High-Energy 
             Astrophysics, P.O.~Box 94249, 1090 GE, Amsterdam, The Netherlands
	  \and
	     International Space Science Institute, Hallerstrasse 6, CH-3012,  Bern, Switzerland
	  \and
	      UMD/CRESST/GSFC, Greenbelt, MD 20771, USA
	  \and
	      Astrophysics Science Division, NASA/GSFC, Greenbelt, MD 20771, USA
          \and
              Physics Department, McGill University, 3600 rue University, Montreal, QC, H3A 2T8, Canada	      
	  \and
	      ISS Science Project Office ISAS, JAXA, 2-1-1 Sengen, Tsukuba, Ibaraki 305-8505, Japan
             }

   \date{Received; accepted}

\abstract{The low-mass X-ray binary 4U\,0614+091 is a source of
sporadic thermonuclear (type~I) X-ray bursts.  
We find bursts with a wide variety of characteristics in 
serendipitous wide-field X-ray observations by the WATCH on {\it EURECA}, the ASM
on {\it RXTE}, the WFCs on {\it BeppoSAX}, the FREGATE on {\it HETE-2}, 
the IBIS/ISGRI on {\it INTEGRAL}, and the BAT on {\it Swift}, 
as well as pointed observations with the
PCA and HEXTE on {\it RXTE}.
Most of the bursts are bright, i.e., they reach a peak flux of
about 15 Crab, but a few are weak and only reach a peak flux below
a Crab.  One of the bursts shows a very strong photospheric
radius-expansion phase.  This allows us to evaluate the
distance to the source, which we estimate to be 3.2\,kpc. 
The burst durations vary generally from about 10\,sec to 5\,min. 
However, after one of the intermediate-duration bursts, a faint tail 
is seen to at least about 2.4 hours after the start of the burst.
One very long burst was observed, which lasted for several hours.  
This superburst candidate was followed by a normal type-I burst only 19 days later.
This is, to our knowledge, the shortest burst-quench time among the
superbursters. The observation of a superburst in this system 
is difficult to reconcile if the system is accreting at about 1\%\ of the Eddington limit.
We describe the burst properties in relation to
the persistent emission.  No strong correlations are apparent,
except that the intermediate-duration bursts occurred
when 4U\,0614+091's persistent emission was lowest and calm, and when bursts were infrequent
(on average roughly one every month to 3 months). The average burst rate increased significantly after this period. The
maximum average burst recurrence rate is about once every week to 2 weeks.
The burst behaviour may be partly understood if
there is at least an appreciable amount of helium present in the accreted
material from the donor star. If the system is an ultra-compact X-ray binary
with a CO white-dwarf donor, as has been suggested, this is unexpected.
If the bursts are powered by helium, we find that
the energy production per accumulated mass is about 2.5 times less
than expected for pure helium matter.
}


   \maketitle
%

\section{Introduction}
\label{introduction}

Type I X-ray bursts (Grindlay et al.\ 1975, Belian et al.\ 1976,
Hoffman et al.\ 1978; hereafter bursts) result from thermonuclear shell flashes on a
neutron star, which is caused by the ignition of either He and/or H-rich material supplied by a binary companion star
(Hansen \&\ van Horn 1975, Woosley \&\ Taam 1976, Maraschi \&\ Cavaliere 1977, Lamb \&\ Lamb 1978; 
for reviews see Lewin et al.\ 1993, Strohmayer \&\ Bildsten 2006).  
Bursts generally appear as short transient events wherein the X-ray intensity rises
rapidly on a time scale of seconds, and decays in an exponential fashion
back to the pre-burst level.  The decay lasts almost always longer than the rise.
Burst durations range from several seconds up to half an hour. 
The burst spectra generally harden during the rise and soften during the decay.  
This has been attributed to the heating and cooling of the
uppermost layers of the neutron star.
The spectra can be satisfactorily described by black-body emission from spherical
regions with radii of around 10\,km at inferred temperatures up to
$kT$$\simeq$3\,keV. The burst-to-burst time intervals are typically of
the order of hours to days. They are thought to be determined by the
time for the neutron star to accumulate enough fuel to power another
burst.

Under certain conditions, the local luminosity may reach or exceed the
Eddington limit and matter may be pushed
outward.  As a result, the neutron star photosphere also moves
outward.  Consequently, the emitting area increases and the observed
inferred black-body temperature drops.
When the surge of energy release is over, the
photosphere gradually returns to its pre-burst radius.  During
this phase, the emitting area decreases and the inferred temperature increases. 
During the expansion and contraction phase, the
luminosity is expected to be close to the
Eddington limit. After the photosphere recedes to its pre-burst radius (called `touch-down')
cooling is typically observed. Such bursts are referred to as
photospheric radius-expansion type~I X-ray bursts
(e.g., Lewin et al.\ 1984, Tawara et al.\ 1984; hereafter radius-expansion bursts).

A few bursts last hours to half a day (Cornelisse et al.\ 2000, Strohmayer \&\ Brown
2002; for a review see, e.g., Kuulkers 2004).  Such bursts show the same
characteristics as the typical bursts described above,
except that the energy output is about a thousand times higher. 
They recur on time scales of months to years. Their origin is
different from the typical bursts, because they are thought to be
due to unstable burning of C deeper in the neutron star envelope
(Cumming \&\ Bildsten 2001, Strohmayer \&\ Brown 2002). These long events are called
superbursts (Wijnands 2001).  Typical bursts are seen
up to the time of the superbursts.  However, they
cease to occur after the superburst, for up to about a month (see, e.g., 
Cornelisse et al.\ 2002a, Kuulkers et al.\ 2002a).  This is thought to
be due to the superburst heating up the surface layer, and therefore
preventing the unstable ignition of H and/or He (Cumming \&\ Macbeth 2004).

There are also bursts that have durations and energy releases
intermediate between typical bursts and superbursts.
They are referred to as intermediate-duration X-ray bursts (Cumming et al.\ 2006; hereafter intermediate-duration bursts).
They show decay times ranging from several to tens of minutes and energy outputs
of about 10$^{41}$\,erg.  
Most of these events occur in sources with low persistent luminosity
($L_{\rm pers}$$\la$0.01$L_{\rm Edd}$, where $L_{\rm Edd}$ represents the Eddington luminosity) and are thought to be due to 
flashes of relatively thick He layers in ultra-compact X-ray binaries (UCXBs; 
i.e., low-mass X-ray binaries with an orbital period less than an hour), see
in 't Zand et al.\ (2005), Cumming et al.\ (2006), Falanga et al.\ (2008),
and references therein.
The intermediate-duration bursts in the H-rich and luminous system GX\,17+2 (Kuulkers et al.\ 2002b) 
probably have a different origin.

The low-mass X-ray binary 4U\,0614+091 has long been
known to be a source of bursts, but this identification
was based on the detections of only several of them, with relatively
poor, $\sim$1$\degr$, position determinations\footnote{Note that spatially the
  closest known X-ray burster to 4U\,0614+091 is MXB\,0513$-$40, with a distance of 51$\degr$.}, as well as coarse timing and/or X-ray
spectral information (Lewin 1976, Swank et al.\ 1978, Brandt et
al.\ 1992, 1993a,b, Brandt 1994, Brandt \&\ Lund 1995).
More recently, an hours-long flare was reported, which showed characteristics resembling
superbursts (Kuulkers 2005; see Sect.~\ref{superburst}).
Brandt et al.\ (1992) used the observation with {\it GRANAT}/WATCH
of a peak burst flux of $\sim$10\,Crab (6--20\,keV) to infer,
assuming this peak flux does not exceed the Eddington limit for a
1.4\,M$_{\odot}$ neutron star with a He-rich atmosphere, that the
source distance must be about 3.2\,kpc.  
4U\,0614+091 is, therefore, one of the closest to Earth of the known actively bursting systems
(see, e.g., Jonker \&\ Nelemans 2004, Galloway et al.\ 2008). 

In the last decade, {\it HETE-2} triggered thirteen times on
bursts from 4U\,0614+091. Moreover, three
strong bursts from 4U\,0614+091 triggered the $\gamma$-ray burst monitors on
board {\it INTEGRAL} (see Chelovekov et al.\ 2007) and {\it Swift}. 
Strohmayer et al.\ (2008) discovered burst oscillations near 415\,Hz in 
one of the brightest {\it Swift}/BAT bursts.

It is difficult to understand 4U\,0614+091 as the origin of bursts. Since it
is thought to be an UCXB (in 't Zand et al.\ 2007, Shahbaz et al.\ 2008, and 
references therein), it is supposed to be so compact that the donor star can 
only be a non-degenerate H-deficient star or a white dwarf. 
Evolutionary models suggest that the accreted material could be rich in C/O or He 
(Nelemans et al.\ 2009). However, no evidence of He (or H) 
has been found so far in optical spectra, with rather stringent upper
limits: He and/or H are at most present at the 10\%\ level (Werner et
al.\ 2006). These spectra suggest the donor star to be a C/O white dwarf
(Nelemans et al.\ 2003, 2006; Werner et al.\ 2006). 
This seems to be at odds (see Juett et al.\ 2001, Juett \&\ Chakrabarty 2003)
with the fact that we see bursts in 4U\,0614+091,
which are usually understood as being due to unstable ignition of He or H. 
Moreover, 4U\,0614+091 is an interesting case
among the sources showing superbursts.  Superbursts are expected to
occur only in systems with accretion rates in excess of about 10\%\ of the
Eddington rate, i.e., the rate which yields emission at the Eddington limit 
(Cumming \&\ Bildsten 2001, Strohmayer \&\ Brown 2002). 
However, the low persistent X-ray luminosity of 4U\,0614+091
indicates an accretion rate which is an order of magnitude lower
(e.g., Ford et al.\ 2000).

In this paper we report on the detection of bursts from
4U\,0614+091 made with various instruments.
In Sects.~\ref{our_obs} and \ref{spectra} we describe the instruments and the
data analyses of the bursts seen between 1992 and 2007, as well
as the analysis of the persistent emission. We complement this with information
extracted for the bursts seen before 1992 (Sect.~\ref{previous}).
An overview of all the typical and intermediate-duration bursts 
seen appears in Sect.~\ref{overview}. We collectively refer to both types of bursts 
as normal bursts. One of them is a radius-expansion
burst; it is discussed in Sect.~\ref{intermediate}. We then describe the results of 
our analysis of the superburst (Sect.~\ref{superburst}), our
burst oscillation search (Sect.~\ref{oscillation}), 
and the characteristics of the persistent emission of the source (Sect.~\ref{persistent}).
Finally, we show the long-term X-ray behaviour of 4U\,0614+091
(Sect.~\ref{longterm}) and close the paper with a discussion of
our results (Sect.~\ref{discussion}).

\section{Observations and data analysis}
\label{data_analysis}

Almost all the observations between 1992 and 2007 described in this paper were serendipitously obtained using six
different X-ray instruments on just as many space-borne observatories,
see Table~\ref{table_obslog}. Only the observations with the {\it RXTE}/PCA and {\it RXTE}/HEXTE
were purposely dedicated to 4U\,0614+091. In the next subsections we describe these
instruments and the analysis of the data in more detail. We usually refer to the instruments by acronyms only.
In Table~\ref{table_obslog} we also give the total time of exposure for
observations with 4U\,0614+091 in the field of view and the total number of bursts detected from 
this source in these observations.

\begin{table}
\caption{Overview of instruments that have detected bursts from 4U\,0614+091 between 1992 and 2007.}
\begin{tabular}{c@{\,}c@{\,}c@{}c@{}c@{}}
\hline
\multicolumn{1}{c}{Time span} &
\multicolumn{1}{c}{Satellite/Instrument} &
\multicolumn{1}{c}{$E_{\rm sens}$$^a$} & 
\multicolumn{1}{c}{$t_{\rm exp}$$^b$} & 
\multicolumn{1}{c}{$n$$^c$} \\
\multicolumn{1}{c}{} &
\multicolumn{1}{c}{} &
\multicolumn{1}{c}{(keV)} &
\multicolumn{1}{c}{(days)} &
\multicolumn{1}{c}{} \\
\hline
Aug 1992 -- Jun 1993 & {\it EURECA}/WATCH & 6--150 & $\sim$50 & 3 \\
Dec 1995 -- Aug 2007 & {\it RXTE}/ASM & 1.5--12 & $\simeq$58 & 7 \\ 
Dec 1995 -- Aug 2007 & {\it RXTE}/PCA & 2--60 & $\simeq$23 & 2 \\ 
Dec 1995 -- Aug 2007 & {\it RXTE}/HEXTE & 15--250 & $\simeq$23 & 2 \\ 
Apr 1996 -- Apr 2002 & {\it BeppoSAX}/WFC & 2--28 & $\simeq$27 & 1 \\
Oct 2000 -- Mar 2007 & {\it HETE-2}/FREGATE & 6--400 & $\sim$194 & 13 \\
Oct 2002 -- Aug 2007 & {\it INTEGRAL}/ISGRI & 15--1000 & $\simeq$22 & 2 \\
Oct 2002 -- Aug 2007 & {\it INTEGRAL}/JEM-X & 3--35 & $\simeq$0.8 & 0 \\
Nov 2004 -- Aug 2007 & {\it Swift}/BAT & 15--150 & $\simeq$26$^d$ & 2 \\
\hline
\multicolumn{5}{l}{\footnotesize $^a$\,Nominal sensitive energy range.} \\
\multicolumn{5}{l}{\footnotesize $^b$\,Total observation net exposure time.} \\
\multicolumn{5}{l}{\footnotesize $^c$\,Number of bursts seen.} \\
\multicolumn{5}{l}{\footnotesize $^d$\,Total effective exposure time (corrected for off-axis response).} \\
\end{tabular}
\label{table_obslog}
\end{table} 

\subsection{Observations and instrument-specific data analysis issues}
\label{our_obs}

\subsubsection{{\it EURECA} WATCH}
\label{eureca}

{\it EURECA} ({\bf EU}ropean {\bf RE}trievable {\bf CA}rrier)
carried the {\bf W}ide {\bf A}ngle {\bf T}elescope for {\bf C}osmic 
{\bf H}ard X-rays (WATCH; Lund 1985, Brandt et al.\ 1990).
WATCH operated from August 1992 to June 1993. 
It comprised a rotating modulation collimator (RMC) and yielded
images through a cross-correlation method.  It viewed a circular
field with a radius of about 65$\degr$, was sensitive 
between 6 and 150\,keV, and had an effective area of
about 45\,cm$^2$.  For events with a duration longer than the
rotation period of the RMC ($\simeq$1\,s), the relative position
accuracy was better than 1$\degr$, and, in favourable cases, was as
good as about 0.1$\degr$.  The flux sensitivity was about 100\,mCrab in
one day.

4U\,0614+091 was located in the $\simeq$1 steradian field
of view from mid January 1993 until the end of April 1993, with an
observing efficiency of a little more than 40\%.
{\it EURECA}/WATCH observed a total of three bursts 
during that time. Preliminary
reports of these events can be found in Brandt et al.\ (1993a,b),
Brandt (1994) and Brandt \&\ Lund (1995).

For our analysis, we used count rates as a function of time
along with vignetting estimates for the source.  
Background rates were determined by polynomial fits to the count rates outside
the intervals of each burst. We define the 3$\sigma$ confidence level above which
we regard a signal as significant as 3 times the square
root of the number of background counts in each time bin.  

\subsubsection{{\it RXTE} All-Sky Monitor}

\paragraph{2.1.2.1 Scanning Shadow Cameras.}
The {\bf R}ossi {\bf X}-ray {\bf T}iming {\bf E}xplorer ({\it RXTE};
launched December 1995) {\bf A}ll-{\bf S}ky {\bf M}onitor (ASM)
consists of three {\bf S}canning {\bf S}hadow {\bf C}ameras (SSCs,
hereafter called SSC0, SSC1 and SSC2; Levine et al.\ 1996). 
Each SSC views a 12$\degr$$\times$110$\degr$ (full-width at zero
response, FWZR) field through a random-slit coded mask.  
The field of view of one SSC is $90\degr$ from the colocated field centres of the other
two SSCs. The assembly holding the three SSCs is generally held
stationary for a 90\,s ``dwell''.  A drive then rotates this
assembly through 6$\degr$ between dwells. This yields good
sky coverage: as much as 80\%\ of the X-ray sky is covered
every 90\,min orbit around the Earth.

For each dwell, histograms of counts (`raw' counts, i.e., 
not corrected for transmission, sources
in the field of view, background, etc.) 
as a function of position in
each detector are recorded for three energy bands (roughly 1.5--3,
3--5, 5--12\,keV). Also, the total number of counts registered in
each SSC are recorded in 1/8\,s time-series bins in the same
energy bands.  For observations done before March 2001 (MJD
51970), there is imaging data integrated over entire dwells, and
count rate data with no imaging information.  Since March 2001
event-by-event data are telemetered so that both temporal and
imaging information can be extracted for portions of dwells.  For
sources away from bright sources, the sensitivity is roughly
10--15\,mCrab ($1\sigma$ in the 1.5--12 keV band) for a reliable
single-dwell source detection; multi-day averaging improves the
sensitivity down to about a few mCrab.  The average intensity of
each known source in the field of view is determined by the instrument
team through an analysis of the coded-aperture data and is made
available via the internet, on a dwell by dwell basis as well as 1-day
averages.\footnote{{\tt \tiny http://heasarc.gsfc.nasa.gov/docs/xte/asm\_products.html}
{\tiny and} {\tt \tiny http://xte.mit.edu/ASM\_lc.html}.}

The intensity history of 4U\,0614+091 up to September 2007 was obtained from each of 62720
dwells, more or less evenly spread over the years (see also Sect.~\ref{ASMlongterm}). We used the following
method to search for bursts: 1) identify
candidate dwells in which the source intensity minus its
uncertainty was larger than the overall mean intensity of
4U\,0614+091 by 4 times the rms uncertainty in the overall mean
intensity; 2) visually inspect raw count rates at 1-s time resolution
within the candidate dwells for evidence of burst-like
temporal behaviour; 3) use the hardness ratios, defined as the
time-bin-wise ratios of the count rates in the 5--12\,keV band to
those in the 1.5--5\,keV band, to verify the typical hard-to-soft behaviour of bursts.
Note that this technique is not suitable for bursts that last substantially longer
than a dwell or for bursts that do not start within a dwell. Our
search yielded 6 bursts and one superburst.

\paragraph{2.1.2.2 (Time-resolved) X-ray spectral analysis of ASM data.}
The 3-channel ASM data can constrain simple spectral models
like power laws and black-body radiation (see, e.g., Ford et al.\ 1996,
Kuulkers 2002, Keek et al.\ 2008). In our spectral analyses we
employed the same method as in Keek et al.\ (2008). 
This includes calibrations of the source flux with average ASM data of the Crab
over 200-day intervals around each data point.

The spectral analysis can also be satisfactorily applied to raw
ASM data when dealing with bursts.  If one assumes that the
burst emission is not influenced by persistent emission (see,
e.g., Kuulkers et al.\ 2002b), net-burst count rates can be obtained by
subtracting the average raw pre-burst count rates from the raw burst count rates.
If one also knows the celestial position
of the burst, the effective mask/collimator transmission can be
computed.  This allows the estimation of equivalent on-axis net
burst count rates for the given energy bands and these, after
renormalization with the Crab as described above, can be fit with
a black-body spectrum model.  Whenever a burst was simultaneously observed by two
SSCs we combined the renormalized net count rate information from
the two cameras.  We derived the three-channel transmission-corrected
net-burst spectra from the 1-s raw count rates. We calculated the errors 
by taking square roots of the numbers of counts per each time bin.  
We started with a spectral time bin size of 1\,s
(2S\,0918$-$549, see below) or 2\,s (4U\,0614+091). Each
time the net-burst rates following the peak decreased by a factor
of $\sqrt{2}$, we doubled the spectral time-bin size.

Since all the corrections may introduce errors in our analysis of the raw ASM data, 
we first applied our method to bursts from 2S\,0918$-$549 that, like
4U\,0614+091, lies in a relatively empty field in the sky (spatially the
closest bursters to 2S\,0918$-$549 are EXO\,0748$-$676 at
$\simeq$16$\degr$ away and the transient GS\,0836$-$429 at
$\simeq$14$\degr$ away).  2S\,0918$-$549 is similar to 4U\,0614+091
in that it is also thought to be an UCXB which shows low persistent accretion and infrequent
bursts (see, e.g., in 't Zand et al.\ 2005).  
Most of the 2S\,0918$-$549 bursts show a radius-expansion phase.  By comparing the ASM time-resolved X-ray
burst spectral analysis to that derived from high-quality X-ray
spectra observed with well-calibrated detectors, one can verify our
ASM burst-spectral analysis method.  

We searched the ASM data on 2S\,0918$-$549 for bursts using the same algorithm
as that used for 4U\,0614+091. We found 5 bursts (see Table~\ref{table_0614}) 
over the more than $\simeq$11 year time span of observations, with a net
effective exposure time of 47.8~days between 1996 January and
2007 August. Three of them were already noted by in 't Zand et al.\ (2005); our third X-ray
burst, which occurred on UT 2002 August 23, was not noted by them
probably due of its weakness (because of relatively low
transmission, see Table~\ref{table_0614}).
By comparing our ASM time-resolved spectral fit results
with those for the bursts seen with the WFC (see in 't Zand et al.\ 2005) and
PCA (see Galloway et al.\ 2008), we find comparable values for all spectral parameters.
The ASM spectral fits show similar kind of radius expansion phases which reach similar derived peak fluxes
as the other instruments. We, therefore, conclude that our time
resolved X-ray spectral analysis of the raw ASM data gives consistent
results with that seen from other instruments, and therefore our
method can be trusted.

\subsubsection{{\it RXTE} PCA and HEXTE}
\label{pca}

The {\bf P}roportional {\bf C}ounter {\bf A}rray (PCA; Bradt et
al.\ 1993, Jahoda et al.\ 2006) onboard {\it RXTE} provides a
large collecting area (maximum net geometric area of about 8000\,cm$^2$) and high
time resolution (down to $\mu$s). It consists of 5 proportional
counter units (PCUs) behind 1$\degr$ (full-width at half maximum,
FWHM) collimators.  It is sensitive in the 2--60\,keV range 
down to $\simeq$0.2\,mCrab.

The {\bf H}igh-{\bf E}nergy {\bf X}-ray {\bf T}iming {\bf E}xperiment 
(15--250\,keV; Rothschild et al.\ 1998) onboard the same satellite
consists of 8 detectors with a total area of about
1600\,cm$^2$. The 8 detectors are split up in two clusters
of 4 each. During normal operations, each cluster is alternately
pointed on and off the source generally every 16 or 32\,s, to
provide near real-time background measurements (note that at the
end of 2006 cluster A was fixed to always view the source). HEXTE
can measure a typical X-ray source down (at the 3$\sigma$ level)
to about 1\,mCrab up to $\simeq$100\,keV in 10$^5$\,s.

We first inspected the light curves at 1\,s time resolution of all the publicly available PCA data 
up to September 2007 on 4U\,0614+091 that was collected in the Standard 1 mode. 
A couple of strong burst-like events were seen; 
all except one were due to detector-related events. The one celestial event 
occurred simultaneously with the burst seen by FREGATE,
on UT 2001 February 4 (MJD\,51944; see Sect.~\ref{fregate}). Only the
first 30\,s of the event were covered, after which the PCA was
shut off because the count rate ($>$55\,kct\,s$^{-1}$) exceeded
the `High Rate Monitor' safety threshold.  
Apart from the first 30 seconds of the burst, the
PCA also observed the tail of the burst. In observations taken on UT 2000 September 2 (MJD\,51789) we found the tail 
of another burst for which the onset was missed.

We used version 6.5 of the HEASOFT software suite for our
{\it RXTE} data analysis.  For the time-resolved analysis of the
prompt burst on MJD\,51944 we used the PCA data from either Event
mode or Burst Catcher mode when available (the high count rates
reached during the maximum of the burst resulted in data losses
near the ends of the 1\,s buffers of the Event mode).  These modes
provided numbers of counts in 64 channels covering the PCA energy
range at time resolutions of 122\,$\mu$s and 8\,ms, respectively.
At the time the prompt burst occurred, 4 PCUs (PCU 0--3) were on;
the data from all layers and all PCUs were automatically combined
in these modes.  During the spike (see Sect.~\ref{intermediate}) we
created time-resolved spectra at 2\,ms resolution.  During the
rest of the prompt burst, we used a time resolution of 0.125\,s.  
Although the majority of the prompt burst was observed by
HEXTE, we do not use the HEXTE data for the burst
spectral analysis because data from a broader energy range
(7--400\,keV) was obtained by FREGATE (see Sect.~\ref{fregate}). 
For the tail of the bursts on MJD\,51789 and 
MJD\,51944 we used the PCA data from the Standard~2
mode; this mode provides counts in 129 channels covering the PCA
energy range with a time resolution of 16\,s.  We formed spectra
averaged over 112\,s intervals from the events occurring in all layers
of the PCUs which were on during and before or
after the burst; this was PCU 3 on MJD\,51944 and PCUs 2,3 on
MJD\,51789.  

Dead-time correction is only possible for all spectra
with a time resolution equal or higher than 0.125\,s.  We subtracted the
pre-burst persistent emission from the burst emission. 
Our time-resolved burst spectral fits of PCA data covered
the 3 to 20\,keV energy band.  We included a 1\%\ systematic
uncertainty in each spectral bin in addition to the usual statistical
uncertainties.

We extracted information on the persistent emission only when {\it RXTE} 
data was available within 1~day of a burst (see Sect.~\ref{persistent}). We
did this for data from both the PCA and HEXTE by applying standard
criteria, i.e., by filtering out data taken at elevations less than
10$\degr$ or with an offset from the source greater than 0.02$\degr$.
Also, we created instrument response and background files following the standard
analysis threads using the latest information 
available.\footnote{{\tt \tiny http://heasarc.nasa.gov/docs/xte/data\_analysis.html}.}
We corrected the count-rate spectra for dead-time losses.
For the PCA we only used PCU\,2 data (across the {\it RXTE} mission this PCU has the highest duty cycle)
obtained with the Standard~2 mode. Pulse-height spectra were formed
from the events from all layers and we included a 1\%\ systematic uncertainty in each spectral bin.
We performed fits to the persistent emission in the 3--30\,keV and
17--100\,keV bands for the PCA and HEXTE, respectively. A
multiplicative constant was applied to the HEXTE spectra to account
for the uncertainty in the relative normalizations of the instruments,
i.e., both HEXTE Cluster A and B were allowed to vary with respect to
the PCA. This multiplicative factor was found to be between 0.6 and 0.8.

For the comparison (see Sect.~\ref{intermediate}) of the burst observed on MJD\,51944 with that of the superbursts
of 4U\,1820$-$303 (Strohmayer \&\ Brown 2002) and 4U\,1636$-$536 (Strohmayer \&\ Markwardt 2002)
at hard X-ray energies, we extracted light curves from data obtained by HEXTE
in the 15--60\,keV band.  In this process we applied the same standard filtering
criteria as those described above.
We corrected the light curves for dead time and background following the standard 
procedures\footnote{{\tt \tiny http://heasarc.gsfc.nasa.gov/docs/xte/recipes/hexte.html}.}, and normalized them to the Crab
count rate in the same energy band. For this analysis we only used the 
Standard Modes (Archive Spectral Bin, 64-bin spectra produced every 16\,s) data from Cluster A.

To characterise the overall spectral behaviour of 4U\,0614+091 we created so-called 
colour-colour and hardness-intensity diagrams (CD and HID, respectively). All 
publicly available pointed PCA observations of 4U\,0614+091
up to September 2007 were used. We extracted background and dead-time corrected count rates from the Standard 2 
data using the same standard procedures and filtering criteria as applied to the spectral data, 
with 16\,s time resolution. We used the following energy bands: 
2.0--3.5\,keV (A), 3.5--6.0\,keV (B), 6.0--9.7\,keV (C), and 9.7--16.0\,keV (D).
We define the soft and hard colours (SC, HC) as the ratios of the count rates in the various bands: 
SC=B/A and HC=D/C; the intensity (Int) is defined as the sum of
the count rates in the four energy bands: Int=A+B+C+D. We normalized both the colours and intensity
to the Crab values nearest in time to account for changes in detector gas gain
(see, e.g., Kuulkers et al.\ 1994, van Straaten et al.\ 2003).
Since the values do not change significantly within an observation (i.e., on time scales of 
order an hour), they were averaged over one observation.

\subsubsection{{\it BeppoSAX} Wide-Field Cameras}

The {\bf W}ide-{\bf F}ield {\bf C}ameras (WFCs; Jager et al.\ 1997)
were two identical coded-aperture instruments onboard the {\it BeppoSAX}
satellite ({\bf S}atellite per {\bf A}stronomia {\bf X}; Boella et
al.\ 1997), which were operated between April 1996 and April 2002.
The field of view was 40$\degr$$\times$40$\degr$
(FWZR), the angular resolution 5$\arcmin$
(FWHM) and the source-location accuracy was generally better than
1$\arcmin$ (99\%\ confidence).  The detectors were sensitive to the
energy range 2 to 28\,keV
and had a net collecting area of 140\,cm$^2$. The on-axis detection
threshold was of the order of 0.3 Crab for a 1\,s observation and
a few mCrab for a 10$^5$\,s observation.

The WFCs pointed in opposite directions with respect to each other and
perpendicular to the {\bf N}arrow-{\bf F}ield {\bf I}nstruments (NFIs) on the same
satellite.  Since the pointing directions
of the WFCs were usually governed by the observations of the NFIs,
the WFC sky coverage was not uniform.
4U\,0614+091 was mostly seen at large
off-axis angles.

We extracted the reconstructed source flux in the 2--25\,keV
bandpass with a time resolution of 2\,s whenever 4U\,0614+091 was in
the field of view. We also extracted information on all photons
detected on the appropriate detector with varying time resolutions
between 0.5 and 8\,s. Light curves were formed from both the fluxes
and the photon rate data. Bursts were searched for by eye as well
using an automatic algorithm (see Cornelisse et al.\ 2003). We found
only one burst.

For the generation of WFC X-ray spectra we first
cross-correlated the detector data with
the expected imaging response of WFC unit 2 for 4U\,0614+091
(see in 't Zand 1992 for more detail). Background radiation is
then automatically subtracted. We then extracted 2--28\,keV spectra from the
resulting imaging data. Burst time bins were defined such that
the significance of the source (i.e., the photon flux divided by the
standard deviation expected from all other photon sources) is at least
10. Experience shows that spectra are ill defined for lower
significances. The X-ray spectrum of the persistent emission around the burst was determined 
from the whole observation in which the burst occurred (with an exposure
time of 14.3\,ksec). This spectrum was subtracted from the burst emission in
our time-resolved burst spectral analysis.
The burst and persistent spectra were fit in the 2--28\,keV band.

\subsubsection{{\it HETE-2} FREGATE}
\label{fregate}

One of the instruments onboard the {\bf H}igh {\bf E}nergy {\bf T}ransient {\bf E}xplorer
satellite ({\it HETE-2}; Ricker et al.\ 2003) that operated between 2000 and 2007,
was the omnidirectional $\gamma$-ray
spectrometer named {\bf Fre}nch {\bf Ga}mma {\bf Te}lescope (FREGATE;
Atteia et al.\ 2003). FREGATE consisted of 4 detectors, mounted in two pairs.
It was sensitive in the 6--400\,keV range, with a maximum effective area of 158\,cm$^2$.
The field of view was 70$\degr$ (half-width at zero response, HWZR), but it had no imaging
capacities. It provided continuous 128 channel energy spectra with 5\,s time resolution
(except in the beginning of the mission when it was 10\,s), as well as
continuous four-channel energy spectra with 0.164\,s time resolution
(0.327\,s in the beginning). 

A burst trigger occurred when there were two coincident 6$\sigma$ or higher excesses on two of the 
FREGATE detectors within a time bin; this corresponds to an increase by 1.2~Crab or higher in 
5.24\,sec for a burst observed on-axis in a source-free region.
When this happened, 256000 individual photons were time- and energy-tagged.
FREGATE triggered 13 times on bursts coming from the direction of 4U\,0614+091
(see Table~\ref{tableburstproperties}).  Three of them
were weak and exhibited low signal to noise (on MJD\,52961, MJD\,53074 and MJD\,53740).
The first two FREGATE events (MJD\,51944 and MJD\,52322) 
were already reported by Barraud (2002); they occurred when the 
{\bf W}ide Field {\bf X}-ray {\bf M}onitor (WXM, imager onboard {\it HETE-2}, 
2--25\,keV; Shirasaki et al.\ 2003) was not operating.
The remaining bursts were detected by the WXM, and some of these were also detected by the 
{\bf S}oft {\bf X}-ray {\bf C}amera (SXC, CCD-based imager onboard {\it HETE-2},
0.5--2\,keV; Villasenor et al.\ 2003).
The WXM and SXC, when operating, were able to get
precise burst localizations (see, e.g., Suzuki et al.\ 2004).
Two additional bursts were seen by the WXM, but FREGATE was not operating at those times
(MJD\,53041 and MJD\,54101).  
Due to the anti-solar pointing of HETE-2, 4U\,0614+091 was within the
field-of-view of FREGATE during nearly 4 months every year.
We estimate the total exposure on 4U\,0614+091 from 2001 to 2006 to be about 194~days.

The extraction of burst spectra relies on a quadratic fit of the `background' (i.e., true background 
plus any emission from sources in the field of view) during tens of seconds before and after
a burst in all the 128 energy channels. The availability of this simple
background spectrum allows for the subtraction of the background channel by channel.
We used the counts above the fitted background to build the count spectrum of the burst, which we then 
deconvolved using the response matrix constructed with the known gain and angular response of the detector.
We performed the X-ray spectral fits to the integrated and time-resolved 
(10\,s resolution for the burst on MJD\,51944, 5\,s for the rest) burst emission in the well-calibrated 7--40\,keV range.
For several of the bursts we also extracted spectral information at a higher time resolution in a similar way.

\subsubsection{{\it INTEGRAL} IBIS/ISGRI and JEM-X}

One of the two main instruments onboard {\it INTEGRAL} 
({\bf Inte}rnational {\bf G}amma-{\bf R}ay {\bf A}strophysics 
{\bf L}aboratory; Winkler et al.\ 2003; launched October 2002) is IBIS
({\bf I}mager on {\bf B}oard the {\bf I}NTEGRAL {\bf S}atellite;
Ubertini et al.\ 2003). It comprises two detector planes that view
the sky through a coded mask.  The field of view is
29$\degr$$\times$29$\degr$ (FWZR) and the angular resolution is
12$\arcmin$ (FWHM).  We use data collected with one of its 
detectors: the {\bf I}NTEGRAL {\bf S}oft {\bf G}amma-{\bf R}ay 
{\bf I}mager (ISGRI) which is sensitive in the $\simeq$15\,keV to 1\,MeV
range with a total effective area of about 2600\,cm$^2$ (Lebrun et al.\ 2003).

One of the two supplementing monitors onboard {\it INTEGRAL} is the
{\bf J}oint {\bf E}uropean X-ray {\bf M}onitor (JEM-X; Lund et al.\ 2003).
JEM-X consists of 2 identical units, which are both sensitive in the 3--35\,keV band.
Most of the time only one unit is operating.
The angular resolution is 3$\arcmin$ (FWHM); one single unit has a detector area of about 500\,cm$^2$.
The units have a circular view with diameter of about 13$\degr$ (FWZR), i.e., narrower than ISGRI.
In practice, the transmission of the collimator beyond an off-axis angle of 5$\degr$ is so 
low that only the brightest sources can be observed at larger angles.
JEM-X is sensitive to X-ray bursts seen on-axis down to about 0.1\,Crab at the 5$\sigma$ level in 5\,s.

The {\bf I}NTEGRAL {\bf B}urst {\bf A}lert {\bf S}ystem 
(IBAS; Mereghetti et al.\ 2003)\footnote{{\tt \tiny http://ibas.iasf-milano.inaf.it/}.} is the
automatic on-ground software that in near-real time
searches for $\gamma$-ray bursts in IBIS data and promptly publicly
distributes results.
For a trigger time interval of 1\,s and a certain
threshold value (currently about 8$\sigma$), a minimum flux of about
0.5-0.75 Crab (20--200\,keV) is required to trigger a typical
$\gamma$-ray burst (and to produce enough counts to locate the position
in the deconvolved image). One burst from
4U\,0614+091 triggered IBAS on UT 2005 March 31 (nr.~2441 on MJD\,53460).  

To search offline for bursts and to
study the hard X-ray long-term persistent behaviour of
4U\,0614+091, we analysed all public {\it INTEGRAL} pointing data 
in the ISOC Science Data Archive.\footnote{{\tt \tiny http://integral.esac.esa.int/isda/}.}  
Up to September 2007 there
were $\simeq$840 pointings available where 4U\,0614+091 was within the ISGRI
field of view with a total exposure time of about 25 days. Since most of these observations were
taken with the prime objective of instrument calibration on the Crab,
many pointings (about 200) were carried out in non-standard modes.  The
total exposure time of all ISGRI standard pointings is about 22 days.
All public JEM-X data whenever 4U\,0614+091 was in the field of view (i.e., $<$6$\degr$
off-axis) amounts to a total exposure of only about 70\,ksec.
In our analysis we use the high-energy source
ISDC reference catalog (see Ebisawa et al.\ 2003; v.~27) as input
source catalog. We processed the ISGRI and JEM-X data using the latest
available Off-line Scientific Analysis software ({\sc OSA}; see
Courvoisier et al.\ 2003), v.~7.0.  The description of the algorithms used in the
ISGRI and JEM-X scientific analyses can be found in Goldwurm et
al.\ (2003) and Westergaard et al.\ (2003), respectively.

We 1) processed the ISGRI 15--20\,keV
band data using {\sc OSA} with default parameters, 2) analysed 
the data through to the imaging step, and 3) extracted light curves for
all detected sources with 10\,s time resolution.\footnote{{\tt \tiny http://www.isdc.unige.ch/integral/download/osa\_doc}.}
Next, a potential onset of a burst was flagged when, in a time
bin, the difference between the source count rate and the average
source count rate in the whole pointing exceeded four times the
standard deviation of the count rates in the whole pointing. The
count rate versus time was then examined around each flagged time to
check for the presence of a shape consistent with that of a type~I
X-ray burst.  If this was the case, we generated reconstructed images
within the good-time interval covering the whole burst, and checked
them visually to verify that the event indeed originated from
4U\,0614+091.

Using the above described procedure we confirmed the burst which triggered IBAS
in 2005 (see above). We also detected another burst from
4U\,0614+091 on UT 2003 August 16 (MJD\,52867). Unfortunately, the latter burst
was seen far off-axis, about 16.5$\degr$, near the edge of the detector (see
Fig.~\ref{ISGRI_ima_lc}), and thus again outside the field of view
of JEM-X. We found no bursts from 4U\,0614+091 in the JEM-X data.

For the time-resolved spectral analysis of the 2005 burst we divided the burst only in three parts so that the
rise, the top and the decay were covered with satisfactorily spectral quality, in 5 spectral energy bins
covering 17 to 40\,keV.
The statistical quality of the 2003 burst precludes a time-resolved analysis. 
The ISGRI spectrum of the persistent emission around the 2005 burst
is derived from the 1800\,s single pointing during which the burst occurred. 
The net-burst spectra were calculated by subtracting the persistent spectrum
from the burst spectra. X-ray spectral fits to the burst and persistent ISGRI spectra were
performed in the 17--40\,keV and 20--200\,keV bands, respectively.
During the INTEGRAL observations around the 2003 burst 4U\,0614+091
was always far off-axis ($>$10$\degr$), and therefore we did not
include these ISGRI data in our analysis of the persistent emission.

For the long-term light curves we used {\sc OSA} through the
production of ISGRI images per single pointing in the 15--50\,keV range.
This energy band was chosen to supplement the ISGRI light curves with the publicly 
available BAT long-term light curves (see Sect.~\ref{swift}). 
We force the flux extraction of each of the catalog
sources, regardless of the detection significance of the source (see,
e.g., Kuulkers et al.\ 2007).  The correction for off-axis response
is generally good up to $\simeq$10$\degr$ from the centre of the
field of view. We therefore selected only those pointings where
4U\,0614+091 was less than 10$\degr$ off axis.  This led to a
total of 115 pointings with different exposure times spanning the
time interval from 2003 February 18 to 2006 April 18.  

\subsubsection{{\it Swift} BAT}
\label{swift}

The {\bf B}urst {\bf A}lert {\bf T}elescope (BAT; Barthelmy et
al.\ 2005) onboard {\it Swift} (launched November 2004; Gehrels et
al. 2004) is a coded-aperture imager with a very wide field of view of
about 2 steradians, which operates in the 15--150\,keV band. The
detector plane covers a net collecting area of 5200\,cm$^2$;
the BAT angular resolution is 22$\arcmin$ (FWHM).

As soon as the BAT triggers on a $\gamma$-ray burst, {\it Swift} automatically slews to
the $\gamma$-ray burst position and starts to observe the source 
with its more sensitive instruments. However, strong type~I X-ray bursts may
also trigger the BAT (but do not lead to automatic slews if
they come from a known source). Assuming a black-body source with $kT$$\simeq$2.5\,keV, 
the 5$\sigma$ detection limit for the BAT in 5\,s is approximately 
0.7\,Crab in the 15--25\,keV band.

Two bursts from 4U\,0614+091 triggered the BAT 
into a $\gamma$-ray burst data collecting mode but did not result in automated slews.
The triggers occurred on UT 2006 October 21 (nr.~234849 on MJD\,54029) and 2007 March 31 (nr.~273106 on MJD\,54189).  
For these triggers, about 45\,s
of BAT event data were produced.  A first account of these bursts has been
given by Strohmayer et al.\ (2008).
We ran the standard complete $\gamma$-ray burst processing script 
on the BAT burst products of the two triggers to obtain a first impression
of the burst behaviour, and to derive an updated position of the
origin of the burst (see Appendix~\ref{appendix}).  We then applied the latest calibration available
(as of August 2007) to these data (energy calibration, detector quality
map, mask weighting), following the analysis threads provided by the
{\it Swift} Science Center\footnote{{\tt \tiny http://swift.gsfc.nasa.gov/docs/swift/analysis/}.}.  
From these calibrated data we produced light curves, images and spectra.
The first burst is strong enough to support a detailed time-resolved
spectral analysis in the 15--30\,keV band, starting with a time
resolution of 1\,s. For the second, weaker burst we started with a
time resolution of 2\,s.  We doubled the spectral time bin
size, whenever the net-burst rates following the peak decreased by a
factor of $\sqrt{2}$.

The BAT continually monitors the sky; more than about 70\%\ of the sky
is observed on a daily basis. Results from this survey mode are publicly
available in the form of light curves covering the 15--50\,keV energy
band on two time scales: a single Swift pointing ($\simeq$20\,min) and
the weighted average for each day.\footnote{{\tt \tiny http://swift.gsfc.nasa.gov/docs/swift/results/transients/\ index.html}.}
In the latter a 6\,mCrab source
typically can be detected at the 3$\sigma$ level (Krimm et al.\ 2006).
We used the daily average light curve for 4U\,0614+091 to study its long-term hard
X-ray behaviour.  

\subsection{Further data analysis issues}
\label{spectra}

\subsubsection{X-ray burst properties}
\label{xrbprop}

The band passes we used for the investigation of the burst light curves and
extraction of some of the burst properties (see below)
are 6--15\,keV, 2--28\,keV, 1.5--12\,keV, 2--60\,keV, 7--40\,keV, and
15--30\,keV for the {\it EURECA}/WATCH, WFC, ASM, PCA, FREGATE, and both the ISGRI and 
BAT, respectively. 

For each of the ASM bursts, we derived the 
start time by taking the time when the count rate rose more
than 5$\sigma$ above the pre-burst level, while, for each
FREGATE burst, the start time is the actual
trigger time. For the other bursts we determined the start times 
by eye, because most of these bursts rise more slowly
at higher energies.  We define the burst rise time, $t_{\rm rise}$,
as the difference between the time of the start and the
time of the peak of the burst; the derived values were verified 
by visual inspection of the light curves.
We estimate the error on the rise time to be about half a second (except
during the spike of the burst observed on MJD\,51944, see
Sect.~\ref{intermediate}). The decay times, $t_{\rm decay}$, 
are derived from fits
of an exponential function plus a constant to the light curves. 
We fit from the maximum of the burst up to about 100\,s after
burst onset, if possible. For the longer bursts a longer
time base was used when possible. When no post-burst information is
available, we first fit the pre-burst rate with a constant and fix
that when fitting the decay portion of the burst light curve.  
The range of values for which $\chi^2$ remains within 1 of the
minimum value (i.e., $\Delta\chi^2$=1) is used to set the uncertainties of the decay times
(assuming the fit parameters are uncorrelated).

Further burst properties include the bolometric fluence ($E_{\rm b}$) and bolometric black-body
peak flux ($F_{\rm peak}$), where the underlying assumption is that the burst
emission is on top of unchanged persistent emission
(see, e.g., Kuulkers et al.\ 2002b), as well as information on the out-of-burst,
i.e., persistent, emission (see below). Estimates of
$E_{\rm b}$ were obtained by integrating the fitted black-body models of the
time-resolved spectra over photon energy and time (i.e., we do not
take into account any residual burst emission outside the time range in which we did the 
time-resolved spectral fits). Since
bursts decay more rapidly at higher energies, i.e., $t_{\rm decay}$
depends on the bandpass used, we also determine the characteristic
decay time $\tau$=$E_{\rm b}/F_{\rm peak}$ which is not dependent on
bandpass. Uncertainties in
$E_b$ and $\tau$ were estimated by propagating the errors derived from the individual
time-resolved burst spectral fits and by assuming that these errors
are symmetric. For the WATCH bursts and the 3 weak bursts observed by FREGATE, 
time-resolved X-ray spectral analysis is not possible.  Instead,
we derive $F_{\rm peak}$ from the peak count rates by scaling
from the count rates and fluxes of the Crab.  The Crab flux
is taken to be 1.03$\times$10$^{-8}$\,erg\,cm$^{-2}$\,s$^{-1}$ in
the 6--15\,keV WATCH band and 1.89$\times$10$^{-8}$\,erg\,cm$^{-2}$\,s$^{-1}$ in the
7--40\,keV FREGATE band.  We then converted these values 
to estimated bolometric values assuming the emission
comes from a 3\,keV black body (which is fine for a large part of a
burst seen at hard energies, see, e.g., Fig.~1). For the 3 weak FREGATE bursts we derived
E$_{\rm b}$ from the average burst-integrated spectra; 
for the WATCH bursts we derived $E_{\rm b}$ from the burst integrated number of counts
and by performing a correction similar to that described above for $F_{\rm peak}$.
We define the duration of the burst ($t_{\rm dur}$) as
$t_{\rm rise}$+$2\tau$, since not all bursts are fully
covered. Inspection of the light curves show that these values are
consistent with the total durations determined by eye. 

Burst oscillations at 414.7\,Hz were  present during a 5\,s time
interval in the brightest BAT burst cooling tail at the 4$\sigma$
significance level (Strohmayer et al.\ 2008).  We verified the
presence of burst oscillations during this burst and searched for
burst oscillations during the FREGATE and the 2005 ISGRI  bursts where
data were available at a high time resolution.  We selected the
13--20\,keV band for BAT, 15--30\,keV for  ISGRI and 7--40\,keV for
FREGATE, mainly to have the highest possible signal-to-noise ratio to
search for a timing signature.  We applied the Z$^{2}_{1}$-statistic
(standard Rayleigh statistic, see Buccheri et al.\ 1983) to the photon
event distributions for trial frequencies in a narrow window centred
on the expected frequency of 414.7\,Hz (i.e., 413--416\,Hz). The dynamical
power spectra were computed using the same method as in Strohmayer et al.\
(2008). We employed intervals of 4\,s or 8\,s, stepping through the
BAT, FREGATE and ISGRI bursts with intervals of 0.25\,s; for the calculation
of upper limits we used intervals of 10\,s.

We searched data from both the PCA and HEXTE for  oscillations during
the prompt burst seen on MJD\,51944 using fast Fourier transforms (FFTs).  We used the single
bit mode data available  (full 2--60\,keV PCA energy range taken at 
122\,$\mu$s time resolution).
FFTs of data from 4\,s-long intervals within the burst were performed keeping the original time resolution.
For HEXTE we computed FFTs of 1\,s-long sets of 15--40\,keV band data with 
122\,$\mu$s time resolution.
For the oscillation search in the long-lasting faint tails 
we employed the PCA data only.

\subsubsection{X-ray spectral fits}
\label{xray_spectra}

In all our spectral fits (using {\sc XSPEC 11.3.0}) we fix the
low-energy absorption equivalent to a hydrogen column density of
$N_{\rm H}$=3$\times$10$^{21}$\,cm$^{-2}$ (e.g., Piraino et al.\ 1999;
M\'endez et al.\ 2002) for a composition of the absorbing matter as
given by Wilms et al.\ (2000), employing absorption cross sections
as provided by Verner et al.\ (1996). We derived errors in the fit
parameters by using $\Delta\chi^2=1$ (assuming the fit parameters are uncorrelated).  
We use a distance of 3\,kpc (see Sect.~\ref{intermediate}) when reporting black-body radii or emission areas, 
X-ray luminosities and absolute fluences.

To describe the persistent emission we used ASM
spectra around the ASM, FREGATE, 2003 ISGRI and 
BAT bursts, ASM plus 
ISGRI spectra around the 2005 ISGRI
burst and WFC spectra around the 
WFC burst.  We also used PCA+HEXTE spectra,
whenever available within about 1 day of a burst.
All spectra, except the PCA+HEXTE spectra,
could be well modelled by an absorbed power-law
(power-law parameters: index, $\gamma_{\rm pl}$ and
normalization at 1\,keV, $N_{\rm pl}$).
For the PCA+HEXTE spectral fits a soft component
had to be included (we used a black body), as well as an Fe\,K line
in the form of a Gaussian (line parameters: line energy, $E_{\rm Fe}$ (keV), 
and photon flux, $N_{\rm Fe}$ (photons\,cm$^{-2}$\,s$^{-1}$); line 
width arbitrarily fixed at 0.1\,keV).
The time ranges over which the persistent
ASM or WFC emission was integrated are given in
Table~\ref{tableburstproperties}.  For the ASM spectra we
generally used all data from about 1.5~days before the burst up to the
last dwell before the start of the burst.  This leads to reasonably
good spectra except for a few cases.  

To get an estimate of the
bolometric (unabsorbed) persistent luminosity, we extrapolate the
spectral fit results to the 0.1--200\,keV band (arbitrarily chosen).
Note that these values should be taken with some caution: 
power-law emission can contribute significantly to the total spectrum, at the lowest energies
for high power-spectral indices (about 80\%\ and 8\%\ of the 0.1--200\,keV emission is at
0.1--2\,keV and 10--200\,keV, respectively, for $\gamma_{\rm pl}$=2.5),
and at the highest energies for low power-spectral indices (about 8\%\ and 80\%, respectively, for $\gamma_{\rm pl}$=1.5).

{\sc XSPEC} cannot provide errors on the integrated flux in an 
energy band.  We therefore randomized the spectral parameters using
the fit values and the derived 1$\sigma$-errors; this was done 10000
times and we recorded the resulting integrated 2--10\,keV and
0.1--200\,keV fluxes. The flux distributions are significantly skewed
towards larger fluxes. We therefore fitted the flux distributions
below and above the peak of the distribution with Gaussians with
different widths. 

\begin{figure*}
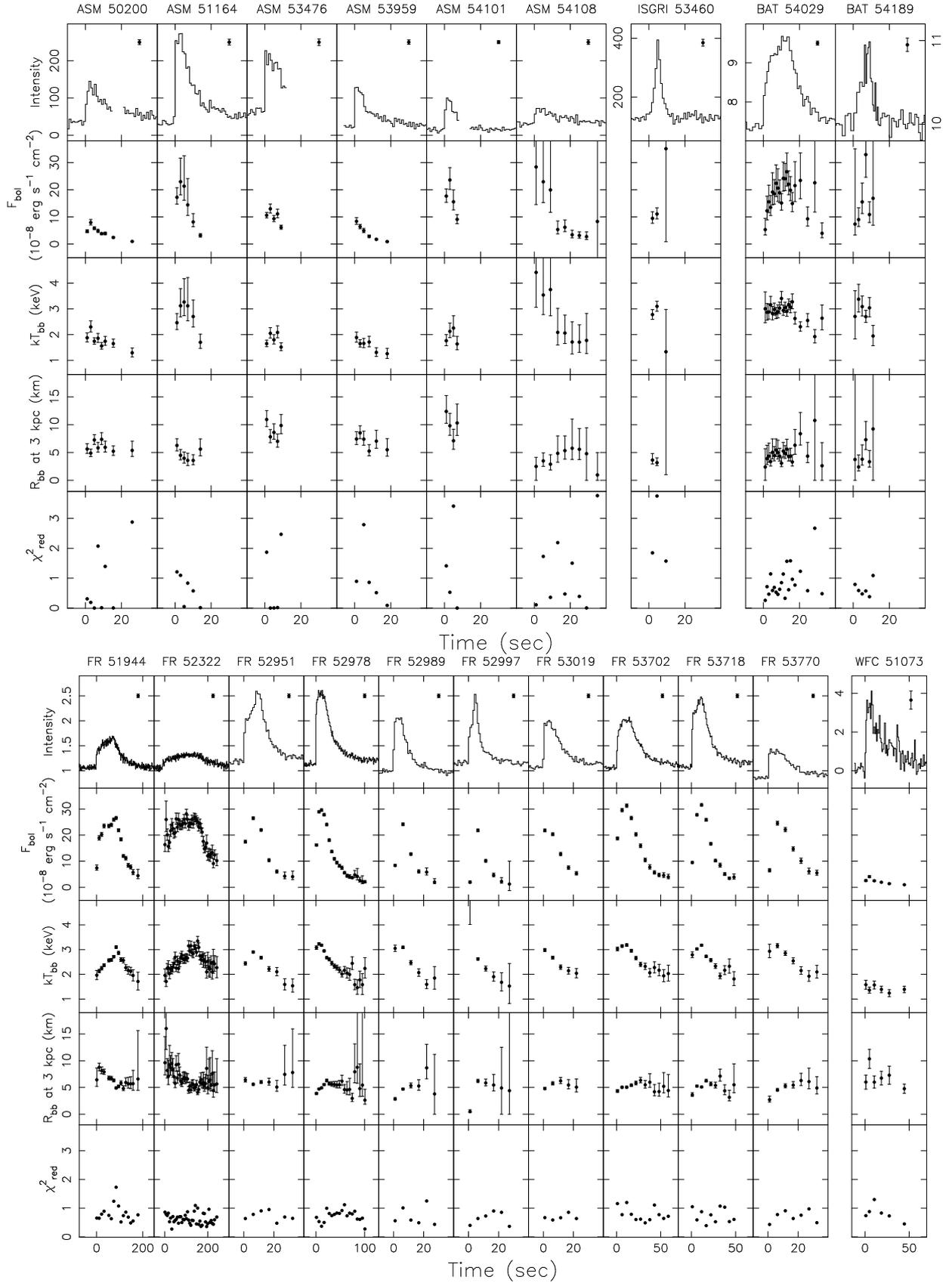

\centering
  \includegraphics[width=1.27\columnwidth,angle=-90]{13210f1a.ps}
  \includegraphics[width=1.22\columnwidth,angle=-90]{13210f1b.ps}
  \caption{Time profiles of raw full-bandpass photon rates (top
    panels), bolometric black-body flux (2nd panels),
    black-body temperature (3rd panels), black-body radius for 3\,kpc
    distance (4th panels) and goodness of fit (bottom panels).
The unit of intensity (top panels) is cts\,s$^{-1}$ for the 
6 ASM bursts, cts\,s$^{-1}$\,bin$^{-1}$ for the 
ISGRI burst, kcts\,s$^{-1}$\,bin$^{-1}$ for
the 2 BAT bursts, kcts\,s$^{-1}$ for the 10 FREGATE
bursts and cts\,s$^{-1}$ for the WFC burst.
The bursts are denoted by instrument (FR=FREGATE) and MJD (see Table~\ref{tableburstproperties}). For the ASM bursts which occurred
on MJD\,50200 and MJD\,51164 we show the light curves from SSC1 and SSC0 data, respectively
(see Table~\ref{table_0614}). }
\label{figcombined}
\end{figure*}

\begin{landscape}
\begin{table}[t]
\caption{Burst and persistent emission properties for all bursts from 4U\,0614+091.$^1$ 
\vspace{-0.3cm}
}
\centering
\begin{tabular}{l@{\,\,}l@{}c@{}c@{\,}c@{ }c@{}c@{}c@{}c@{}c@{\,}c@{}c@{ }c@{\,\,\,}c@{\,\,\,}rl}
\hline \hline
\multicolumn{1}{c}{MJD} &
\multicolumn{1}{c}{Start time} &
\multicolumn{1}{c}{$\Delta t$} &
\multicolumn{1}{c}{$t_{\rm rise}$} &
\multicolumn{1}{c}{$t_{\rm decay}$} &
\multicolumn{1}{c}{$\chi^2_{\rm red}$} &
\multicolumn{1}{c}{$F_{\rm peak}$ (10$^{-8}$} &
\multicolumn{1}{c}{$E_{\rm b}$ (10$^{-8}$} &
\multicolumn{1}{c}{$\tau$} &
\multicolumn{1}{c}{$t_{\rm dur}$} &
\multicolumn{1}{c}{Time span} &
\multicolumn{1}{c}{$\chi^2_{\rm red\,,pl}$} &
\multicolumn{1}{c}{$\gamma_{\rm pl}$} &
\multicolumn{1}{c}{$N_{\rm pl}$} &
\multicolumn{2}{c}{$F_{\rm X,l}$} \\
\multicolumn{1}{c}{} &
\multicolumn{1}{c}{(UTC)} &
\multicolumn{1}{c}{(day)} &
\multicolumn{1}{c}{(s)} &
\multicolumn{1}{c}{(s)} &
\multicolumn{1}{c}{/dof} &
\multicolumn{1}{c}{erg\,s$^{-1}$\,cm$^{-2}$)} &
\multicolumn{1}{c}{erg\,cm$^{-2}$)} &
\multicolumn{1}{c}{(s)} &
\multicolumn{1}{c}{(s)} &
\multicolumn{1}{c}{(MJD$-$50000)} &
\multicolumn{1}{c}{/dof} &
\multicolumn{1}{c}{} &
\multicolumn{1}{c}{} &
\multicolumn{2}{c}{(10$^{-8}$\,erg\,s$^{-1}$\,cm$^{-2}$)} \\
\hline
\multicolumn{16}{l}{{\it OSO-8}/GCSXE} \\
42680 & 1975-09-25 22:48 & -- & 7 & 32$^{+6}_{-5}$ & 0.7/73 & $\simeq$3.3/$\simeq$0.8$^e$ [1] & $\simeq$24$^e$ [1] & $\simeq$30$^e$ & 71 & -- & -- & -- & -- & \multicolumn{2}{c}{0.05--0.09$^f$, 0.02--0.13$^g$ [2]} \\
\multicolumn{16}{l}{{\it SAS-3}} \\
42820 & 1976-02-12 20:17 & 140 & -- & -- & -- & $\gtrsim$0.5$^h$ [3] & $\gtrsim$38$^h$ [3] & $\sim$76 & -- & -- & -- & -- & -- & \multicolumn{2}{c}{$\lesssim$0.11$^h$ [2]} \\
\multicolumn{16}{l}{{\it GRANAT}/WATCH} \\
47908 & 1990-01-17 17:42:28 & 5088 & 40 & 23.0$^{+4.2}_{-3.5}$ & 1.0/49 & $\simeq$19 & $\simeq$907 & $\simeq$48 & 136 &  -- & -- & -- & -- & \multicolumn{2}{c}{$\lesssim$0.33$^i$, $\lesssim$0.34$^j$ [4,5]} \\
\multicolumn{16}{l}{{\it EURECA}/WATCH} \\
49019 & 1993-02-01 06:27:39 & 1111 & 21 & 15.2$^{+2.9}_{-2.3}$ & 1.0/53 & $\simeq$19 & $\simeq$352 & $\simeq$19 & 59 & -- & -- & -- & -- &  \multicolumn{2}{c}{0.020$\pm$0.008$^k$, 0.09$\pm$0.01$^l$} \\
49035 & 1993-02-17 11:09:27 & 16 & 9 & 11.7$^{+2.7}_{-2.3}$ & 1.2/54 & $\simeq$21 & $\simeq$258 & $\simeq$12 & 33 & -- & -- & -- & -- & \multicolumn{2}{c}{"} \\
49048 & 1993-03-02 14:15:27 & 13 & 18 &  6.0$^{+0.9}_{-0.8}$ & 0.9/53 & $\simeq$28 & $\simeq$493 & $\simeq$18 & 54 & -- & -- & -- & -- & \multicolumn{2}{c}{"} \\
\multicolumn{14}{l}{{\it RXTE}/ASM} &
\multicolumn{1}{c}{$F_{\rm X}$} &
\multicolumn{1}{l}{$F_{\rm X\,,bol}$} \\
50200 & 1996-04-27 10:25:37 & 1152 & 3 & 13.0$^{+1.5}_{-2.0}$ & 0.7/8 & 7.9$^{+1.1}_{-0.9}$ & 100$\pm$4 & 12.6$\pm$1.7 & 28 &  198.90--200.44 &  0.04/1 & 1.96$\pm$0.28 & 0.37$\pm$0.12 & 0.10$^{+0.06}_{-0.03}$ & 0.49$^{+0.17}_{-0.14}$ \\
51164 & 1998-12-17 01:29:22 & 91 & 1 &  7.8$\pm$0.5 & 1.1/64 & 23$^{+9}_{-5}$ & 207$\pm$29 & 9$\pm$3 & 19 & 1162.50--1164.06 & 0.04/1 & 2.27$\pm$0.21 & 0.43$\pm$0.12 & 0.08$^{+0.03}_{-0.02}$ & 0.42$^{+0.11}_{-0.12}$ \\
53441$^b$ & 2005-03-12 16:52:59 & 367 & -- & 7571$^{+995}_{-819}$ & 2.4/29 & $\gtrsim$0.67$\pm$0.04 & $\gtrsim$5525 & $\simeq$8194 & -- & 3438.00--3441.50 &  0.2/1 & 2.06$\pm$0.31 & 0.31$\pm$0.09 & 0.07$^{+0.03}_{-0.02}$ & 0.35$^{+0.12}_{-0.10}$ \\
53476 & 2005-04-16 15:35:09 & 16 & 1 & 11.0$^{+1.9}_{-1.4}$ & 2.1/7 & 13.0$^{+1.8}_{-1.5}$ & 132$\pm$6 & 10.2$\pm$1.4 & 21 & 3475.00--3476.65 &  3.8/1 & 1.59$\pm$0.37 & 0.11$\pm$0.06 & 0.05$^{+0.04}_{-0.02}$ & 0.35$^{+0.17}_{-0.09}$ \\
53959 & 2006-08-12 19:00:54 & 189 & 1 &  8.0$\pm$0.7 & 1.0/78 & 8.4$^{+1.3}_{-1.1}$ & 69$\pm$4 & 8.2$\pm$1.3 & 17 & 3959.80--3962.00 &  0.1/1 & 2.53$\pm$0.41 & 0.62$\pm$0.34 & 0.08$^{+0.05}_{-0.03}$ & 0.63$^{+0.41}_{-0.36}$ \\
54101$^c$ & 2007-01-01 11:32:49 & 72 & 2 &  5.1$^{+1.2}_{-0.8}$ & 1.3/3 & 23.6$^{+4.5}_{-3.5}$ & 156$\pm$12 & 6.6$\pm$1.2 & 15 & 4099.50--4101.48 &  0.07/1 & 2.37$\pm$0.33 & 0.41$\pm$0.17 & 0.06$^{+0.04}_{-0.02}$ & 0.39$^{+0.21}_{-0.16}$ \\
54108 & 2007-01-08 18:10:51 & 7 & 1 & 11.0$^{+2.3}_{-1.8}$ & 1.0/57 & 28$^{+192}_{-14}$ & $\simeq$400 & $\simeq$14 & 29 & 4106.50--4108.76 & 2.31/1 & 2.12$\pm$0.27 & 0.31$\pm$0.11 & 0.07$^{+0.04}_{-0.02}$ & 0.33$^{+0.13}_{-0.12}$ \\
\multicolumn{16}{l}{{\it BeppoSAX}/WFC} \\
51073 & 1998-09-17 07:51:37 & 873 & 3 & 18.2$^{+4.1}_{-3.1}$ & 1.0/89 & 4.1$\pm$0.4 & 111$\pm$5 & 27$\pm$3 & 57 & 1073.03--1073.66 & 0.9/25 & 2.17$\pm$0.07 & 0.47$\pm$0.05 & 0.09$^{+0.02}_{-0.01}$ & 0.48$^{+0.06}_{-0.05}$ \\
\multicolumn{16}{l}{{\it RXTE}/PCA \&\ HEXTE} \\
51789 & 2000-09-01 21:21 -- 02 01:20 & 625 & -- & \multicolumn{2}{l}{Tail info only:} & $\gtrsim$0.123$\pm$0.002 & $\gtrsim$149 & $\simeq$1211 & -- & 1786.50--1788.90 & 1.4/1 & 2.05$\pm$0.22 & 0.31$\pm$0.10 & 0.07$^{+0.04}_{-0.02}$ & 0.36$^{+0.14}_{-0.12}$ \\
51944 & 2001-02-04 21:52:42.8 & 155 & 0.005 & \multicolumn{2}{l}{Tail info only:} & $\gtrsim$0.033$\pm$0.003 & $\gtrsim$69 & $\simeq$2090 & -- & \multicolumn{6}{l}{See {\it HETE-2}/FREGATE burst on MJD\,51944}\\
\multicolumn{16}{l}{{\it HETE-2}/FREGATE} \\
51944 & 2001-02-04 21:52:44 [H1488] & 155 & 70 & 39.7$\pm$1.5 & 1.2/226 & 26.6$\pm$0.6 & 3170$\pm$43  & 119$\pm$3 & 308 & 1943.00--1944.90 & 0.2/1 & 2.21$\pm$0.12 & 0.38$\pm$0.06 & 0.07$^{+0.02}_{-0.01}$ & 0.38$^{+0.06}_{-0.07}$ \\
52322 & 2002-02-17 12:40:04 [H1926] & 378 & 137 & 89$\pm$5    & 1.1/366 & 28.1$\pm$1.7 & 6075$\pm$145 & 216$\pm$14 & 569 & 2321.00--2322.52 &  0.02/1 & 1.94$\pm$0.24 & 0.24$\pm$0.08 & 0.07$^{+0.04}_{-0.02}$ & 0.32$^{+0.13}_{-0.10}$ \\
52951 & 2003-11-08 16:23:18 [H2915] & 84 & 8  & 6.7$\pm$0.2  & 1.4/88  & 26.5$\pm$0.5 & 493$\pm$14   & 18.6$\pm$0.6 & 45 & 2950.00--2951.68 & 1.8/1 & 1.91$\pm$0.27 & 0.22$\pm$0.10 & 0.07$^{+0.04}_{-0.02}$ & 0.31$^{+0.04}_{-0.02}$ \\
52961 & 2003-11-18 07:34:46 [H2935] & 10 & 2  & 4.4$\pm$0.8  & 0.9/97  & $\simeq$12 & 80$\pm$11 & $\simeq$7 & 16 & 2959.50--2961.30 &  2.0/1 & 1.93$\pm$0.19 & 0.31$\pm$0.09 & 0.09$^{+0.04}_{-0.02}$ & 0.43$^{+0.16}_{-0.12}$ \\
52978 & 2003-12-05 23:03:43 [H2959] & 78 & 5  & 20.4$\pm$0.5 & 1.2/182 & 29.6$\pm$0.5 & 1240$\pm$21  & 42.0$\pm$1.0 & 89 & 2977.00--2978.95 &  0.9/1 & 1.95$\pm$0.31 & 0.19$\pm$0.09 & 0.05$^{+0.03}_{-0.02}$ & 0.26$^{+0.14}_{-0.11}$ \\
52989 & 2003-12-16 19:34:07 [H2974] & 11 & 5  & 6.7$\pm$0.3  & 1.3/93  & 24.1$\pm$0.5 & 318$\pm$10   & 13.2$\pm$0.5 & 31 & 2988.00--2989.81 &  0.1/1 & 2.20$\pm$0.34 & 0.62$\pm$0.30 & 0.12$^{+0.08}_{-0.04}$ & 0.61$^{+0.34}_{-0.31}$ \\
52997 & 2003-12-24 11:50:29 [H2980] & 8 & 4  & 3.7$\pm$0.2  & 1.4/94  & 24.0$\pm$1.0 & 199$\pm$8    & 8.3$\pm$0.5 & 21 & 2997.50--2999.00 & 1.6/1 & 2.06$\pm$0.14 & 0.68$\pm$0.14 & 0.16$^{+0.05}_{-0.04}$ & 0.76$^{+0.20}_{-0.16}$ \\
53041$^d$ & 2004-02-06 10:04:45 [H3040] & 22 & -- & -- & -- & -- & -- & -- & -- & 3040.00--3041.40 & 0.2/1 & 2.39$\pm$0.48 & 0.35$\pm$0.23 & 0.05$^{+0.04}_{-0.02}$ & 0.34$^{+0.27}_{-0.23}$ \\
53019 & 2004-01-15 21:28:27 [H3012] & 22 & 4  & 9.6$\pm$0.4  & 1.0/93  & 21.8$\pm$0.5 & 389$\pm$7    & 17.9$\pm$0.5 & 40 & 3018.00--3019.88 &  0.4/1 & 1.97$\pm$0.21 & 0.21$\pm$0.07 & 0.06$^{+0.03}_{-0.02}$ & 0.27$^{+0.11}_{-0.08}$ \\
53074 & 2004-03-10 10:06:21 [H3103] & 33 & 2  & 3.5$\pm$0.7  & 0.8/68  & $\simeq$29 & 113$\pm$15 & $\simeq$4 & 10 & 3072.50--3074.41 & 1.9/1 & 2.20$\pm$0.19 & 0.32$\pm$0.09 & 0.06$^{+0.03}_{-0.02}$ & 0.32$^{+0.07}_{-0.09}$ \\
53702 & 2005-11-28 21:27:17 [H3972] & 226 & 8  & 14.2$\pm$0.5 & 1.1/134 & 31.3$\pm$0.6 & 1015$\pm$16  & 32.4$\pm$0.8 & 73 & 3699.00--3702.88 &  1.8/1 & 2.33$\pm$0.45 & 0.44$\pm$0.28 & 0.05$^{+0.04}_{-0.01}$ & 0.38$^{+0.17}_{-0.10}$ \\
53718 & 2005-12-14 18:45:31 [H3982] & 16 & 11 & 11.3$\pm$0.3 & 1.4/107 & 31.5$\pm$0.5 & 805$\pm$24   & 25.5$\pm$0.9 & 62 & 3717.00--3718.77 &  0.1/1 & 1.95$\pm$0.30 & 0.26$\pm$0.11 & 0.07$^{+0.04}_{-0.02}$ & 0.33$^{+0.16}_{-0.13}$ \\
53740 & 2006-01-05 05:18:53 [H3994] & 22 & 1  & 2.7$\pm$0.7  & 0.8/73  & $\simeq$8 & 27$\pm$6 & $\simeq$3 & 7 & 3738.50--3740.22 &  0.6/1 & 1.71$\pm$0.38 & 0.16$\pm$0.10 & 0.06$^{+0.04}_{-0.02}$ & 0.37$^{+0.19}_{-0.12}$ \\
53770 & 2006-02-04 17:48:46 [H4020] & 30 & 2  & 10.9$\pm$0.6 & 1.3/91  & 24.5$\pm$0.8 & 525$\pm$15   & 21.4$\pm$0.9 & 45 & 3769.00--3770.74 &  0.1/1 & 1.62$\pm$0.25 & 0.17$\pm$0.06 & 0.08$^{+0.04}_{-0.02}$ & 0.50$^{+0.24}_{-0.11}$ \\
\multicolumn{16}{l}{{\it INTEGRAL}/IBIS/ISGRI} \\
52867 & 2003-08-16 21:31:26     & 545 & 14 & 4.3$^{+1.4}_{-1.0}$ & 0.8/34 & 30$\pm$4 & 179$^{+81}_{-50}$  & $\simeq$6  & 26 & 2866.00--2867.89 & 0.01/1 & 2.16$\pm$0.35 & 0.27$\pm$0.15 & 0.05$^{+0.04}_{-0.02}$ & 0.27$^{+0.18}_{-0.15}$\\
53460 & 2005-03-31 07:12:18 [2441]& 19 & 5 & 2.0$^{+0.3}_{-0.2}$ & 0.3/38 & 11$\pm$2 & 342$^{+126}_{-85}$ & $\simeq$31 & 67 & 3455.00--3460.25 & 1.5/10 & 2.01$\pm$0.10 & 0.18$\pm$0.05 & 0.05$^{+0.01}_{-0.01}$ & 0.22$^{+0.04}_{-0.06}$ \\
\multicolumn{16}{l}{{\it Swift}/BAT} \\
54029 & 2006-10-21 09:01:54 [234849] & 70 & 14 & 6.4$\pm$0.5 & 1.2/22 & 27$^{+7}_{-5}$ & 545$\pm$49 & 20$\pm$5 & 54 & 4028.00--4029.38 & 1.1/1 & 2.06$\pm$0.23 & 0.34$\pm$0.11 & 0.08$^{+0.04}_{-0.03}$ & 0.38$^{+0.13}_{-0.12}$ \\
54189 & 2007-03-30 08:53:15 [273106] & 81 &  9 & 3.2$^{+0.5}_{-0.4}$ & 1.3/28 & 33$^{+12}_{-8}$ & 187$\pm$65 & 6$\pm$3 & 21 & 4189.00--4190.37 & 0.04/1 & 2.12$\pm$0.16 & 0.46$\pm$0.10 & 0.10$^{+0.03}_{-0.02}$ & 0.49$^{+0.12}_{-0.10}$ \\
\hline
\multicolumn{16}{l}{\tiny $^a$\,References: [1] Swank et al.\ (1978), [2] Mason et al.\ (1976),
[3] Lewin (1976), [4] S.~Kitamoto (priv.\ comm.), [5] Brandt et
al.\ (1992), [6] {\tt \tiny http://space.mit.edu/HETE/Bursts/summaries.html}.} \\
\multicolumn{16}{l}{\tiny $^b$\,Superburst; $^c$\,Also seen by {\it HETE-2}/WXM (H4087) [6]; $^d$\,Only seen by {\it HETE-2}/WXM [6]; $^e$\,2--20\,keV; $^f$\,2--6\,keV; $^g$\,2.5--7.5\,keV; $^h$\,1.3--5\,keV; $^i$\,1--20\,keV; $^j$\,6--150\,keV; $^k$\,6--12\,keV; $^l$\,20--100\,keV.}\\
\end{tabular}
\vspace{-0.4cm}
\note{
\tiny
The first 3 columns give the day of the burst, its start time (trigger numbers in brackets)
and time since the last observed burst ($\Delta t$).  
For the burst parameters ($t_{\rm rise}$, $t_{\rm decay}$, $F_{\rm peak}$, $E_{\rm b}$,
$\tau$ and $t_{\rm dur}$) see Sect.~\ref{xrbprop}; $\chi^2_{\rm red}$/dof gives
the goodness of the fit of the decay to an exponential and the degrees of freedom (dof).
The rest of the columns refer to the average ASM persistent emission (Sect~\ref{xray_spectra}):
time span over which it was averaged, goodness of fit of the emission to a
power law and dof ($\chi^2_{\rm red\,,pl}$/dof) and parameters $\gamma_{\rm pl}$ and $N_{\rm pl}$.
$F_{\rm X,l}$, $F_{\rm X}$ and $F_{\rm X,bol}$ refer to literature$^a$ values of the absorbed persistent flux in
various energy bands before 1994$^{d-k}$, and the unabsorbed persistent fluxes at 2--10\,keV and 0.1--200\,keV after 1994, respectively.
References are given between brackets, see footnote $a$.
}
\label{tableburstproperties}
\end{table}
\end{landscape}

\noindent
The plus and minus errors in the integrated flux
derived from the spectral fits were then taken as the 1$\sigma$ widths.

\subsection{Data and properties of previously reported bursts}
\label{previous}

We complete the information on the properties of the bursts seen after 1992 with that of the
bursts reported by Lewin (1976), Swank et al.\ (1978) and Brandt et al.\ (1992).
We digitized the 2--60\,keV light curve of the {\it OSO-8}/GCXSE burst from figure~2 in
Swank et al.\ (1978) and also obtained further information on the event from that paper.
Data on the {\it GRANAT}/WATCH burst (6--20\,keV; Brandt et al.\ 1992) is
available in a form similar to that of the {\it EURECA}/WATCH bursts (see Sect.~\ref{eureca}). 
We analysed this {\it GRANAT}/WATCH burst in a way similar to our analysis of the {\it EURECA}/WATCH bursts.
Further information on the burst seen with the Horizontal Tube collimator 
(see Lewin et al.\ 1976) onboard SAS-3 was taken from Lewin (1976).
Information about the persistent emission around the times of all these bursts, including those seen by {\it EURECA}/WATCH, were 
extracted from other literature (see Table~\ref{tableburstproperties}).

\section{Results}

\begin{figure}[top]
\centering
  \includegraphics[height=.35\textheight,angle=-90]{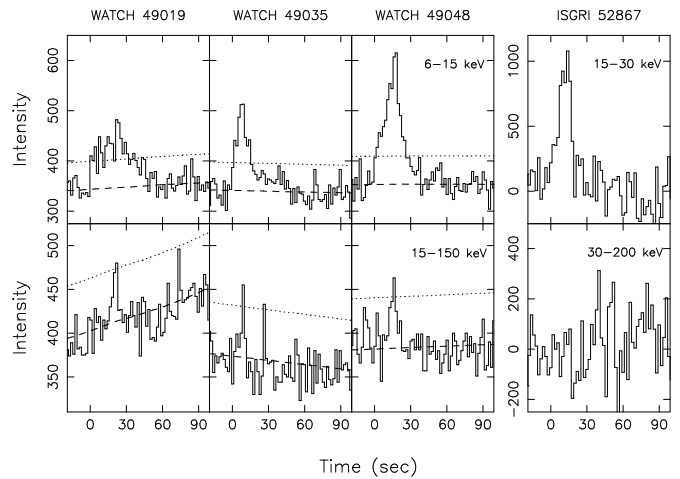}
  \caption{From left to right: The 3 bursts observed with {\it EURECA}/WATCH
in the 6--15\,keV ({\it top}) and 15--150\,keV ({\it bottom}) bands,
and the burst observed with ISGRI in 2003
in the 15--30\,keV ({\it top}) and 30--200\,keV ({\it bottom}) bands.
Zero time corresponds to the start of each burst (see Table~\ref{tableburstproperties}).
For {\it EURECA}/WATCH detector count rates (in cts\,bin$^{-1}$) are shown, i.e., no correction has been made for
background and vignetting. The dashed line indicates the background
level and the dotted line marks the 3$\sigma$ excess level.
For ISGRI the off-axis and background corrected count rates (in cts\,s$^{-1}$) are shown.
The bursts are denoted by instrument and MJD (see Table~\ref{tableburstproperties}). 
}
\label{plot_eureca_raw_integral}
\end{figure}

We report on 33 bursts from 4U\,0614+091, including the bursts
previously reported and the superburst. They are listed in
Table~\ref{tableburstproperties}. 
Figs.~\ref{figcombined} (top panels) and \ref{plot_eureca_raw_integral} 
present the raw light curves (for a perspective on the short time-scale behaviour of the bursts)
of, respectively, 20 unpublished bursts, and the 3 previously reported {\it EURECA}/WATCH and the 2003 ISGRI bursts. 
Time-resolved spectroscopy for the 20 bursts is also presented separately in
Fig.~\ref{figcombined}. An overview of the normal bursts is given in
Sect.~\ref{overview}. The burst that occurred on MJD\,51944 and that
shows evidence of radius expansion as well as a long-lasting faint tail is described in detail in Sect.~\ref{intermediate}.
Our analysis of the superburst is presented in Sect.~\ref{superburst}.

\subsection{Overview of all normal bursts}
\label{overview}

Apart from the fact that 4U\,0614+091 is not bursting as 
seldomly as previously thought, Table~\ref{tableburstproperties} shows that
the recurrence time between bursts can be on the order of a week,
when burst active (see also Sect.~\ref{ASMlongterm}).  
The bursts generally last between about 10\,s and 10\,min (ignoring long-lasting faint tails, see below). Based on the
FREGATE bursts, we designate bursts as short bursts if the burst
duration is less than about 100\,s, and intermediate-duration bursts if
they last between about 5 and 10\,min. The intermediate-duration bursts 
thus refer to the two bursts that were seen by FREGATE in 2001 and 2002. 
The short bursts will be discussed in Sect.~\ref{short} and the 
intermediate-duration bursts in Sect.~\ref{intermediate}.

Based on the facts that for all bursts the emission can be well described by
black-body emission with canonical type~I X-ray burst parameter
values, that their profiles generally show an initial fast
rise (at low energies) with a final slower decay which is
exponential-like, and that the effective black-body temperature
decreases during the decay for most of the bursts, we infer that
the events are indeed type~I X-ray bursts.  

\subsubsection{Short bursts}
\label{short}

For the short bursts the rise times to maximum of the bursts as recorded by ASM, WFC and 
FREGATE are relatively fast, i.e., about 1--10\,s. 
For the bursts seen
by WATCH, ISGRI and BAT the rise times
tend to be a bit longer, i.e., about 5--20\,s; this can be attributed
to these instruments being sensitive primarily at higher energies than the former
instruments. Also, many of the bursts from these hard X-ray
instruments show a two-stage rise, i.e., first a fast rise within
seconds, then a slower rise to the peak intensity. The slow rise may
be attributed to the slow increase in $kT_{\rm bb}$ (see, e.g.,
Fig.~\ref{figcombined}), and therefore to the peak of the X-ray
spectrum shifting towards the centre of the instrument bandpass.  

The short bursts have e-folding decay times between 2 and 20\,s; the
shortest decay times are seen at the highest energies 
(ISGRI and BAT), as expected if the burst
spectrum softens during the decay. The characteristic decay time, $\tau$,
is in the range 6--42\,s. 

\begin{figure*}
\centering
  \includegraphics[height=.7\textheight,angle=-90]{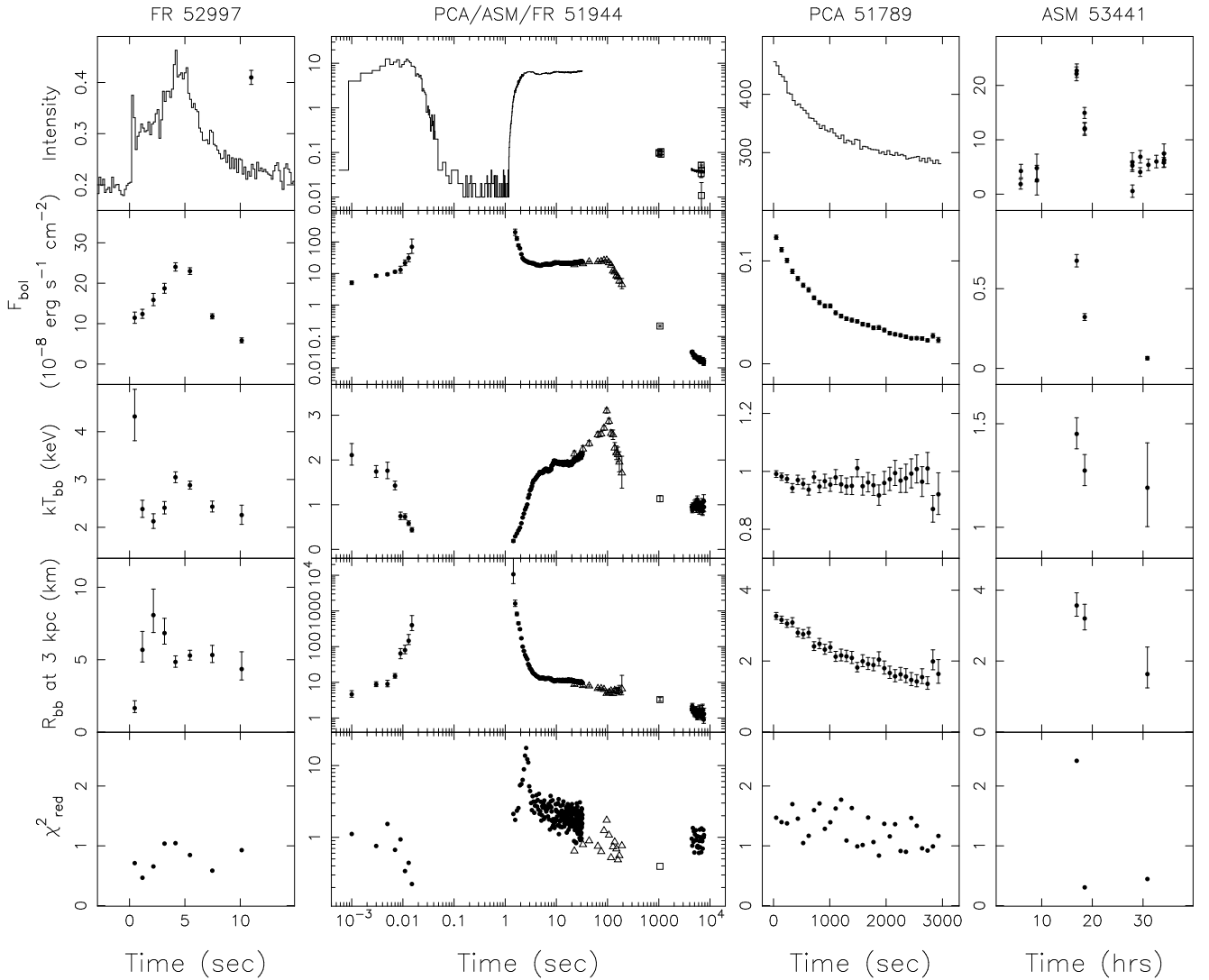}
  \caption{
Light curves (top row) and time-resolved spectral fit results (lower four rows) for selected
bursts shown with higher time resolution than in Fig.~\ref{figcombined}.  See the
caption to Fig.~\ref{figcombined} for brief explanations of the spectral parameters.
The date (MJD) of each burst and the instrumental origin (FR=FREGATE)
of the data are indicated.
{\it Left:} 
FREGATE light curve and time-resolved spectroscopy for the burst on MJD\,52997 
with a time resolution of 0.16\,s.  The unit of intensity is kcts\,s$^{-1}$.
{\it Middle left:} 
(top) PCA full band-pass light curve of the MJD\,51944 burst 
with a time resolution of 0.001\,s (from $t$=0--0.05\,s), 0.01\,s (from $t$=0.05--2.25\,s),
0.125\,s (from $t$=2.25--30\,s), and 16\,s (from $t$=4000--8000\,s). 
The unit of intensity is 10$^4$\,c\,s$^{-1}$ per 4 PCUs. The PCA (PCU3) rates after 
$t$=4000\,s have been multiplied by four.
The ASM data are indicated with open squares; they have been scaled to the PCA rates
using the Crab as a reference. Time $t$=0\,s corresponds to UT 2001 Feb 4 21:52:42.8.
(rest) Time-resolved spectral fit results using the data from the PCA
(filled circles), ASM (open squares) and FREGATE (open triangles).
{\it Middle right:} 
PCA full band-pass light curve (top) and time resolved spectroscopy 
of the long-lasting faint tail of a possible burst
observed on MJD\,51789. For this burst we do not have an exact start time, so $t$=0\,s corresponds
to the start of the data sequence (UT 2000 Sep 2 at 01:19:47).
Intensity is in units of c\,s$^{-1}$ for PCUs 2 and 3 combined.
{\it Right:} 
ASM full-bandpass light curve of the superburst on MJD\,53441
from 4U\,0614+091 (top) and time resolved spectroscopy.  Time is given in hours
since UT 2005 March 12, 0h, and the intensity is given in cts\,s$^{-1}$.
}
\label{plot_pca_all_hete_sb}
\end{figure*}

The WFC burst (on MJD\,51073) and two FREGATE bursts (on MJD\,52961 and
MJD\,53740) are intrinsically weak; they reach peak intensities of
less than a few Crab. The WFC event showed peak temperatures of about 1.5\,keV,
markedly below that seen during maximum of most of the other bursts
($\sim$3\,keV).  These weak events may be similar to the burst
reported from OSO-8, which reached a peak of $\simeq$0.36\,Crab and
showed $kT_{\rm bb}$=0.8$\pm$0.1\,keV during the first 20\,s (Swank et
al.\ 1978; see also Table~\ref{tableburstproperties}).  The event seen
by SAS-3 (Lewin 1976) was rather weak too (see
Table~\ref{tableburstproperties}), but only part of a burst was
seen, so one cannot infer its true peak intensity.

The fluxes of the strongest bursts (including the intermediate-duration bursts)
reach peak values of about 15~Crab.  Both this and the properties of the time-resolved
spectral fits (see Fig.~\ref{figcombined} and
Table~\ref{tableburstproperties}) suggest they reach a
limiting bolometric flux between about 21 and
32$\times$10$^{-8}$\,erg\,cm$^{-2}$\,s$^{-1}$.  When taking into
account only those bursts which reached a peak flux of more than
15$\times$10$^{-8}$\,erg\,cm$^{-2}$\,s$^{-1}$, the average peak value
is about 27$\times$10$^{-8}$\,erg\,cm$^{-2}$\,s$^{-1}$ (fitting the 
observed values with a constant results, however, in a very high value of
$\chi^2_{\rm red}$, i.e., 23, for 14 dof\footnote{The goodness of fit is expressed by the
reduced chi-squared, or $\chi^2_{\rm red}$.
It is the sum of the weighted square deviations between
the data and the model, divided by the number of degrees of
freedom (dof).}).  The highest bolometric peak flux measured 
is 31.5$\pm$0.5$\times$10$^{-8}$\,erg\,cm$^{-2}$\,s$^{-1}$.

The observed fluences, $E_{\rm b}$, of the intrinsically weak bursts are
between 7 and 13$\times$10$^{-7}$\,erg\,cm$^{-2}$.  The values for the
other short bursts range from about 16 to about
124$\times$10$^{-7}$\,erg\,cm$^{-2}$.  
For a source distance of 3\,kpc we derive
absolute fluences from the above quoted values between 0.8 and
13$\times$10$^{39}$\,erg.

We also inspected the burst light curves at higher time
resolutions. One of the FREGATE bursts (MJD\,52997) showed an
interesting spike at the start of the burst.  The high-time resolution
light curve is shown in Fig.~\ref{plot_pca_all_hete_sb} (top left).  The spike
lasts for about 0.3\,s; after that the flux slowly rises to maximum
for about 4\,s.  Double peaked light curves are typically seen in hard X-ray light curves
when photospheric radius expansion is important (see, e.g., Lewin
et al.\ 1984, Tawara et al.\ 1984, Cocchi et al.\ 2000). 
We performed time-resolved X-ray spectral fits
at higher time resolution, to follow in more detail the evolution of
the burst, especially the spike (Fig.~\ref{plot_pca_all_hete_sb}, left, bottom panels).  
During the spike, the spectrum is quite hard 
compared to that generally observed near maximum of type~I X-ray bursts
($kT_{\rm bb}$=4.3$^{+0.6}_{-0.5}$\,keV versus $\simeq$3\,keV),
and $R_{\rm bb}$ is quite small.
During the first few seconds $R_{\rm bb}$ rises, after
which it drops slightly and then levels off. $kT_{\rm bb}$ quickly drops after the spike, but
then it increases up to $\simeq$3\,keV at the peak of the burst. 
This may be a hint of a short radius-expansion phase, but the statistics are somewhat poor. 
Note that $F_{\rm bol}$ does not stay constant but slowly rises during that phase, 
casting some doubt on the radius-expansion interpretation.

\subsubsection{Intermediate-duration bursts and evidence of photospheric radius expansion}
\label{intermediate}

\begin{figure}[top]
\centering
  \includegraphics[height=.35\textheight,angle=-90]{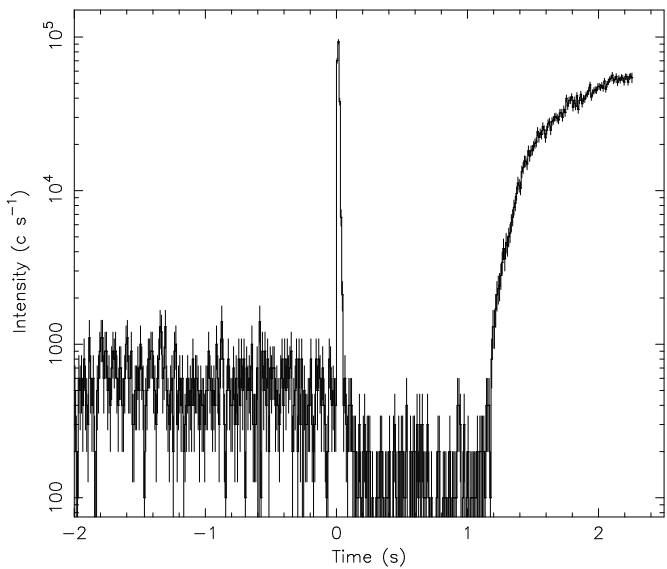}
  \caption{PCA 2--60\,keV light curve centred around the burst onset on MJD\,51944
at a 10\,msec time resolution.
The intensity is as measured by 4 PCUs combined. Time $t$=0\,s corresponds to UT 2001 Feb 4 21:52:42.8.}
\label{burstonset}
\end{figure}

The PCA light curve around the time of the start of burst MJD\,51944 
is shown in Fig.~\ref{burstonset}. One may see a strong, very short,
spike followed by a large drop in the flux, and then the onset of the main burst.
In the top part of Fig.~\ref{plot_pca_all_hete_sb} (middle left), the first part of the 
burst is plotted again, but the spike is shown at higher time resolution. During the spike the intensity
reaches maximum within 5\,ms. Then, within 0.04\,s, the intensity
drops to a level consistent with the background and certainly below
that seen before the spike. About 1.1\,s later emission reappears
again and continues as the main burst\footnote{FREGATE triggered on
the main burst.}. A local maximum is then reached within 2\,s. In
the spike, higher count rates are observed than during the first 30\,s
of the main burst.

The spike, drop below pre-burst levels and main burst indicate
strong radius-expansion (see, e.g., Molkov et al.\ 2000, Strohmayer \&\
Brown 2002; see also in 't Zand \&\ Weinberg, in preparation).
The results of the PCA time-resolved spectral fits are shown in the bottom panels of
Fig.~\ref{plot_pca_all_hete_sb} (middle left), together with spectral fits to the average of the
ASM measurements about 0.3\,hrs after the start of the burst. We also overlaid the 
spectral fit results from FREGATE (see Fig.~1, most left of bottom panels). 
Indeed, the inferred emitting area clearly 
increases during the spike and decreases during the very first part of the main burst.
The expansion is more than 3 orders of magnitude. The peak of the emission (as parametrized by $kT_{\rm bb}$)
therefore shifts outside the PCA observable energy band, 
causing the dramatic drop in intensity. 

After the fast contraction, it takes about 100\,s for the burst
to reach a maximum inferred temperature.
During that time the emission area slowly decreases further
(the end of this phase is usually referred to as `touch-down'). 
During the maximum of the main burst, between about 5 and 100\,s after the start, the flux 
varies only slightly between about 18 and 26$\times$10$^{-8}$\,erg\,s$^{-1}$\,cm$^{-2}$.
This is consistent with the peak flux values reached during the other strong bursts (see Sect.~\ref{short}).
If we assume that during that time the flux is close to the Eddington value, the distance
to the source may be inferred.
By using the highest measured peak flux (Sect.~\ref{short}) and the empirically derived Eddington luminosity of
3.79$\times$10$^{38}$\,erg\,s$^{-1}$ appropriate for a pure He
accretor (Kuulkers et al.\ 2003), we infer a distance to the 
source of 3.2\,kpc. Since the estimated uncertainty in the empirically derived Eddington luminosity
is about 15\%\ (see Kuulkers et al.\ 2003), we use for simplicity a distance value of 3\,kpc in all
our calculations.
We note that during the strong expansion and contraction phase the apparent flux
increases by more than an order of magnitude with respect to that described above. This would mean that super-Eddington
fluxes are reached; however, since most of the emission is outside the 
PCA sensitive energy band and the spectra in this phase are highly non-Planckian
(see below), these flux values must be regarded with caution.

Because of the very high count rates, the spectra constructed from
the PCA data have small statistical errors, and this often means that
simple spectral models are not sufficient to fit the spectra within
the uncertainties.
We find $\chi^2_{\rm red}$ values as high as 20, especially
during the first part of the contraction phase (see bottom panel of Fig.~\ref{plot_pca_all_hete_sb}, middle left).
Deviations from a pure black body are seen frequently during radius-expansion bursts
(see, e.g., Kuulkers et al.\ 2002b, 2003, and references therein). Since the temperature
changes are fast during the expansion/contraction phase, a single black body may not be adequate in 
the time bins used. However, 10--20\,s after the start of the bursts the temperature
stays more or less constant, while still the spectra fits are not ideal.
The deviations are thus intrinsic to the source.

Both intermediate-duration bursts show very long rise times during the main burst part, i.e.,
$\simeq$70\,s and $\simeq$140\,s, on MJD\,51944 and MJD\,52322, respectively.
They also have long exponential and characteristic decay times, i.e., $t_{\rm decay}$$\simeq$40\,s and
$\simeq$90\,s, respectively, and $\tau$$\simeq$119\,s and $\simeq$216\,s, respectively..  

The burst on MJD\,51944 showed a long and faint tail, lasting for
hours: after the main burst, during the next pass of PCA observations (between about
1.2 and 2.2\,hrs after the start of the burst), the intensity was still slowly decaying 
but had not yet reached pre-burst levels.
Serendipitously, the ASM scanned over 4U\,0614+091 about
0.3\,hrs (i.e., in between the two above described PCA passes) and 1.9\,hrs after the 
start of the burst. The trend seen in the ASM data points is consistent with that seen in the 
second PCA pass (see Fig.~\ref{plot_pca_all_hete_sb}, top middle left). 
The e-folding decay as measured from the bolometric black-body flux, $F_{\rm bol}$, light 
curves, $\tau_{\rm exp,bb}$, is 1143$^{+2119}_{-492}$\,s,
i.e., a factor of about 35 longer than that measured during 
the first part of the decline from maximum of the main burst ($\tau_{\rm exp,bb}$=33$^{+7}_{-5}$\,s).
During the first part
of the decline $kT_{\rm bb}$ drops quickly (see Fig.~\ref{figcombined}), but after the flux has dropped by
more than an order of magnitude, $kT_{\rm bb}$ does not decrease anymore, but stays 
more or less constant near 1\,keV; the decrease in flux is then due to an inferred
decrease in emitting area. About 5.2\,hrs after the start of the burst the flux still
had not yet reached pre-burst levels; the steady flux decay suggests that it is still
due to remaining burst emission. However, at that low flux level we cannot be certain
whether this is indeed remaining burst emission or due to a slight change in the persistent flux.

We found one other clear example of a long and faint tail in the PCA
data, on MJD\,51789, between UT 01:19:34 and 02:10:35
(see Fig.~\ref{plot_pca_all_hete_sb}, top, middle right).
In the bottom panels of Fig.~\ref{plot_pca_all_hete_sb} (middle right) we show the results of the time-resolved fits. The 
behaviour is similar to that seen in the long-lasting faint tail
of the burst on MJD\,51944, albeit at a slightly
higher flux level: $kT_{\rm bb}$ stays constant near
$\simeq$1\,keV, while the inferred emitting area decreases by a
factor of about 2.  The exponential decay time as measured from
the bolometric black-body flux light curves is $\tau_{\rm exp,bb}$=868$\pm$32\,s.
No {\it RXTE} observations of 4U\,0614+091 were obtained for
about 5 hours prior to the start of the observation at 
UT 01:19:47 and none for about 2.5 hours following the end of this observation.
Therefore, an intermediate-duration burst like that seen
on MJD\,51944 may have preceded the tail here.

\begin{figure}
\centering
  \includegraphics[height=.33\textheight,angle=-90]{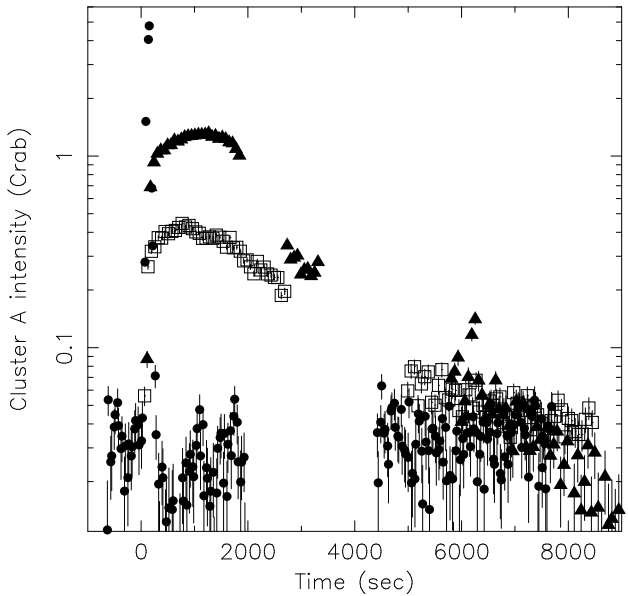}
  \caption{HEXTE hard X-ray (15--60\,keV, cluster A) light curve of 
the burst on MJD\,51944 of 4U\,0614+091 (filled circles), as well as the superburst light curves
of 4U\,1820$-$303 (filled triangles) and 4U\,1636$-$536 (open squares). 
The time resolution is 16\,s for 4U\,0614+091 and 32\,s for 4U\,1820$-$303 and 4U\,1636$-$536.
Time = 0\,s sec corresponds to UT 2001-02-04 21:52:43, 1999-09-09 01:46:54 and 2001-02-22 16:52:11,
for 4U\,0614+091, 4U\,1820$-$303 and 4U\,1636$-$536, respectively.}
\label{0614_1820_1636}
\end{figure}

Although the long-lasting faint tails last as long as various superbursts,
their properties are different. 
Superbursts still show evidence of cooling, hours after they started. We show this 
in Fig.~\ref{0614_1820_1636} by
comparing the hard X-ray (15--60\,keV) emission during the
intermediate-duration burst with that seen during the two
superbursts observed from 4U\,1820$-$303 (Strohmayer \& Brown 2002)
and 4U\,1636$-$536 (Strohmayer \&\ Markwardt 2002). Clearly, hard
X-ray persists during most of the superburst, while it only lasts for
minutes during the intermediate-duration burst.

For the two intermediate-duration bursts the observed fluences are approximately
320$\times$10$^{-7}$\,erg\,cm$^{-2}$ (MJD\,51944) and
610$\times$10$^{-7}$\,erg\,cm$^{-2}$ (MJD\,52322), corresponding to
absolute fluences of 34$\times$10$^{39}$\,erg\,s$^{-1}$ and 
67$\times$10$^{39}$\,erg\,s$^{-1}$, respectively, at 3\,kpc. We have ignored here any
long-lasting faint tails; we expect that corrections for contributions from the tails would change the
above values by only a few per cent (see Table~\ref{tableburstproperties}).  

\subsection{Superburst}
\label{superburst}

\begin{figure}
\centering
  \includegraphics[height=.33\textheight,angle=-90]{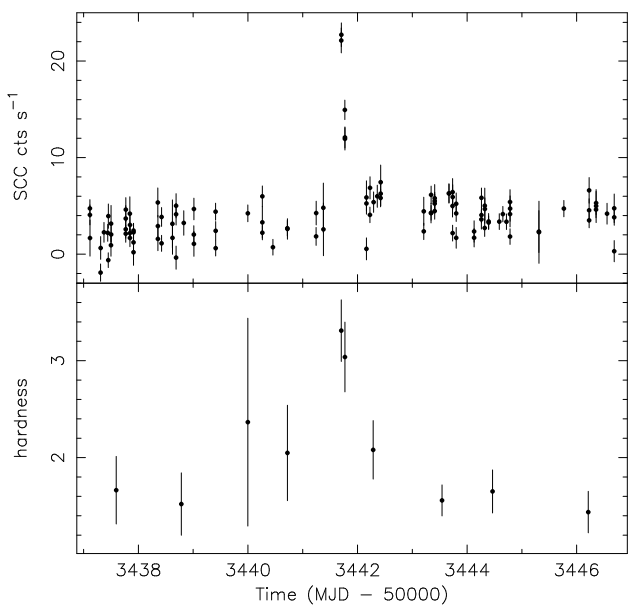}
\caption{ASM individual dwell intensity measurements
  (1.5--12\,keV) and average hardness ratios around the time of the
  superburst from 4U\,0614+091.  Hardness is defined as the ratio of
  the intensity in the 3--12\,keV band to that in the 1.5--3\,keV
  band.  }
\label{superburstplot}
\end{figure}

Inspection of the ASM light curve of 4U\,0614+091 
revealed a long flare on March 12, 2005 (MJD\,53441; see 
top right of Fig.~\ref{plot_pca_all_hete_sb}, and Fig.~\ref{superburstplot}, see also Kuulkers 2005).  Between
about UT 10:38 and 16:53 the ASM flux increased
by a factor of 9.5 up to 0.3\,Crab. About 1.5 hours later the flux had
dropped to 0.17 Crab; $\simeq$9.5 hours later it was still a factor of
about 2 above the pre-flare flux level. The exponential decay time, t$_{\rm decay}$, of
the flare is $\simeq$2.1\,hr.  We calculated the hardness ratio using
various energy bands. We found the most significant results when we used
the ratio of count rates in the 3--12\,keV to 1.5--3\,keV bands, as shown in
Fig.~\ref{superburstplot}. The hardness was higher during the flare
than in the pre- or post-flare measurements; it softened during the decay.
We investigated the raw dwell light curves during the
flare, and found no evidence of strong variability, anomalous
features such as normal bursts, or for instrumental artifacts. 
We, therefore, confidently attribute the high fluxes to the source.

We grouped the flare into three time bins, and performed a time-resolved 
spectral analysis of the average ASM spectra in
these time bins. 
The spectra of the first two time bins are
inconsistent with an absorbed power law ($\chi^2_{\rm red}$/dof of
6.1/1 and 16.2/1, respectively; for the third time bin 
$\chi^2_{\rm red}$/dof=0.7/1).  
The net-flare spectra can best be modelled by absorbed
black-body spectra. 
The results of this time resolved spectral
analysis are displayed in 
Fig.~\ref{plot_pca_all_hete_sb} (right).  The peak flux is rather low, about
0.7$\times$10$^{-8}$\,erg\,s$^{-1}$\,cm$^{-2}$, and the maximum value for $kT_{\rm bb}$
black-body temperature only about 1.5\,keV. However, we
may not have seen the true peak intensity or black-body temperature due to the sparse
coverage by the ASM.  The inferred 
emission areas are rather low (2--4\,km).  They are similar to those inferred
for the long-lasting faint tails of the bursts observed by the PCA
(Sect.~\ref{intermediate}).
However, $kT_{\rm bb}$ is somewhat higher, and the time scales
involved about a factor of 7 longer, than those seen in these tails. 
Note that the inferred radii are not inconsistent with
those seen during some of the superbursts from other LMXB burst sources (e.g., Kuulkers et al.\ 2002a).
Since our observed flare resembles other superbursts,
we suggest the flare to be a superburst.

\subsection{Search for burst oscillations}
\label{oscillation}

\begin{table*}
\caption{Results of spectral fits of PCA+HEXTE data taken outside of but within 
one day of bursts, i.e., representing persistent emission only.$^2$
}
\begin{tabular}{c@{}c@{}c@{}c@{\,\,\,}c@{}c@{}c@{\,\,\,}c@{\,\,}c@{}c@{}c@{}c@{}c@{\,\,}c}
\hline
\multicolumn{1}{c}{MJD} &
\multicolumn{1}{c}{$\Delta t$} & 
\multicolumn{1}{c}{$\chi^2_{\rm red}$/dof} &
\multicolumn{1}{c}{$kT_{\rm bb}$} &
\multicolumn{1}{c}{$R_{\rm bb}$} &
\multicolumn{1}{c}{$F_{\rm bb,bol}$ (10$^{-10}$} &
\multicolumn{1}{c}{$\gamma_{\rm pl}$} & 
\multicolumn{1}{c}{$N_{\rm pl}$} & 
\multicolumn{1}{c}{$E_{\rm Fe}$} & 
\multicolumn{1}{c}{$N_{\rm Fe}$ (10$^{-3}$} &
\multicolumn{1}{c}{$F_{\rm X}$} & 
\multicolumn{1}{c}{$F_{\rm X,bol}$} \\ 
\multicolumn{1}{c}{} &
\multicolumn{1}{c}{(days)} &
\multicolumn{1}{c}{} &
\multicolumn{1}{c}{(keV)} &
\multicolumn{1}{c}{(km)} &
\multicolumn{1}{c}{erg\,s$^{-1}$\,cm$^{-2}$)} &
\multicolumn{1}{c}{} &
\multicolumn{1}{c}{} &
\multicolumn{1}{c}{(keV)} &
\multicolumn{1}{c}{ph\,s$^{-1}$\,cm$^{-2}$)} &
\multicolumn{2}{c}{(10$^{-10}$\,erg\,s$^{-1}$\,cm$^{-2}$)} \\
\hline
51789 &  $<$0.27 & 1.81/108 & 0.93$\pm$0.01 & 1.45$\pm$0.01 & 1.92$\pm$0.03 & 1.841$\pm$0.003 & 0.206$\pm$0.001 & 6.9$\pm$0.1 & 0.52$\pm$0.05 & 8.37$\pm$0.04 & 35.9$\pm$0.2 \\
51944 &     0.09 & 1.17/108 & 1.02$\pm$0.02 & 1.18$\pm$0.01 & 1.75$\pm$0.04 & 1.988$\pm$0.004 & 0.289$\pm$0.001 & 6.6$\pm$0.1 & 0.58$\pm$0.08 & 9.10$\pm$0.06 & 37.7$\pm$0.3 \\
53040 &     1.05 & 1.14/108 & 1.09$\pm$0.01 & 1.12$\pm$0.01 & 2.19$\pm$0.04 & 2.162$\pm$0.004 & 0.269$\pm$0.002 & 6.8$\pm$0.2 & 0.34$\pm$0.11 & 7.36$\pm$0.06 & 29.7$\pm$0.2 \\
54188 &     1.02 & 1.35/79$^a$ & 0.90$\pm$0.02 & 1.51$\pm$0.02 & 1.86$\pm$0.04 & 1.957$\pm$0.003 & 0.256$\pm$0.001 & 6.7$\pm$0.1 & 0.68$\pm$0.07 & 8.44$\pm$0.05 & 35.3$\pm$0.2 \\
\hline
\multicolumn{12}{l}{\footnotesize $^a$\,No HEXTE Cluster A spectrum.} \\
\end{tabular}
\vspace{-0.2cm}
\note{
\footnotesize
Given are the day of the observation (MJD),
the time to the closest burst ($\Delta t$), and the spectral parameter values.
The latter are the goodness of fit degrees of freedom ($\chi^2_{\rm red}$/dof), the black-body parameters
(temperature $kT_{\rm bb}$, radius $R_{\rm bb}$ at 3\,kpc, bolometric black-body flux $F_{\rm bb,bol}$),
the power-law parameters (power-law index $\gamma_{\rm pl}$, normalization at 1\,keV $N_{\rm pl}$),
the Gaussian line (presumably Fe\,K) parameters (line energy $E_{\rm Fe}$, line normalization
$N_{\rm Fe}$; line width was fixed at 0.1\,keV), and the
unabsorbed fluxes in the 2--10\,keV ($F_{\rm X}$) and 0.1--200\,keV ($F_{\rm X,bol}$) bands.
}
\label{pcahexte}
\end{table*} 

We confirm the presence of burst oscillations
during the brightest BAT burst, but do not detect any
significant signal in any of the other bursts.  This is no surprise:
compared to the bright burst BAT light curve, the 
2005 ISGRI burst light curve has a factor 6 lower
signal-to-noise ratio, and the FREGATE
light curves have a factor 1.6 to 2 lower signal-to-noise ratio.
Therefore, a 4$\sigma$ detection by BAT would correspond to non-detection by 
both ISGRI and FREGATE, assuming that the fractional amplitude was equal to that measured in the BAT data. 
We derive 3$\sigma$ upper limits on the fractional rms 
(e.g., van der Klis 1995) in the 413--416\,Hz range of about 4--6\%\ at the peaks of the 
bursts observed by FREGATE, while they are about 9--12\%\ in the burst tails.

The 3$\sigma$ upper limits on the fractional rms amplitude of burst oscillations during the prompt burst 
on MJD\,51944 are about 0.3\%\ and 2\%\ in the 100--1000\,Hz range
as measured by the PCA and HEXTE, respectively.
Thus, if there were oscillations during the prompt burst
on MJD\,51944 they were much weaker than those observed during the 
brightest BAT burst on MJD\,54029.
No evidence of oscillations was found during the faint long-lasting tails
either, with 3$\sigma$ upper limits of about 1.2\%.

\subsection{Persistent emission}
\label{persistent}

Moderate resolution X-ray spectra of the persistent emission from 4U\,0614+091 above typically 1\,keV 
have been satisfactorily fit by a
soft component plus a hard, power-law component, both subjected to interstellar
absorption. 
Generally, for 4U\,0614+091 photon indices for the power-law
component are found to be between 2 and 3 and temperatures for the
soft black-body component are between $kT$=0.5 and 1.5\,keV.  Total
fluxes (1--20\,keV) are typically between 1 and
5$\times$10$^{-9}$\,erg\,s$^{-1}$\,cm$^{-2}$ (Barret \&\ Grindlay
1995, Ford et al.\ 1996, 1997; see also Piraino et al.\ 1999, Fiocchi
et al.\ 2008). Our four most accurate measurements for these
components are provided in Table~\ref{pcahexte}. Our other lower-quality measurements based on
power-law fits only are provided in Table~\ref{tableburstproperties}.
Our derived values of the power-law index, $\Gamma$, vary between $\simeq$1.6
and $\simeq$2.5. The persistent 2--10\,keV flux
is around 0.5--2$\times$10$^{-9}$\,erg\,s$^{-1}$\,cm$^{-2}$, with 
an average of 0.66$\pm$0.06$\times$10$^{-9}$\,erg\,s$^{-1}$\,cm$^{-2}$.
Our spectral fits are consistent with that derived previously.  

The unabsorbed bolometric flux (as estimated from the 0.1--200\,keV flux estimates in 
Table~\ref{tableburstproperties}, see Sect.~\ref{xray_spectra})
is about a factor of 5 higher than the 2--10\,keV flux, i.e., on average 
3.4$\pm$0.2$\times$10$^{-9}$\,erg\,s$^{-1}$\,cm$^{-2}$.
The best estimates of the bolometric correction factor from the
2--10\,keV flux are obtained from a joint analysis of the PCA and
HEXTE spectra,
because of the quality and broad-band coverage. They are between 4.0 and 4.3 with an average of 4.16$\pm$0.05
(see Table~\ref{pcahexte}), which is consistent with that derived from the ASM plus ISGRI
spectra (4.4$\pm$1.3, see Table~\ref{tableburstproperties}).
These factors are compatible with that derived by in 't Zand et al.\
(2007) from a broad sample of LMXBs, i.e., a factor 2.9$\pm$1.4
between the 2--10\,keV flux and the bolometric flux as estimated from
the extrapolated 0.1--100\,keV flux, but are a bit higher than the
value estimated by Migliari \&\ Fender (2006) for an atoll source in
the hard state,
i.e., 2.5. Our bolometric fluxes derived from the ASM spectral fits are consistent
with the fluxes derived from the broad-band PCA plus HEXTE spectral fits. 
Also, our fluxes are comparable to that measured using
broad-band {\it BeppoSAX} data on 4U\,0614+091 in the 0.1--200\,keV band
(Piraino et al.\ 1999; see also Fiocchi et al.\ 2008), as well as simultaneous 
{\it Swift}/XRT and PCA plus HEXTE data in the 0.6--100\,keV band (Migliari et al.\ 2009). This lends
confidence to the bolometric correction and the interpretation that the extrapolated 0.1--200\,keV flux
represents the bolometric flux.

If the source emits isotropically during and between bursts, and
if the peak flux during the bursts $F_{\rm peak}$ represents
emission at the Eddington limit, then the parameter $\gamma$
($\equiv$$F_{\rm pers}/F_{\rm peak}$, where both fluxes are
bolometric) gives the source persistent luminosity in terms of the
Eddington limit (see, e.g., Cornelisse et al.\ 2002b).  Using the
values from Table~\ref{tableburstproperties} and using only those
bursts where $F_{\rm peak}$$>$15$\times$10$^{-8}$\,erg\,s$^{-1}$\,cm$^{-2}$
(see Sect.~\ref{short}), we find
values for $\gamma$ between about 0.0032 and 0.016, with an
average of $\gamma$=0.013$\pm$0.001.  This is consistent with the
estimates by, e.g., Ford et al.\ (2000) and in 't Zand et al.\
(2007).

\subsection{Long-term X-ray behaviour}
\label{longterm}

\begin{figure*}
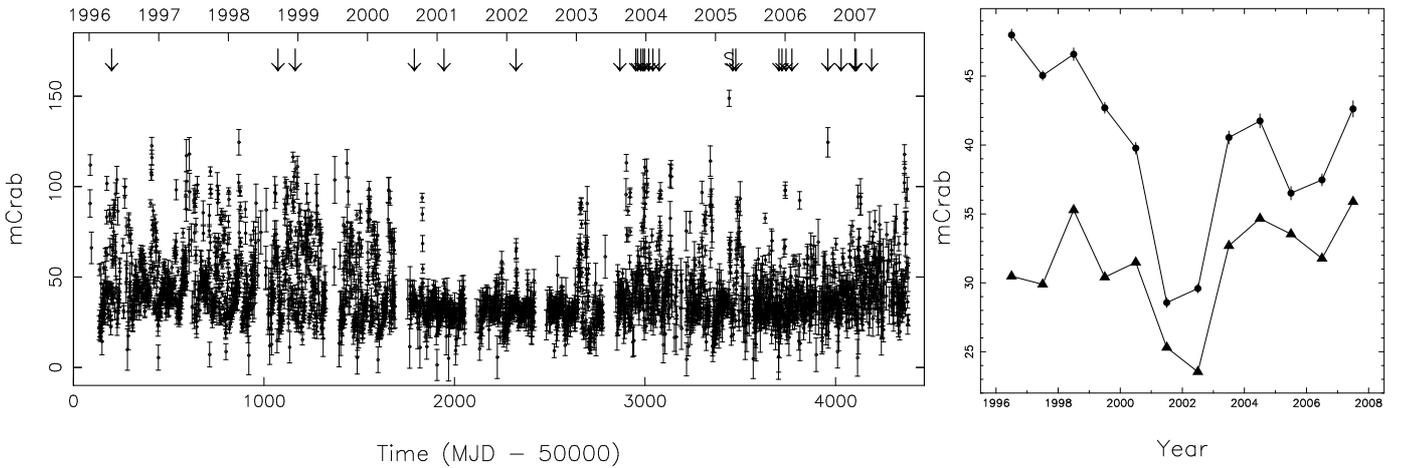

\centering
  \includegraphics[height=.49\textheight,angle=-90]{13210f7a.ps}
  \includegraphics[height=.24\textheight,angle=-90]{13210f7b.ps}
  \caption{{\it Left}: Daily averaged ASM light curve (1.5--12\,keV). The times of the
    normal bursts (see Table~\ref{tableburstproperties}) are indicated with arrows at the
    top. The superburst occurrence is marked with an `S'.
{\it Right}: The yearly averaged ASM light curve (1.5--12\,keV; filled circles)
and the measure of the variability (rms) in the yearly time bins (filled triangles).}
\label{plot_asm}
\end{figure*}

We investigated the long-term X-ray behaviour of 4U\,0614+091 to see
whether it affected its bursting behaviour using results from the {\it RXTE}/ASM
(Sect.~\ref{ASMlongterm})
and instruments on a number of other spacecrafts (Sect.~\ref{further}).

\subsubsection{ASM long-term persistent flux and burst recurrence times}
\label{ASMlongterm}

\begin{table*}
\caption{Average burst recurrence time estimates for various instruments during three periods.}
\begin{tabular}{ccccccccccccccccccc}
\hline
\hline
\multicolumn{4}{c}{1996--2000 flaring period (I)} & 
\multicolumn{4}{c}{2001--2002 calm period (II)} & 
\multicolumn{4}{c}{2003--2007 flaring period (III)} \\ 
\hline
\multicolumn{1}{c}{Inst} & \multicolumn{1}{c}{$t_{\rm exp}$} & \multicolumn{1}{c}{$n^a$} & \multicolumn{1}{c}{$\delta t_{\rm b}$$^b$} &
\multicolumn{1}{c}{Inst} & \multicolumn{1}{c}{$t_{\rm exp}$} & \multicolumn{1}{c}{$n$} & \multicolumn{1}{c}{$\delta t_{\rm b}$} &
\multicolumn{1}{c}{Inst} & \multicolumn{1}{c}{$t_{\rm exp}$} & \multicolumn{1}{c}{$n$} & \multicolumn{1}{c}{$\delta t_{\rm b}$} \\
\multicolumn{1}{c}{~} & \multicolumn{1}{c}{(d)} & \multicolumn{1}{c}{~} & \multicolumn{1}{c}{(d)} &
\multicolumn{1}{c}{~} & \multicolumn{1}{c}{(d)} & \multicolumn{1}{c}{~} & \multicolumn{1}{c}{(d)} &
\multicolumn{1}{c}{~} & \multicolumn{1}{c}{(d)} & \multicolumn{1}{c}{~} & \multicolumn{1}{c}{(d)} \\
\hline
WFC     & 24.9 & 1   & 11--83 & WFC     &  2.3       & 0   &  $>$5   & WFC     &  --         & --   &  --   \\
ASM     & 27.5 & 2   & 7--31  & ASM     &  9.6       & 0   &  $>$19  & ASM     &  20.5       &  4   & 3--9  \\ 
PCA     &  9.6 & 0   & $>$19  & PCA     &  3.7       & 1   &  2--12  & PCA     &   9.4       &  0   & $>$19 \\
FREGATE &  --  & --  & --     & FREGATE & $\simeq$65 & 2   & 17--72  & FREGATE & $\simeq$129 & 11   & 9--16 \\
ISGRI   &  --  & --  & --     & ISGRI   & --         & --  & --      & ISGRI   & 25.3        &  2   & 7--28 \\
\hline
All     & $\simeq$62 & 3 & 12--39 & All  & $\simeq$81 & 2$^c$ & 22--90 & All    & $\simeq$184 & 17   & 9--14 \\
\hline
\hline
\multicolumn{12}{l}{\footnotesize $^a$\,$n$ = number of bursts observed in the considered period.} \\
\multicolumn{12}{l}{\footnotesize $^b$\,$\delta t_{\rm b}$ = expected average burst recurrence time based on Poisson likelihood 1$\sigma$ limits, see text.} \\
\multicolumn{12}{l}{\footnotesize $^c$\,Note that PCA and FREGATE observed one of the bursts simultaneously, see Sect.~\ref{intermediate}.} \\
\end{tabular}
\label{burstrate}
\end{table*}

The ASM data provide a uniquely long and homogeneous history of the
persistent flux.
In Fig.~\ref{plot_asm} (left) we show the ASM (1.5--12\,keV)
long-term light curves as daily averages.
The corresponding average flux and variability (rms) per year are shown in 
Fig.~\ref{plot_asm} (right). From 1996 to mid 2000 the source was
flaring on time scales of days to weeks from about 25--50\,mCrab up to
about 100--125\,mCrab, with yearly averages above 30--35\,mCrab.  Then the source stayed at a relatively quiet
level near 30\,mCrab for about two and a half years (mid 2000 to
2003). Flaring behaviour similar to before mid 2000 was seen from
2003 to mid 2005. From mid 2005 to mid 2007 4U\,0614+091 was still
flaring, but less intensely, i.e., it varied from 25--50\,mCrab up to
about 75--100\,mCrab and the yearly averages were again above
30--35\,mCrab.  The figure suggests that the degree of flaring and the
long-term average (months to years) intensity seem to be related;
a high degree of flaring is seen when the yearly average
persistent flux is above about 30--35\,mCrab and a low degree of
flaring is seen otherwise.  A flux of 30--35\,mCrab (1.5--12\,keV)
corresponds to about 1\%\ of the Eddington limit (see
Sect.~\ref{persistent}). For the following discussion we roughly
divide the ASM history of 4U\,0614+091 into three periods: (I)
1996--2000 flaring period, (II) 2001--2002 calm period, and (III)
2003--2007 flaring period.

In Fig.~\ref{plot_asm} (left) we mark the times of the normal
bursts, as well as the superburst. The timing of burst
detections is related, in part, to the times of year when the
Sun is not near the source and the latter is more easily observed.
Moreover, bursts could have occurred during many of the data gaps. 
It is also possible that with the ASM we missed the bursts which last significantly longer
than the dwell duration (like the two intermediate-duration bursts).  
Also, weak bursts like that seen by the WFC may have been gone unnoticed with the other
instruments; moreover, FREGATE, ISGRI and BAT operate in a higher bandpass than the ASM
and WFC, and have to cope with higher backgrounds and so tend to trigger on brighter bursts.
Therefore, we cannot rule out that we did not see all bursts which happened in our time frame of interest.

Fig.~\ref{plot_asm} (left) suggests, however, 
that bursts appear more frequently during the 2003--2007 flaring period with respect to the previous periods.
This is especially apparent if we take
into account that during the 2000--2002 period there were 3
surveying X-ray instruments active (WFC, ASM and FREGATE). 
To support this suggestion, we determined the average burst recurrence times
during the three above described periods for each instrument, as well as all together.
We excluded results from the BAT (we only have information on triggered bursts, i.e., we did not perform a 
systematic search through the BAT archive) and JEM-X (very low source exposure time).
We also did not take into account either the tail of a possible burst on MJD\,51789 or the superburst.
When considering all instruments, we assumed that the exposure intervals did not overlap,
and that they have the same efficiency in detecting bursts.
We also assumed that the exposure time of a particular instrument is randomly distributed in each of the three periods. 
If $n$ is the number of bursts seen during a certain period, then the average burst recurrence time 
is defined as $t_{\rm exp}$ divided by $n$, where $t_{\rm exp}$ is the total exposure time on the source in question
in that period (note that this does not necessarily mean that the bursts recur periodically). 
Since we are dealing with relatively low number of observed bursts, Poisson statistics apply.
For the expected range in average burst recurrence times ($\delta t_{\rm b}$) we use the Poisson likelihood 
1$\sigma$ lower and upper limits on the expected number of events when $n$ events are observed (where $n$=0,1,2,3,$\cdots$).
We verified our method by doing Monte Carlo simulations.
The net source exposure times, number of bursts observed, and resulting average burst recurrence times per period of
interest are shown in Table~\ref{burstrate}.
For the individual instruments, the ASM and FREGATE give the best constraints on the average burst recurrence times, 
mainly because of the number of bursts observed. 
The estimates of the PCA average burst recurrence times differ significantly from those for the other instruments; we attribute 
this is to the fact that the data are not taken serendipitously, as well as to the relatively low exposure times.

The average burst recurrence times in periods I and II are not significantly different; the average over the two periods combined
is 29$^{+17}_{-10}$~days (taking into account all considered instruments). Period III shows a significant change in the average burst recurrence time with
respect to the previous period: 11$^{+3}_{-2}$~days (taking into account all considered instruments); it is just consistent with
period I. The lower limit on the average burst recurrence time in period III is close to the shortest observed 
value of $\Delta t$ in that period (7~days, see Table~\ref{tableburstproperties}).
During the EURECA/WATCH period in 1993 the burst average recurrence time is 17$^{+15}_{-7}$~days, consistent with
values of $\Delta t$ in that period (see Table~\ref{tableburstproperties}). It is consistent also with the values 
of all three ASM periods I, II, and III.

Interestingly, the only two bursts that were observed in 2001 and 2002 are both intermediate-duration bursts. 
They thus occurred during a period when 4U\,0614+091 was not only less burst active, but also when the persistent emission was rather calm
and had the lowest averaged values.
All other bursts (22, including the BAT bursts) are shorter and occurred during the two flaring periods. The burst duration, therefore, seems to be related
to the level of the persistent emission, and possibly to the degree of flaring.

\begin{figure}
\centering
  \includegraphics[height=.33\textheight,angle=-90]{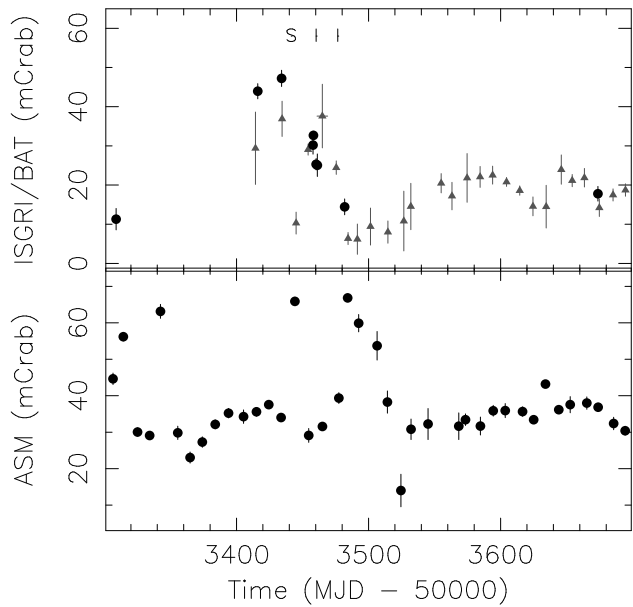}
  \caption{Zoom in on the long-term hard X-ray (15--50\,keV) light
    curve ({\it top}) and soft X-ray (1.5--12\,keV) light curve ({\it
      bottom}), of 4U\,0614+091.  Shown are the 1-day and 10-day
    averages for ISGRI (filled circles) and {\em
      Swift}/BAT (grey filled triangles), respectively ({\it top}),
    and 10-day averages for the ASM. Time in years is given above the plot.
    The occurrence times of the normal bursts are indicated by arrows at the top. The
    superburst occurrence is marked with an `S'.  }
\label{isgri_bat_asm_zoom}
\end{figure}

\subsubsection{Further long-term broad-band X-ray behaviour}
\label{further}

4U\,0614+091 was discovered by {\it UHURU} with variable flux levels
between about 15 and 70\,mCrab (2--6\,keV; Giacconi et al.\ 1972).
Other early measurements of the flux, besides those from {\it UHURU}, 
may be found in, e.g., Mason et al.\ (1976), Swank et al.\ (1978),
Parsignault \&\ Grindlay (1978), Markert et al.\ (1979), and Warwick et
al.\ (1981). In that time it was always seen at a flux of about 10--100\,mCrab 
in roughly the $\sim$2 to $\sim$10\,keV band.
During sporadic measurements by the {\it Ginga}/All Sky Monitor
between February 1987 and November 1991 4U\,0614+091 was never detected in the
1--20\,keV band, with an upper limit on the source flux of about
100\,mCrab (S.~Kitamoto, private communication).  

4U\,0614+091 was too weak to be detected by {\it EURECA}/WATCH on a
daily basis, but data from the entire observation period can be
combined.  By adding the skymaps (see Brandt 1994 for a description of
the technique), 4U\,0614+091 was detected at an average
persistent flux of 25$\pm$10\,mCrab in the 6--12\,keV energy band
(see also Fig.~\ref{WATCH_ima}, right). 

We investigated the {\it CGRO}/BATSE occultation measurements (see, e.g., Harmon et al.\ 2004)
in the period April 1991 and November 2000; they show 10-day average 20--100\,keV 
fluxes from non-detections (with typical 10-day 3$\sigma$ upper limits between about 50--70\,mCrab) 
up to about 150\,mCrab. During the period of the
{\it EURECA}/WATCH observations when 4U\,0614+091 was in the field of
view, the 20--100\,keV flux was on average 57$\pm$6\,mCrab; 
close to the time of the bursts seen by {\it EURECA}/WATCH it was around
100\,mCrab.

The BAT and ISGRI 15--50\,keV
long-term light curves are consistent with each other. The 15--50\,keV
flux in the February 2003 to September 2007 time frame ranges from a
few mCrab to 50\,mCrab, but is most frequently around 20\,mCrab. A zoom in on the
15--50\,keV light curve and the corresponding part of the ASM light curve is shown in
Fig.~\ref{isgri_bat_asm_zoom}. Three interesting features may be seen
in this figure.
First, there is a rapid
increase to about 50\,mCrab for about a month, followed by a drop
within weeks to about 10\,mCrab at MJD\,53500. After about 2 months
it goes back up again to the average level.  Second, whenever the
hard X-ray flux drops below roughly 10\,mCrab, the soft X-ray
intensity increases, up to a factor of 2 or so. 
Such an anti-correlation between soft and hard X-rays on these long time scales was
previously discussed by Barret \&\ Grindlay (1995) and Ford et
al.\ (1996). On the other hand,
hard X-ray flares are {\it not} accompanied by changes in the soft
X-ray emission. Third, the superburst occurred around the time when the
15--50\,keV flux was at its highest level (unfortunately no simultaneous BAT measurements 
exist at the time of the superburst).  
Just after the superburst the 10-day average 15--50\,keV flux is low, while
the 1.5--12\,keV flux is at high levels. 
The ASM data show that this high average
flux is not only due to the superburst itself, but also to a
somewhat increased persistent level just after the superburst (see
Fig.~\ref{superburstplot}): before the superburst the 1.5--12\,keV
flux was 31$\pm$1\,mCrab, whereas for several days after the
superburst it was at an elevated level of 53$\pm$2\,mCrab. About
10~days after the superburst it was back at the pre-superburst
level. A normal burst (MJD\,53460) was seen only 19 days after the
superburst (see Table~\ref{tableburstproperties}); note that this
is only slightly longer than the typical 
average burst recurrence time when 4U\,0614+091 is burst active (see Sect.~\ref{longterm}).

\begin{figure}
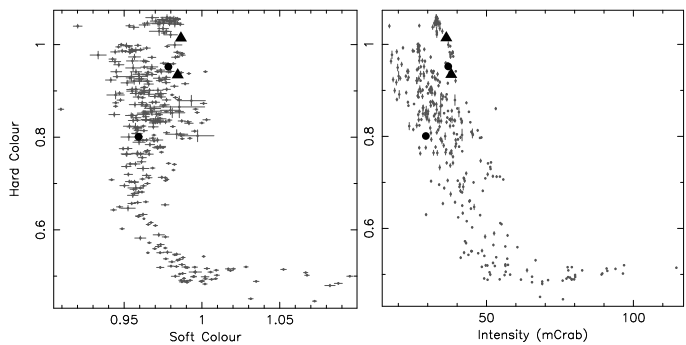

\centering
  \includegraphics[height=.185\textheight,angle=-90]{13210f9a.ps}
  \includegraphics[height=.170\textheight,angle=-90]{13210f9b.ps}
  \caption{PCA colour-colour diagram ({\it Left})
and hardness-intensity diagram ({\it Right}) of all public data
of 4U\,0614+091. Plotted are all persistent data points averaged per observation,
and normalised using the Crab (see Sect.~\ref{pca}).
Data taken close to a burst are marked with a black filled triangle
($\Delta t$$<$0.3\,days) or a black filled circle ($\Delta t$$\sim$1\,day).
For the definitions of the colours and intensity, see Sect.~\ref{pca}.
  }
\label{ccd_hid}
\end{figure}

The CD and HID (Fig.~\ref{ccd_hid}) are qualitatively similar to those presented for the PCA data obtained in 1996--1998 
by van Straaten et al.\ (2000). The colours
often change on a daily time scale, as is typical 
for atoll sources (e.g., Hasinger \&\ van der Klis 1989). For a wide range in hard colour, the soft
colour does not change noticeably. Only near the lowest hard colours does the soft colour start to change;
then there is an anti-correlation between hard and soft
colour. When the hard colours are lowest, however, the soft colour
changes are large while the hard colour changes are small.  This is
in line with the broad-band behaviour on longer time scales of the
fluxes in the 1.5--12\,keV and 15--50\,keV bands described in the
previous paragraph.  In the HID higher intensities
correspond to lower hard colours. Again, at the highest
intensities (above about 50\,mCrab), however, the hard colour
changes only very little.  The part of the diagram where the
soft colour and intensity increases at constant hard colour (value
of about 0.5) is referred to as the soft state or `banana' branch
(see van Straaten et al.\ 2000).  The rest is referred to as the
intermediate state or `island' branch.

For 4 bursts (i.e., those that occurred on MJD\,51789, MJD\,51944, 53040 and 54188)
we have information about the persistent emission within about 1~day of the burst
(see Table~\ref{pcahexte}).
In Fig.~\ref{ccd_hid} we indicate the position the persistent source was in at these times.
All 4 bursts occurred when the persistent emission was hardest and when the intensity 
was below $\simeq$40\,mCrab, i.e., when the source was in the intermediate or hard state 
(`island' or extreme `island').

\section{Discussion}
\label{discussion}

\subsection{Observational summary}
\label{sec:obssumm}

We have expanded the number of observed bursts that can be attributed to 4U\,0614+091 from a
handful to 33.  These include a superburst. Most of the newly-found bursts are localized within 1\farcm 5 of
the optical position of 4U\,0614+091.  This confirms the type~I X-ray
burster nature of this low-mass X-ray binary. We find a significant change in the average burst recurrence
time in a decade of observations, i.e., from roughly a month before mid-2003 to about one to two weeks
afterwards. When burst active, the bursts have peak luminosities of about 3$\times$10$^{38}$\,erg\,s$^{-1}$, 
last for up to about a minute, and release energies of about 10$^{39}$--10$^{40}$\,erg.
The larger set of bursts shows a wide variety of
characteristics, with dynamic ranges of a factor of 40 in peak flux,
100 in duration, and 230 in absolute fluence.

We found three long-duration bursts. Two of them have initial decay
time scales of about 100\,s and resemble intermediate-duration 
bursts seen from other sources (thought to be due to deep He flashes;
see in 't Zand et al.\ 2005 and Cumming et al.\ 2006). The third has
an initial decay time of about 8000\,s, comparable to the decay
times of superbursts (due to an even deeper C flash contained in
the ocean of the NS envelope; see Cumming \&\ Bildsten 2001 and
Strohmayer \&\ Brown 2002). 
It lasts a factor of 10 longer than the longest
known intermediate-duration burst (for example, see in 't Zand et
al.\ 2007, Keek \&\ in 't Zand 2009). Unfortunately, the coverage of
this burst is incomplete. This prevents a measurement of the true peak
flux and a check for the occurrence of radius expansion.
The average burst recurrence time when the two intermediate-duration bursts were seen,
is notably longer than when the shorter bursts were seen
(40$^{+49}_{-18}$~days during a two-year period versus 11$^{+3}_{-2}$~days for other
bursts in the subsequent 4.7-year period).

One of the intermediate-duration bursts shows clear signs of strong
radius expansion within the first few seconds of the event. The
flux seen at the end of this episode of radius expansion must
correspond to a luminosity near the Eddington limit, and we thereby
confirm the distance to be close to 3\,kpc. The average
accretion rate suggested by the broad-band non-burst flux is
0.32--1.6\%\ of the Eddington rate.

The two intermediate-duration bursts occur in a two-year time
interval when the accretion flux lacks the otherwise so typical
flaring behaviour and is on average about 30\%\ lower than at other times.
Similar effects are seen in other (candidate) UCXBs. A calm component
is present which changes on time scales of a year, while flaring occurs
with a time scale of a week (see in 't Zand et al.\ 2007). Also,
bursts are shorter when a source shows more flaring activity 
(e.g., A\,1246$-$58; in 't Zand et al.\ 2008). 
Moreover, UCXBs accreting above 1\%\ of the Eddington limit show burst recurrence
times on the order of days; those accreting below 1\%\ of the
Eddington limit show recurrence times on the order of weeks (in 't
Zand et al.\ 2007). In this respect it is interesting to note that
during 4U\,0614+091's high flux and flaring period it was detected in the
radio (0.34$\pm$0.02\,mJy at 4.86\,GHz), while it was not detected
during the low flux and calm period (3$\sigma$ upper limit of about
0.1\,mJy at 4.86\,GHz). This supports the existence of different states 
and suggests a connection between the radio and X-ray emission mechanisms 
(Migliari et al.\ 2009).

The wide range of peak fluxes is not unprecedented. Galloway et al.\
(2008) find that roughly 2\%\ of bursts have bolometric peak
fluxes less than 0.1 times the maximum for the same source. Three
sources exhibit dynamic ranges in excess of 100: EXO\,0748$-$676
(236$\pm$36), 4U\,1608$-$522 (446$\pm$115), and 4U\,1636$-$536
(151$\pm$25), although it should be mentioned that the high value in
the last source is due to a weak secondary burst in a double burst
(excluding that diminishes the dynamic range to 23$\pm$1). These three
sources are not UCXBs, implying that the wide dynamic range in
4U\,0614+091 is not necessarily due to extraordinary abundances of H,
He, C or O.

We found two long-lasting faint tails wherein the flux decays exponentially
with a time constant in the range of 1000 to 2000\,s. One of them is the remainder of the
intermediate-duration radius-expansion burst. The tail is seen
clearly up to at least 2.5 hours after the start of the
burst, possibly even up to 5 hours.  During the tail no cooling is
seen; the inferred temperature stays constant near about 1\,keV, while
the apparent emitting area decreases. 
Similar faint tails
have been seen in various other bursters (Linares et al.\ 2009,
Falanga et al.\ 2009, in 't Zand et al.\ 2009), but they were not as long
lasting as those seen in 4U\,0614+091.
A hot underlying neutron star (with temperatures between 0.5--1\,keV,
i.e., slightly lower than that found in the burst tail) can explain
the fact that in the tails the inferred temperatures are effectively
constant (van Paradijs \&\ Lewin 1986, in 't Zand et al.\ 2009). In
this regard, we note that a soft component in our persistent PCA
spectral fits is present, perhaps coming from the neutron star and/or the
inner parts of the accretion disk; when the soft component is modelled
by a black-body we infer temperatures of about 1\,keV, i.e., close to
that found in the long-lasting tails. As noted by in 't Zand et al.\ (2009), 
neutron stars in UCXBs accrete at lower rates and lack the
CNO burning of H in their surface layers, compared to those
in other, ordinary, bursters. This may result in a cooler
neutron star in the former systems, leading to a different 
behaviour in the long-lasting tails during their bursts. 
However, the temperature reached in the
long-lasting faint tail of 4U\,0614+091 (and that of the transient
XTE\,J1701$-$407, see Linares et al.\ 2009, Falanga et al.\ 2009,
presumably also containing a cool neutron star due to its long off-states) is
similar to that seen in ordinary persistent bursters, which may
indicate that a different explanation is required for the constant
temperature in the tail.

\begin{figure}
\centering
\includegraphics[height=.35\textheight,bb=55 200 565 680,clip=]{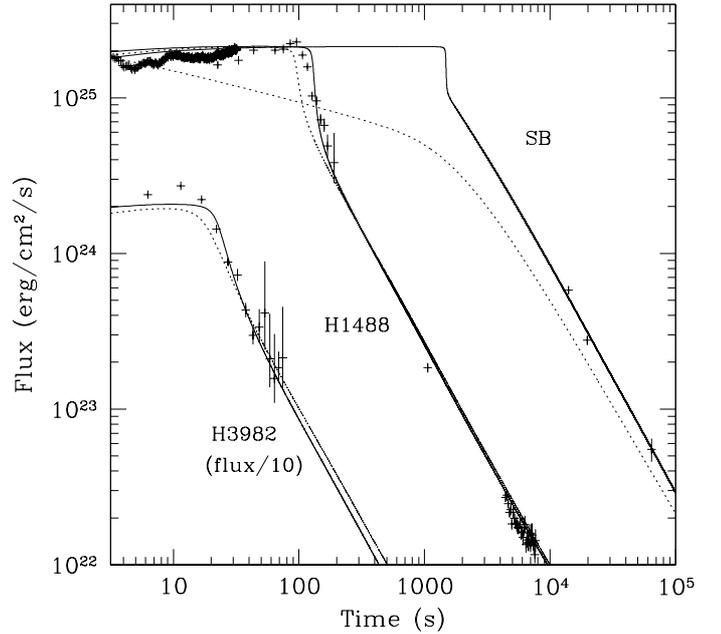}
\caption{Models of the burst light curves for the superburst (SB), the
intermediate-duration burst on MJD\,51944 (H1488), and the short burst on MJD\,53718 (H3982). For the
superburst, the model parameters are $E_{17}$=5, $y_{11}$=2 (solid
curve) and $E_{17}$=2, $y_{11}$=3 (dotted curve). For the intermediate-duration
burst, the model parameters are $E_{17}$=7, $y_{9}$=8 (solid
curve) and $E_{17}$=5, $y_{9}$=10 (dotted curve). For the short burst,
the model parameters are  $E_{17}$=6, $y_{9}$=2 (solid curve) and
$E_{17}$=4, $y_{9}$=3 (dotted curve). Here $E$=$E_{17}\times 10^{17}$\,erg\,g$^{-1}$
and $y$=$y_{9}\times 10^{9}$\,g\,cm$^{-2}$=$y_{11}\times 10^{11}$\,g\,cm$^{-2}$.
The observed fluxes and
model fluxes for the short burst have been divided by a factor of ten
for clarity. The start time of the superburst is
unknown, but we find that choosing the start time to be around 4\,hr
before the time of the first data point gives a slope for the
observed light curve that agrees with the models. Here we adopt
a start time of 3.9\,hr before the first data point.
\label{fig:lc_sb}}
\end{figure}

\subsection{The burst light curves and energetics}
\label{sec:lc}

In an attempt to constrain the energy release and ignition depth of the
bursts, we made some models of the light curves following the approach
of Cumming \&\ Macbeth (2004) for superbursts. We deposit an energy per unit mass
$E$=$E_{17}\times 10^{17}$\,erg\,g$^{-1}$ uniformly in the layer down to a
particular column depth $y$. The cooling of the layer is then followed
by integrating the thermal diffusion equation. We set the composition
after burning to be $^{56}$Fe.
Our models assume neutron star
parameters of 1.4\,$M_\odot$, $R$=10\,km, giving a redshift
factor 1+$z$=1.31 and $g$=2.44$\times$10$^{14}$\,cm\,s$^{-2}$.
To avoid following the expansion of the
outer layers that occurs when the luminosity approaches the Eddington
limit, we set the top of our grid to $y$=10$^8$\,g\,cm$^{-2}$, and
choose our outer boundary condition such that the maximum flux at that point 
is the Eddington flux at the surface of the star
for pure He composition, i.e., $F_{\rm Edd}$=$cg/\kappa$ with
$\kappa$=0.2\,cm$^2$\,g$^{-1}$ (Thomson scattering for pure He), giving
$F_{{\rm Edd},\infty}$=$F_{\rm Edd}/(1$+$z)^2$=2.1$\times$10$^{25}$\,erg\,cm$^{-2}$\,s$^{-1}$.
Because of this crude treatment of the outer layers, 
and the simplistic way in which the energy is initially deposited, 
our models are not valid
for early times, $\lesssim$10\,s (the thermal time scale at a
depth of 10$^8$\,g\,cm$^{-2}$).

We varied the two parameters $E_{17}$ and $y$ to find values that
agree well with the observed light curves. Note that the values of
$E_{17}$ and $y$ are taken to be independent parameters of the model,
which allows us to investigate the ignition depth and energy release
without assuming a particular fuel for the burst. The model light curves
are compared with the observed light curves for the superburst,
the intermediate-duration burst on MJD\,51944, and the short burst on MJD\,53718, in
Fig.~\ref{fig:lc_sb}. We assume a distance to the source of $d$=3\,kpc
and a neutron star radius $R$=10\,km to convert the
observed fluxes into flux at the surface of the neutron star;
these values give good agreement with the maximum flux
from the model.

The best fitting column depths for the three bursts in order of
decreasing duration are
$y$$\simeq$2$\times$10$^{11}$\,g\,cm$^{-2}$,
$\simeq$8$\times$10$^9$\,g\,cm$^{-2}$, and
$\simeq$2$\times$10$^9$\,g\,cm$^{-2}$.
For the intermediate-duration and short bursts, the column
depths inferred from the light curves are consistent with accretion at
about 1\%\ Eddington for the observed times since the previous burst, $\Delta t$=155
and 16~days, respectively (see Table~\ref{tableburstproperties}). 
This can be derived as follows.
The local Eddington accretion rate (accretion rate per unit area) at the
surface of the neutron star is $\dot m_{\rm Edd}$=$c/R\kappa$. We define
the Eddington rate as the value corresponding to $R$=10\,km and
$\kappa$=0.2\,cm$^2$\,g$^{-1}$,
giving $\dot m_{\rm Edd}$$\equiv$1.5$\times$10$^5$\,g\,cm$^{-2}$\,s$^{-1}$,
or 1\%\,$\dot m_{\rm Edd}$=1500\,g\,cm$^{-2}$\,s$^{-1}$.
Accretion at $\dot m$=1\%\,$\dot m_{\rm Edd}$ then gives a
column $y$=$\dot m\Delta t/(1$+$z)$=1.6$\times$10$^9$\,g\,cm$^{-2}$
($\Delta t/16\ {\rm d})(\dot m/1$\% $\dot m_{\rm Edd})$. Matching the
column depths inferred from the light curves assuming that $\Delta t$
is the recurrence time implies $\dot m$=0.52\%, and 1.3\%\,$\dot m_{\rm Edd}$
for the intermediate-duration and short burst, respectively.

\begin{figure}[top]
\centering
\includegraphics[height=.35\textheight,bb= 50 200 555 685,clip=]{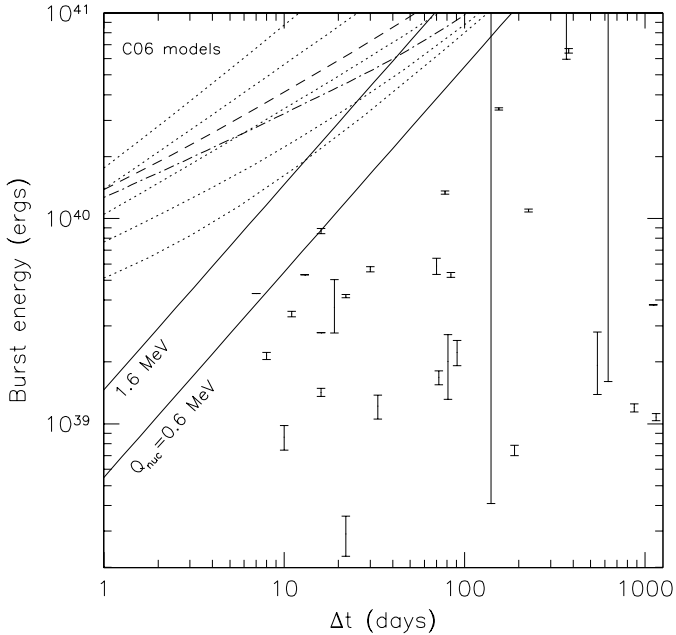}
\caption{The observed burst energies (assuming $d$=3\,kpc)
versus the time since the last observed burst $\Delta t$ from
Table~\ref{tableburstproperties}. Because of possible missed bursts due to the many data gaps, $\Delta t$ is an
upper limit on the average burst recurrence time. The solid lines show the
expected relation for accretion at 1\%\ of the Eddington rate
assuming complete burning of the accumulated fuel for two different
values of nuclear energy release per nucleon, $Q_{\rm nuc}$=1.6\,MeV
corresponding to complete burning of pure He to Fe group, and a
smaller value $Q_{\rm nuc}$=0.6\,MeV. The dotted, dashed, and
dot-dashed curves show the predicted relation for pure He
accretion from Cumming et al.\ (2006). (The different curves are for
different core neutrino emissivities and crust properties; see Cumming et al.\ [2006]
for details.)
\label{fig:et}}
\end{figure}

The time span between the superburst and the previous observed burst is 367~days
(i.e., $\Delta t$=367~days, see Table~\ref{tableburstproperties}),
in which time the accumulated
column is $y=\dot m\Delta t/(1+z)=3.7\times 10^{10}$\,g\,cm$^{-2}$
$(\Delta t/367\ {\rm d})(\dot m/1$\%\,$\dot m_{\rm Edd})$. Therefore, we
require the accretion rate to be $\dot m$$\simeq$5\%\,$\dot m_{\rm Edd}$ for
the column inferred from the superburst light curve to be accreted in
the 367~days leading up to the superburst (assuming no bursts occurred
in between). An alternative explanation
is that the superburst involves a different fuel layer (e.g., He
for the short and intermediate-duration bursts, C for the superburst) in
which case the column for the superburst could accumulate over a
longer time scale than $\Delta t$. At 1\%\ Eddington, the accumulation
time would then be about 5\,yr.

The best fitting values for the energy release are
$E_{17}$$\approx$5--7 (see Fig.~\ref{fig:lc_sb}). In addition, the fact that the light curves reach the
Eddington flux implies a minimum energy release. For the
intermediate-duration burst, we find that the flux does not reach
the Eddington flux for $E_{17}$$\lesssim$4. These values of energy
release are smaller than the values $E_{17}$$\approx$16 for complete
burning of He to Fe group or $E_{17}$$\approx$10 for complete
burning of C to Fe group. 
This implies that the energy release is about a factor of 3 lower
than expected for complete burning of pure He.

The low energies of the bursts can also be seen in the overall
energetics. If all the matter accreted since the previous burst were completely burned
during each burst, one would expect the ratio, $\alpha$, of the
average luminosity in the persistent emission to the time-averaged
luminosity emitted in type-I X-ray bursts to have a value in the range
25--200, depending on the composition of the burning material.
If one assumes that both the persistent and burst emission are
isotropic, that the persistent emission has not varied since the
previous burst, and that $F_{\rm pers}$ is the bolometric persistent
flux, then this statement relating the persistent and burst
luminosities is equivalent to
$\alpha$$\equiv$$\Delta t$$F_{\rm pers}/E_{\rm b}$.
If we take that the minimum average burst
recurrence time is the lowest observed value for $\Delta t$, 
i.e., 7~days, then the corresponding values for
$\alpha$ are in fact lower limits, which are between 125 and 5525
for the short bursts and around 30--70 for the intermediate-duration
bursts. ¬†Calculating an $\alpha$ value by assuming that no bursts were
missed between observed bursts (that is $\Delta t$ is the average burst
recurrence time) gives a range from 500 to 5000, larger than the
canonical value of 100--200 for pure He burning.
Another way to look at this is presented in Fig.~\ref{fig:et} which
shows the burst energy
against time since the last burst $\Delta t$ from Table~\ref{tableburstproperties}.
The solid lines show the expected burst
energy as a function of recurrence time for accretion at 1\% of the
Eddington rate, for two different values of the nuclear energy release
per nucleon $Q_{\rm nuc}$=1.6 and 0.6\,MeV per nucleon
(where $E_{17}$$\sim$$10(Q_{\rm nuc}/{\rm MeV per nucleon})$). The curves at
the top left are pure He ignition models from Cumming et al.\ (2006)
that assume complete burning. The observed bursts lie well
below the expected energies assuming complete burning.

Fig.~\ref{fig:lc_sb} shows also that for the intermediate-duration burst on MJD\,51944,
the model light curve nicely connects the early decay from the peak
luminosity to the long tail lasting thousands of seconds. Therefore
the long tail is naturally explained by the cooling of deep layers heated
by the burst as proposed for tails seen in other sources by in 't Zand et al.\ (2009).
The theoretical expectation is that the luminosity
should decay as a power law in the cooling tail. This is confirmed by our observations
(e.g., Fig.~\ref{fig:lc_sb}).

\subsection{He as a fuel for the bursts}

In Sect.~\ref{sec:lc}, we found that the short and intermediate-duration
burst light curves could be explained by ignition depths of $y$$\simeq$2 and
8$\times$10$^9$\,g\,cm$^{-2}$, respectively, and that these depths are compatible with the amount of mass
accreted at an accretion rate close to $\dot m$$\simeq$1\%\ $\dot m_{\rm Edd}$
in the observed time between bursts, $\Delta t$. These
ignition depths would be naturally explained if the neutron star
accreted a substantial amount of He. At low accretion rates, most of the energy released 
by pycnonuclear and electron capture reactions in the crust flows outwards, because the 
neutron star core is too cold for significant neutrino emission. This results in an outward 
flux from the crust of $F$=$\dot m Q_b$ with $Q_b$$\approx$1\,MeV per nucleon 
(Brown 2000; Cumming et al.\ 2006: their figure 18, top panel). For this value of $Q_b$, figure~22 of
Cumming et al.\ (2006) shows that pure He accretion gives ignition at
$y$$\sim$10$^9$--10$^{10}$\,g\,cm$^{-2}$ at $\dot m$$\sim$1\%\,$\dot m_{\rm Edd}$.

\begin{figure}[top]
\centering
\includegraphics[height=.35\textheight,bb=65 200 570 685,clip=]{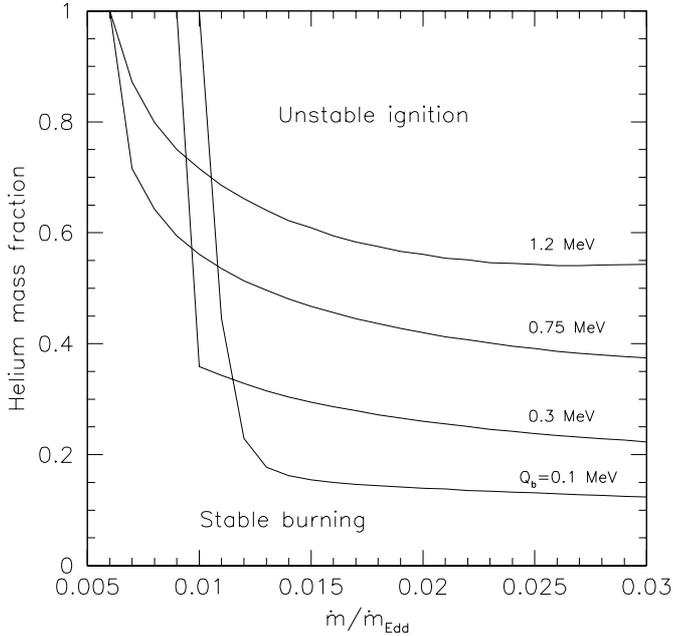}
\caption{The mass fraction of He needed to achieve unstable He
ignition as a function of accretion rate, for accretion of He and
an equal mixture of C and O. The different curves are for
different values of $Q_b$ as labeled.
\label{fig:yc}}
\end{figure}

However, as we noted in the introduction, the lack of He lines in the
optical spectrum of the source implies that the accreted material is
not significantly enriched in He.
To investigate whether a small amount of He in the accreted material
could still explain the observed bursts, we constructed ignition models for
type~I X-ray bursts following Cumming \&\ Bildsten (2000). We assume
that the accreted material consists of He and a 50/50
C-O mixture. We find that for small He mass fractions,
the He burns away stably before reaching the conditions required
for a thermal runaway. In Fig.~\ref{fig:yc} we plot the He mass
fraction required to achieve unstable ignition as a function of $\dot m$
and $Q_b$. It shows that the amount of He inferred
from the fits of disk spectra by Werner et al.\ (2006), i.e., $Y$$\lesssim$10\%,
do not lead to unstable ignition at any accretion rate unless
the neutron star is cold, $Q_b$$\ll$0.1\,MeV per nucleon, and even then
the recurrence times are extremely long. We, therefore, conclude that
matching the observations requires a significant amount of He in
the accreted material.

If we allow a large enough He fraction, we can match the inferred
ignition depth for the short and intermediate-duration bursts. For example,
taking a He mass fraction $Y$=0.5, $Q_b$=0.3\,MeV per nucleon, and 
local accretion rates $\dot m$=3000 and
5000\,g\,cm$^{-2}$\,s$^{-1}$ (2\% and 3.3\% of the Eddington rate)
gives ignition depths of $y$=2 and 8$\times$10$^9$\,g\,cm$^{-2}$, respectively.
The corresponding recurrence times are 7 days and 46 days,
respectively. For pure He and $Q_b$=1\,MeV per nucleon, the
correct ignition depths are obtained for $\dot m$=1\%\ and 1.7\%\,$\dot m_{\rm Edd}$,
with recurrence times of 81 days and 11 days for
the intermediate-duration and short burst, respectively.
These recurrence time estimates are consistent with the observed values.

For each of these examples, a decrease in accretion rate of only 40\%\ is required to change 
the ignition column depth by a factor of 4 from 2 to 8$\times$10$^9$\,g\,cm$^{-2}$. This is 
because of the steep dependence of the ignition
column on temperature -- only a small decrease in temperature, 
and a small corresponding decrease in $\dot m$ is needed. The required decrease in
accretion rate is comparable to the observed difference in persistent
flux between the 2001--2002 calm period when the intermediate-duration bursts
occurred and the 2003--2007 flaring period when more regular bursting resumed: 
as noted in Sect.~\ref{sec:obssumm}, 
the 1-year-averaged ASM fluxes for 4U\,0614+091 in 2001 and 2002 are about 
30\%\ smaller than for the years after this time interval. 

It is not clear why unstable He ignition would give a nuclear
energy release of only $Q_{\rm nuc}$$\approx$0.6\,MeV, as
inferred from the burst energetics and light curve fits (see Sect.~\ref{sec:lc}).
This energy is approximately the energy released in burning He to C. For
a He mass fraction $Y$, setting the total energy release to be  
$Q_{\rm nuc}$=0.6\,MeV per nucleon implies that the energy release from burning beyond C to heavy
elements is $Q_{\rm heavy}$$\approx$0.6$(1-Y)$\,MeV per nucleon. This is much less than the 
$\approx$1\,MeV available for complete burning of C to Fe, unless the helium fraction $Y$ is small
(however, as noted above, the observations require $Y$ to be large).
At a depth of $y$$\sim$10$^9$\,g\,cm$^{-2}$, an energy deposition of only
0.03\,MeV per nucleon is sufficient to raise the temperature to above
10$^9$\,K, at which the nuclear burning would be expected to
proceed beyond C, so a significant energy release from burning beyond C would be expected. 
Numerical models of bursts at low accretion
rates would be useful to follow the nucleosynthesis and determine the expected energy release.

Another possibility is that not all of the accreted fuel burns during the burst. 
For example, the burning may consume only part of the depth of the fuel layer, or may cover 
only part of the stellar surface. Numerical models of bursts at low accretion rates are needed 
to follow the nucleosynthesis and burning dynamics, and determine the expected energy release.

\subsection{The superburst}
\label{superburstdiscussion}

In the previous section we argued that the short bursts and
intermediate-duration bursts could be explained by a substantial
fraction of He in the accreted material, despite the fact that the
optical spectrum of 4U\,0614+091 suggests that very little He is
present in the accreted material. However, the superburst poses a more
severe problem.

The superbursts observed in other sources at accretion rates
$\dot m$$\gtrsim$0.1\,$\dot m_{\rm Edd}$ have been explained as being due to
C ignition (Strohmayer \& Brown 2002, Cumming \& Bildsten
2001). However, there are problems with this scenario: (1) producing
enough C during H/He burning (e.g., Schatz et al.\ 2003,
Woosley et al.\ 2004), (2) heating the neutron star ocean strongly enough to reach
ignition temperature (e.g., Cumming et al.\ 2006, Keek et al.\ 2008), and
(3) accreting rapidly enough for the C to survive to the ignition
depth (Cumming \& Bildsten 2001, Cumming et al.\ 2006).

If there is accretion of a significant amount of C from the CO white dwarf companion in 
4U\,0614+091, the first problem would be eased by removing the need to make the C during 
nuclear burning of H or He. However, if He flashes are responsible for the intermediate-duration 
and short bursts, the C would likely burn away during these flashes. The last two problems are
more difficult to circumvent. First, the ignition temperature for C at the column depth 
inferred from the superburst light curve, $y$$\simeq$2$\times$10$^{11}$\,g\,cm$^{-2}$, is above
6$\times$10$^8$\,K. This temperature is much higher than the ignition
temperatures of He flashes, i.e., about 1$\times$10$^8$\,K.
Achieving this temperature is even more difficult when the accretion
rate and therefore crust heating rate are significantly below those
typical around the times of the superbursts seen in other
sources. Second, we calculated models for C ignition following
Cumming \& Bildsten (2001) and extended their figure 2 to lower
accretion rates, but were not able to find values of $Q_b$ or C
fraction for which the C survives to ignition depth at accretion
rates of $\sim$0.01\,$\dot m_{\rm Edd}$.

Another possibility is that ignition of a thick He layer is
responsible for the observed superburst. Indeed, for the $Q_b$=0.3\,MeV
per nucleon and $Y$=0.5 case considered earlier, we find that reducing
the local accretion rate to 1000\,g\,cm$^{-2}$\,s$^{-1}$, a factor of 3
below the accretion rate that reproduces the ignition column of the
intermediate-duration burst, gives He ignition at 2$\times$10$^{11}$\,g\,cm$^{-2}$. 
The time scale for He burning is longer
than the accumulation time (although close to it), indicating that
He should survive down to the ignition depth, even for such a low
accretion rate (this is not the case for the $Y$=1, $Q_b$=1\,MeV per
nucleon model; there we find that the He burns stably
away). However, although the ignition depth can be achieved, the
expected recurrence time is 9.8\,yr, an order of magnitude longer than the
observed time between the superburst and the previous burst, $\Delta t$=367~days.
On the other hand, this can be reconciled if an appreciable portion of the 
accreted He survives the preceding shorter bursts (see previous Section).

Another constraint on the superburst ignition depth comes from the
quenching time scale for normal bursts following the
superburst. Normal bursting behaviour resumed 19~days following the
superburst, the fastest time scale so far observed (see, e.g., Kuulkers
2004). Cumming \&\ Macbeth (2004) predict that the quench time for
normal bursting behaviour after a superburst is
$t_{\rm quench}$=1.6$y_{12}^{3/4}(\dot{m}/\dot{m}_{\rm Edd})^{-3/4}E_{17}^{3/8}$~days
(where $y$=$y_{12}\times 10^{12}$\,g\,cm$^{-2}$). 
The constraint on the quench time $t_{\rm quench}$$<$19~days implies for $E_{\rm 17}$=6 and
$(\dot{m}/\dot{m}_{\rm Edd})$=0.016 (at the time of the superburst)
that $y_{12}$$<$0.18. This is in good agreement with the ignition
column inferred from the superburst light curve, and consistent with
the normal burst observed 19~days after the superburst being the
first burst to occur following the superburst. The dependencies are
not very strong. For $\dot{M}/\dot{M}_{\rm Edd}$=0.0032--0.016, the
ignition depth is between $y_{12}$$<$0.04--0.18. For $E_{17}$=1--6
this becomes $y_{12}$$<$0.18--0.44. Therefore, the quench time
constraint provides another piece of evidence that the ignition depth
for the superburst was lower than the depths inferred for all or most
previously analysed superbursts in other sources.

\section{Conclusions}

We can understand several aspects of the type I X-ray bursts observed
from 4U\,0614+091. The column depths and energy release per gram in the
bursts can be estimated by comparing the observed light curves with
models, and by considering the burst energetics. Both methods are in
good agreement. Helium ignition naturally explains the observed
ignition depths for accretion at $\dot m$$\sim$1\% $\dot m_{\rm Edd}$.
Furthermore, the sensitive dependence of the He ignition depth on
temperature means that small (factor of two) changes in accretion rate
can lead to an order of magnitude variation among the depths, and can
thereby explain the occurrence of both short bursts and
intermediate-duration bursts. The ignition depth for the superburst
inferred from the light curve is the lowest of the current sample of
superbursts and agrees well with the constraint from the observed
quenching time scale.

However, several puzzles remain. First, the amount of He required
to achieve the required ignition conditions without stably burning
away is significantly larger than the $\lesssim$10\%
limit from the optical spectra.
Recently, however, formation studies using
evolutionary calculations indicate that the donor in 4U\,0614+091
may be a hybrid white dwarf or very evolved helium star
(Nelemans et al.\ 2009). This suggests the
donor still to be a possible supplier of a significant amount
of He to the accretion disk, onto the neutron star.
Further investigations have to be done, why the He does not
show up then in the optical observations.

Second, understanding the superburst remains problematic. 
Unstable C ignition is not known to be
possible at these low accretion rates: the accumulating layer is too
cold, and even if heated the C burns stably. A superburst powered
by a large He pile is possible, but takes several years to
accumulate if the accretion rate is $\sim$1\% Eddington. Such an
accumulation time is much greater than the observed time of one
year. Finally, the reason for the
low energy per gram $Q_{\rm nuc}$$\lesssim$0.6\,MeV per nucleon released
in the bursts is not clear.

It is important to emphasize, however, that the ignition models and light curve models 
used in this paper are simplified. The model light curves assume uniform and 
instantaneous energy deposition in the fuel layer, and do not follow the detailed nucleosynthesis. 
These models cannot address the early part of the light curve, for example the interesting ripples 
in the early light curve of the intermediate duration burst (Fig.\ 10), nor whether the burning 
is expected to be incomplete (as we infer from the burst energetics). 
Incomplete burning, either across the neutron star surface or with depth in the fuel layer, 
may at least partly solve the above puzzles.
Our ignition models rely on 
a simplified one-zone ignition criterion. Numerical models of accumulating and burning of He/CO 
mixtures at low accretion rates are needed to confirm our conclusions, for example, regarding 
the amount of helium needed to avoid stable burning.

It is interesting that at least in principle He can power a
superburst-like event. The requirement is that the accumulating fuel
layer remains cold. Recently, Cooper et al.\ (2009) ruled out He as
a fuel for superbursts, but their argument assumed an ignition
temperature of $T$=5$\times$10$^8$\,K that is much larger than the He
ignition temperature at superburst columns. Even in the superburst
sources with $\dot m$$\gtrsim$0.1\,$\dot m_{\rm Edd}$, He-powered
superbursts could occur if the accumulating layer is cold enough. 
One way to keep the layer cold would be to have direct URCA neutrino 
emission in the neutron star core, so that most of the energy release in 
the crust flows inwards rather than outwards. Triple alpha ignition becomes 
mostly sensitive to density rather than temperature when the column depth reaches 
$y$$\sim$3$\times$10$^{11}$\,g\,cm$^{-2}$, as the ignition becomes pycnonuclear 
and therefore temperature independent. This could potentially explain the narrow 
range of superburst ignition columns. Further work on this possibility is needed.

\begin{acknowledgements}
Partly based on observations with {\it INTEGRAL}, an ESA project with
instruments and science data centre funded by ESA member states
(especially the PI countries: Denmark, France, Germany, Italy,
Switzerland, Spain), Czech Republic and Poland, and with the
participation of Russia and the USA.  The {\it RXTE}/ASM dwell average
results are provided by the ASM/RXTE teams at MIT and at the RXTE SOF
and GOF at NASA's GSFC.  The {\it Swift}/BAT transient monitor results
are provided by the {\it Swift}/BAT team.  We thank Jean Swank and
Lucien Kuiper for discussions regarding the bursts seen with {\it OSO-8} and 
in 2005 with {\it INTEGRAL}, respectively. EK thanks Andy Pollock for 
discussion regarding the burst recurrence times and small-number statistics.
SB was partly supported by the Danish Space Board.  
AC acknowledges support from an NSERC Discovery Grant and the Canadian Institute for Advanced Research (CIFAR). 
This research has made use of the SIMBAD
database, operated at CDS, Strasbourg, France. 
\end{acknowledgements}

\appendix

\section{Localizations of bursts}
\label{appendix}

\begin{figure}[top]
\centering
  \includegraphics[height=.175\textheight]{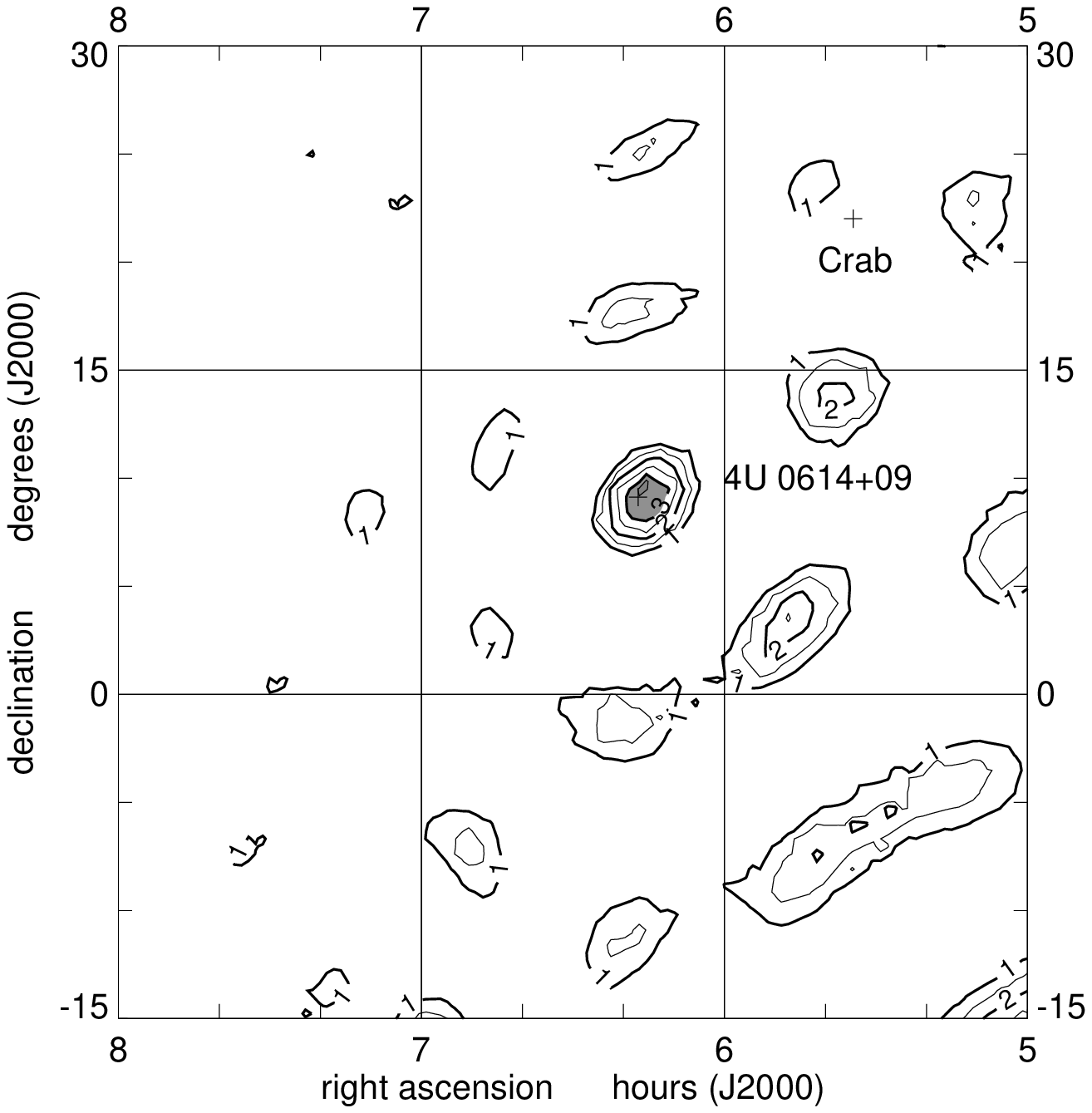}
  \includegraphics[height=.175\textheight]{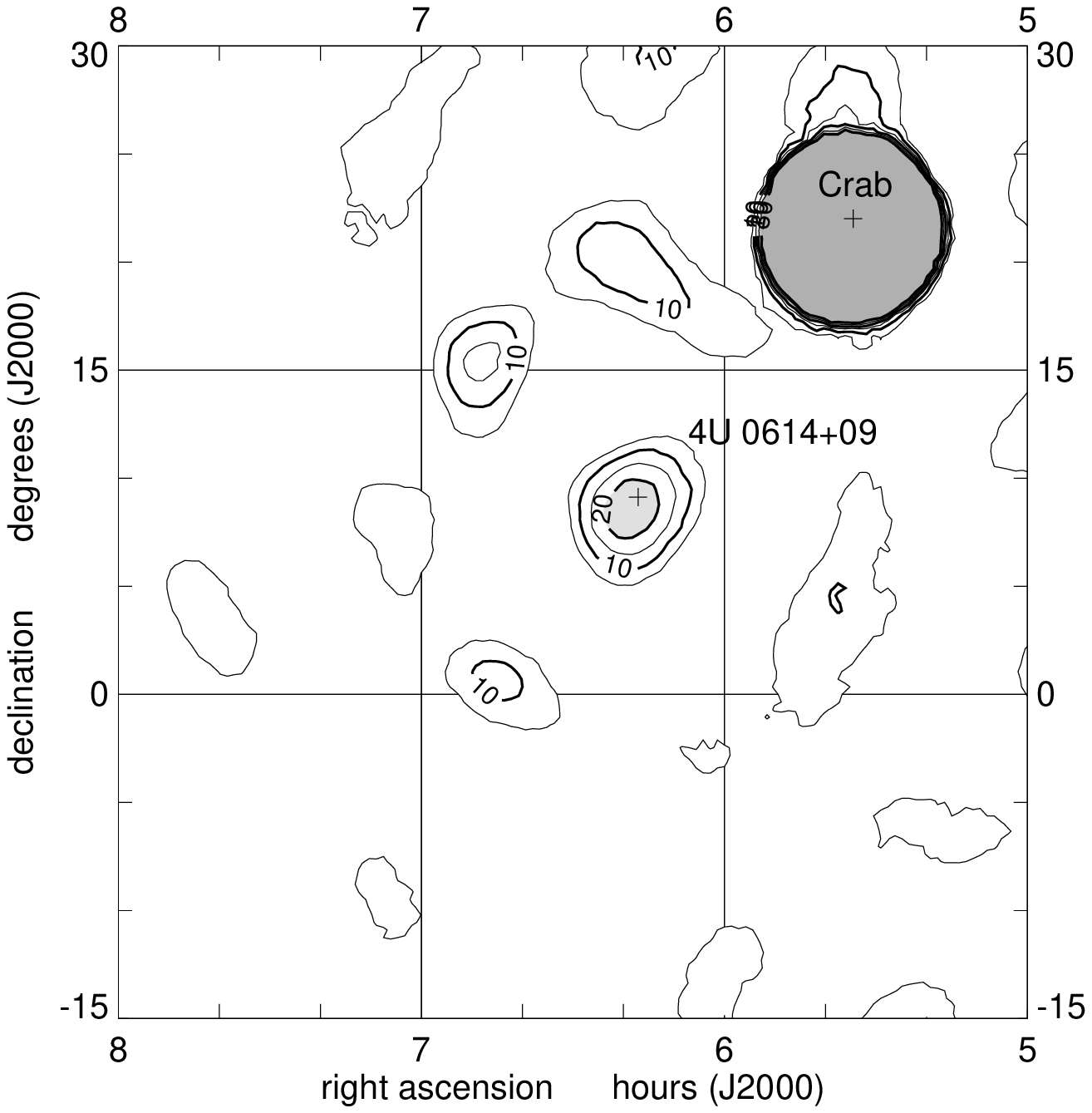}
  \caption{{\it Left:} The cross-correlation map (6--15\,keV)
    containing the burst observed from the direction of
    4U\,0614+091 on February 17, 1993 by {\it EURECA}/WATCH. The burst
    was contained in a modulation pattern integrated for 56.7\,s. The
    contours are shown in Crab units and the averaged correlation at
    the position of 4U\,0614+091 was 3.6~Crab. Note that the
    integration time is too short to give a detection of the Crab.
    {\it Right:} Detection of the persistent emission from 4U\,0614+091 by {\em
      EURECA}/WATCH.  The skymap is based on data from the
    period of February to April 1993 in the energy band
    6--12\,keV. The contours show the cross correlation in units of
    mCrab (only positive correlation is shown).  The ring-like
    structure around 4U\,0614+091 is an artefact caused by the WATCH
    RMC detection principle.  }
\label{WATCH_ima}
\end{figure}

The three bursts observed by {\it EURECA}/WATCH had positional
3$\sigma$ error circles of less than 1$\degr$ radius (taking into
account systematic errors and errors due to uncertainties in the
spacecraft pointing). The optical (Davidsen et al.\ 1974, Murdin et al.\ 1974),
infrared (Migliari et al.\ 2006) and radio (Migliari et al.\ 2009) counterparts to 4U\,0614+091 
were always within these error circles.  The WATCH imaging results for the event which occurred on
February 17 is shown in Fig~\ref{WATCH_ima} (left).\footnote{Note that
  this analysis was done more than 10 years ago. Therefore, the best-fit 
estimates on the position of the various bursts are not
  available; also the regeneration of images is not possible anymore.}


At the time of the sole burst seen by the WFC, the satellite attitude solution was not
optimum. However, the position for the burst is the same to within 0.1 pixel
of the position determined for the persistent emission at times when the
attitude solution was optimum, with an uncertainty of 2$\arcmin$ (99\%
confidence). We conclude that the origin of the burst is coincident
within 2$\arcmin$ of the position of 4U\,0614+091.


The first two bursts (in 1996 and 1998) of the six normal ASM bursts were observed by
multiple SSCs, either because the burst position was in the region
where the fields of view of SSC 0 and SSC 1 overlap, or because the burst was
active across two sequential dwells (or both, as in the first case,
yielding four independent observations).  The four subsequent bursts
were only detected in a single SSC for a single dwell, yielding a
total of eleven observations of the six bursts (see Table~\ref{table_0614}).

\begin{figure}[top]
\centering
  \includegraphics[height=.27\textheight,bb=0 60 504 575,clip=]{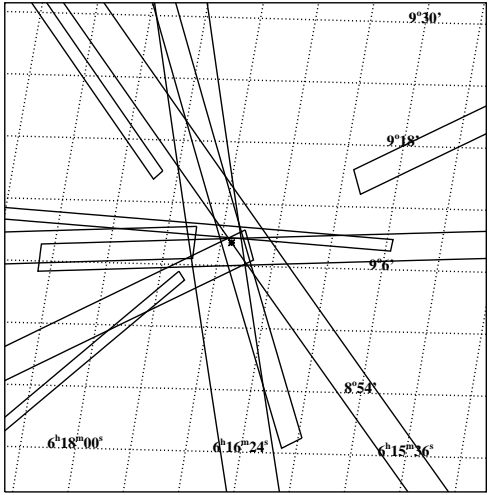}
  \caption{Equatorial (J2000.0) map of ten independent ASM
    localizations of the source of six bursts. The
    asterisk marks the location of the optical counterpart to
    4U\,0614+091 (Davidson et al.\ 1974, Murdin et al.\ 1974).}
\label{skyfig}
\end{figure}

\begin{table}
\caption{Observation log of ASM bursts from 4U\,0614+091
(left) and 2S\,0918$-$549 (right).$^3$}
\begin{tabular}{ccccccc}
\hline
\multicolumn{3}{c}{4U\,0614+091} &
\multicolumn{1}{c}{~} &
\multicolumn{3}{c}{2S\,0918$-$549} \\
\multicolumn{1}{c}{MJD} &
\multicolumn{1}{c}{SSC} &
\multicolumn{1}{c}{f$_X$} &
\multicolumn{1}{c}{~} &
\multicolumn{1}{c}{MJD} &
\multicolumn{1}{c}{SSC} &
\multicolumn{1}{c}{f$_X$} \\
\hline
50200 & 0 & 0.648/0.419 & & 50211 & 0 & 0.418 \\
      & 1 & 0.536/0.522 & & 52181 & 1 & 0.661 \\
51164 & 0,1 & 0.648,0.147 & & 52509 & 0 & 0.208 \\
53476 & 0 & 0.512 & & 52856 & 0,1 & 0.698, 0.230 \\
53959 & 0 & 0.588 & & 53683 & 1 & 0.671 \\
54101 & 2 & 0.141/0.044 & & & & \\
54108 & 1 & 0.142 & & & & \\
\hline
\end{tabular}
\vspace{-0.2cm}
\note{
\footnotesize
Given are the day (MJD) at which the burst occurred, with which SSC it was seen
(0 or 1) and the transmission factors (f$_X$) for the corresponding SSCs
($X$=0 or 1) in the 1.5--12\,keV band. For the first and fifth bursts
of 4U\,0614+091 we give the transmission factors for two sequential dwells, separated by a slash.
}
\label{table_0614}
\end{table}

For each ASM dwell, the intensities of known sources in the field of view
are derived via a fit of model slit-mask shadow patterns to
counts binned by position along each anode in each detector.
The residuals from a successful fit are then cross-correlated with
each of the expected shadow patterns corresponding to a set of
possible source directions which make up a grid covering the field of view.  A
peak in the resulting cross-correlation map indicates the possible
presence and approximate location of a new, uncatalogued X-ray source
(Levine et al.\ 1996, Smith et al.\ 1999).  

To test the hypothesis
that the six ASM bursts were from 4U\,0614+091, we removed this
source from the catalog used for the fitting procedure described. We fit the
eleven observations with this truncated catalog, and we searched the
residuals for evidence of a source of emission
unaccounted for by the catalog.
For bursts that occurred after the ASM began recording in event
mode, we omitted from the fit those time bins that did not show evidence of burst activity,
to increase the signal to noise ratio.  All
eleven dwells showed evidence of a `new' source in the residuals.  If
4U\,0614+091 were the source of the burst in each of these
dwells, the localization of this new source should be consistent with
the location of 4U\,0614+091.

Ten of the eleven localizations are plotted in Fig.~\ref{skyfig}.  The
boxes are determined by the 95\%\ confidence interval in each of the
two dimensions recorded by the SSC, projected onto the sky.  Since the
two dimensions are independent, the total confidence level of the
error box is about 90\%.  These uncertainties are statistical only.
The location of the optical counterpart of 4U\,0614+091 is indicated
by an asterisk. This location cannot be excluded by any of the error
boxes, including an eleventh error box that lies outside the limits of
this plot.

\begin{figure}[top]
\centering
\includegraphics[height=.19\textheight]{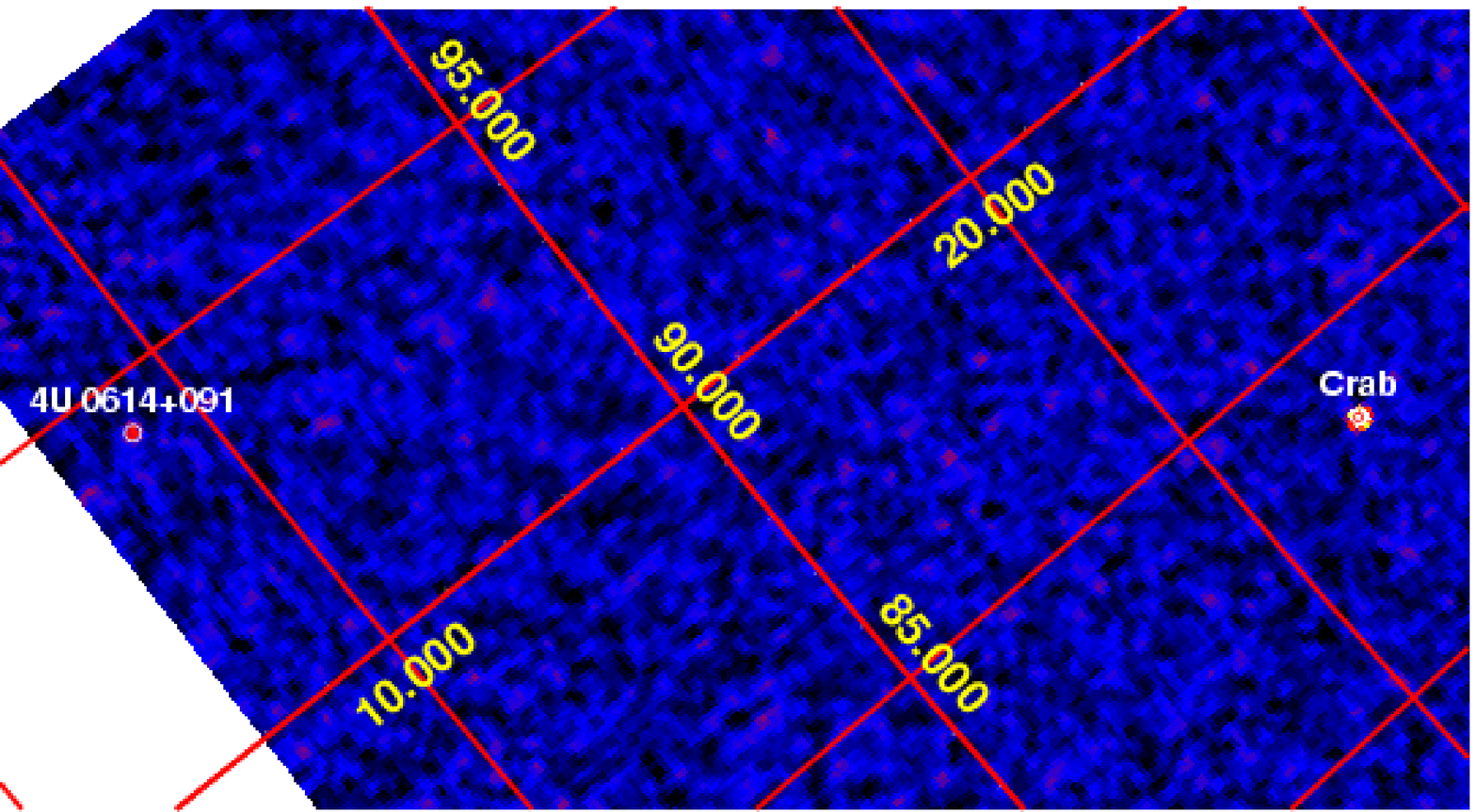}
\includegraphics[height=.19\textheight]{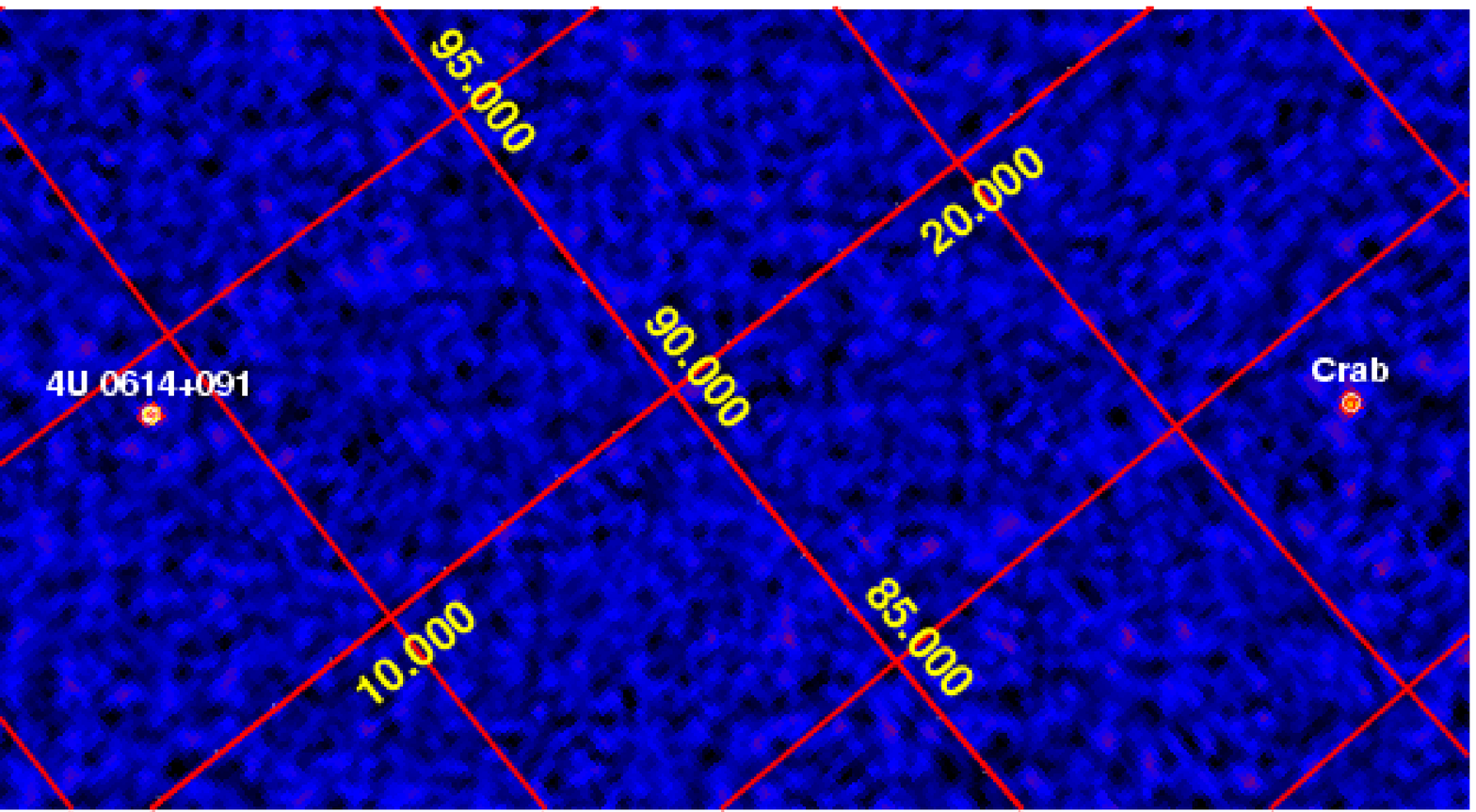}
  \caption{IBIS/ISGRI (15--30\,keV) significance images during the
    bursts detected on 2003 August 16 ({\it top}) and 2005 March 31 ({\it bottom}).  
    Shown is the equatorial J2000.0 grid with 5$\degr$ spacing.  4U\,0614+091 and Crab are the only sources
    detected in the time frames (UT 21:30:18 -- 21:30:43 and UT 07:12:18 -- 07:12:34, respectively).
    The detection significances of 4U\,0614+091 (and Crab) in the 2003 and 2005 time frames are 4.7 (18.3) and 12.7 (6.9), respectively.}
\label{ISGRI_ima_lc}
\end{figure}


For ten of the thirteen FREGATE bursts a precise localization was possible
using the WXM and/or SXC, and they are consistent with coming from 4U\,0614+091
(see Sect.~2.1.5).  For the other three bursts
we can only suppose they are coming from 4U\,0614+091, since the source
was the only persistent burster in the FREGATE field of view.


\begin{table}
\caption{Positions of the bursts seen by ISGRI and BAT.}
\begin{tabular}{cccc}
\hline
\multicolumn{1}{c}{Instrument/} &
\multicolumn{1}{c}{RA} &
\multicolumn{1}{c}{Dec} &
\multicolumn{1}{c}{error$^a$} \\
\multicolumn{1}{c}{MJD} &
\multicolumn{2}{c}{(J2000.0)} &
\multicolumn{1}{c}{~} \\
\hline
\multicolumn{4}{l}{ISGRI} \\
52867 & 94.281$\degr$ & 9.152$\degr$ & 2.1$\arcmin$ \\
53460 & 94.284$\degr$ & 9.133$\degr$ & 7.0$\arcmin$ \\
\multicolumn{4}{l}{BAT} \\
54029 & 94.278$\degr$ & 9.149$\degr$ & 1.5$\arcmin$ \\
54189 & 94.203$\degr$ & 9.155$\degr$ & 3.1$\arcmin$ \\
\hline
\multicolumn{4}{l}{\footnotesize $^a$\,90\%\ confidence error region.} \\
\end{tabular}
\label{localization}
\end{table}

The reconstructed 15--30\,keV images of the field of view around 4U\,0614+091
during the bursts seen by ISGRI are shown in
Fig.~\ref{ISGRI_ima_lc}.  The only two detected sources in the 
2003 (24\,s) and 2005 (15\,s) time frames are indeed 4U\,0614+091, as well as
the persistently bright Crab source.  
The data from the two BAT bursts were subjected to the post `$\gamma$-ray burst' processing script, which led
to refined positions for these bursts. 
The derived ISGRI and BAT bursts coordinates are given in Table~\ref{localization}.
They are also consistent with the optical, IR and radio counterparts to 4U\,0614+091.

We conclude that the 27 bursts seen by {\it EURECA}/WATCH, WFC, ASM, FREGATE, ISGRI and BAT
indeed originated from 4U\,0614+091.


\begin{thebibliography}{}

\bibitem{} 
Atteia, J.-L., Boer, M., Cotin, F., et al. 2003, in Gamma-Ray Burst and Afterglow Astronomy 2001, 
eds.\ G.~Ricker \&\ R.~Vanderspek, AIP Conf.\ Proc.\ 662 (New York: AIP), p.~17

\bibitem{} 
Barraud, C. 2002, presentation given at the AMS Workshop on Sources and GRBs, 6/12/2002, Montpellier, France

\bibitem{}
Barret, D., Grindlay, J.E. 1995, ApJ, 440, 841

\bibitem{}
Barthelmy, S.D., Barbier, L.M., Cummings, J.R., et al. 2005, SSRv, 120, 143

\bibitem{}
Belian, R.D., Conner, J.P., Evans, W.D. 1976, ApJ, 206, L135

\bibitem{}
Bildsten, L., Salpeter, E., Wasserman, I. 1992, ApJ, 384, 143

\bibitem{}
Boella, G., Butler, R.C., Perola, G.C., et al. 1997, A\&A, 122, 299

\bibitem{}
Bradt, H.V., Rothschild, R.E., Swank, J.H. 1993, A\&AS, 97, 355

\bibitem{}
Brandt, S., 1994, PhD thesis, DSRI, Denmark

\bibitem{}
Brandt, S., Castro-Tirado, A.J., Lund, N., Dremin, V., Lapshov, I., Sunyaev, R. 1992, A\&A, 262, L15

\bibitem{}
Brandt, S., Lund, N. 1995, Adv.\ Space Res., 16(8), 37

\bibitem{} 
Brandt, S., Lund, N., Castro-Tirado, A.J. 1993a, IAU Circ.\ 5710

\bibitem{} 
Brandt, S., Lund, N., Castro-Tirado, A.J. 1993b, IAU Circ.\ 5717

\bibitem{}
Brandt, S., Lund, N., Rao, A.R. 1990, Adv.\ Space Res., 10(2), 239

\bibitem{}
Brown, E.F. 2000, ApJ, 531, 988

\bibitem{}
Brown, E.F., Bildsten, L. 1998, ApJ, 496, 915

\bibitem{}
Buccheri, R., Bennett, K., Bignami, G.F., et al. 1983, A\&A, 128, 245

\bibitem{}
Chelovekov, I.V., Grebenev, S.A., Sunyaev, R.A. 2007, Proceedings of the 6th INTEGRAL Workshop "The Obscured Universe", ESA SP-622, p.~445

\bibitem{}
Cocchi, M., Bazzano, A., Natalucci, L., et al. 2000, A\&A, 357, 527

\bibitem{}
Cooper, R.L., Steiner, A.W., Brown, E.F. 2009, ApJ, 702, 660

\bibitem{}
Cornelisse, R., Heise, J., Kuulkers, E., Verbunt, F., in 't Zand, J.J.M. 2000, A\&A, 357, L21

\bibitem{}
Cornelisse, R., in 't Zand, J.J.M., Verbunt, F., et al. 2003, A\&A, 405, 1033

\bibitem{}
Cornelisse, R., Kuulkers, E., in 't Zand, J.J.M., Verbunt, F., Heise, J. 2002a, A\&A, 382, 174

\bibitem{}
Cornelisse, R., Verbunt, F., in 't Zand, J.J.M., et al. 2002b, A\&A, 392, 885

\bibitem{}
Courvoisier, T. J.-L., Walter, R., Beckmann, V., et al. 2003, A\&A, 411, L53

\bibitem{}
Cumming, A., 2004, Nuc.\ Phys.\ B Proc.\ Suppl., 132, 435

\bibitem{}
Cumming, A., Bildsten, L. 2001, ApJ, 559, L127

\bibitem{}
Cumming, A., Macbeth, J. 2004, ApJ, 603, L37

\bibitem{}
Cumming, A., Macbeth, J., in 't Zand, J.J.M., Page, D. 2006, ApJ, 646, 429

\bibitem{}
Davidsen, A., Malina, R., Smith, H., et al. 1974, ApJ, 193, L25

\bibitem{}
Ebisawa, K., Bourban, G., Bodaghee, A., Mowlavi, N., Courvoisier, T. J.-L. 2003,
A\&A, 411, L59

\bibitem{}
Falanga, M., Chenevez, J., Cumming, A., Kuulkers, E., Trap, G., Goldwurm, A. 2008, A\&A, 484, 43

\bibitem{}
Falanga, M., Cumming, A., Bozzo, E., Chenevez, J. 2009,  A\&A, 496, 333

\bibitem{}
Fiocchi, M., Bazzano, A., Ubertini, P., Bird, A.J., Natalucci, L., Sguera, V. 2008, A\&A, 492, 557

\bibitem{}
Ford, E.C., Kaaret, P., Chen, K., et al. 1997, ApJ, 486, L47

\bibitem{}
Ford, E., Kaaret, P., Tavani, M., et al. 1996, ApJ, 469, L37

\bibitem{}
Ford, E.C., van der Klis, M., M\'endez, M., et al. 2000, ApJ, 537, 368

\bibitem{}
Galloway, D.K., Muno, M.P., Hartman, J.M., Savov, P., Psaltis, D., Chakrabarty, D. 2008,
ApJS, 179, 360

\bibitem{}
Gehrels, N., Chincarini, G., Giommi, P., et al. 2004, ApJ, 611, 1005

\bibitem{}
Giacconi, R., Murray, S., Gursky, H., Kellogg, E., Schreier, E., Tananbaum, H. 1972, ApJ, 178, 281

\bibitem{}
Goldwurm, A., David, P., Foschini, L., et al. 2003, A\&A, 411, L223

\bibitem{}
Grindlay, J., Gursky, H., Schnopper, H., et al. 1975, ApJ, 205, L127

\bibitem{}
Hansen, C.J., van Horn, H.M. 1975, ApJ, 195, 735

\bibitem{}
Harmon, B.A., Wilson, C.A., Fishman, G.J., et al., 2004, ApJSS, 154, 585

\bibitem{}
Hasinger, G., van der Klis, M. 1989, A\&A, 225, 79

\bibitem{}
Hoffman, J.A., Marshall, H.L., Lewin, W.H.G. 1978, Nat, 271, 630

\bibitem{}
in ’t Zand, J.J.M. 1992, Ph.D.\ thesis, Utrecht University

\bibitem{}
in 't Zand, J.J.M., Bassa, C.G., Jonker, P.G., Keek, L., Verbunt, F., M\'endez, M., Markwardt, C.B. 2008, A\&A 485, 183

\bibitem{}
in 't Zand, J.J.M., Cumming, A., van der Sluys, M.V., Verbunt, F., Pols, O.R. 2005, 
A\&A, 441, 675

\bibitem{}
in 't Zand, J.J.M., Jonker, P.G., Markwardt, C.B. 2007, A\&A 465, 953

\bibitem{}
in 't Zand, J.J.M., Keek, L., Cumming, A., Heger, A., Homan, J., M\'endez, M. 2009, A\&A, 497, 469

\bibitem{}
Jager, R., Mels, W.A., Brinkman, A.C., et al. 1997, A\&A, 125, 557

\bibitem{}
Jahoda, K., Markwardt, C.B., Radeva, Y., et al. 2006, ApJS, 163, 401

\bibitem{}
Jonker, P.G., Nelemans, G. 2004, MNRAS, 354, 355

\bibitem{}
Juett, A.M., Chakrabarty, D. 2003, ApJ, 599, 498

\bibitem{}
Juett, A.M., Psaltis, D., Chakrabarty, D. 2001, ApJ, 560, L59

\bibitem{}
Keek, L., in 't Zand, J.J.M., in proceedings of the "7th INTEGRAL Workshop - An INTEGRAL View of
Compact Objects", PoS(Integral08)032

\bibitem{}
Keek, L., in 't Zand, J.J.M., Kuulkers, E., Cumming, A., Brown, E.F., Suzuki, M. 2008, 
A\&A, 479, 177

\bibitem{}
Krimm, H., Barbier, L., Barthelmy, S. D., et al. 2006, ATel \#904

\bibitem{}
Kuulkers, E. 2002, A\&A, 383, L5

\bibitem{}
Kuulkers, E. 2004, Nuc.\ Phys.\ B Proc.\ Suppl., 132, 466

\bibitem{}
Kuulkers, E. 2005, ATel \#483

\bibitem{}
Kuulkers, E., den Hartog, P.R., in 't Zand, J.J.M., Verbunt, F.W.M., Harris, W.E., Cocchi, M.
2003, A\&A, 399, 663

\bibitem{}
Kuulkers, E., Homan, J., van der Klis, M., Lewin, W.H.G., M\'endez, M. 2002b, A\&A, 382, 947

\bibitem{}
Kuulkers, E., in 't Zand, J.J.M., van Kerkwijk, M.H., et al. 2002a, A\&A, 382, 503

\bibitem{}
Kuulkers, E., Shaw, S.E., Paizis, A., et al. 2007, A\&A, 466, 595

\bibitem{}
Kuulkers, E., van der Klis, M., Oosterbroek, T., et al. 1994, A\&A, 289, 795

\bibitem{}
Lamb, D.Q., Lamb, F.K. 1978, ApJ, 220, 291

\bibitem{}
Lebrun, F., Leray, J. P., Lavocat, P., et al. 2003, A\&A, 411, L141

\bibitem{}
Levine, A.M., Bradt, H., Cui, W., et al. 1996, ApJ, 469, L33

\bibitem{}
Lewin, W.H.G. 1976, IAU Circ.\ 2914

\bibitem{}
Lewin, W.H.G., Vacca, W.D., Basinska, E.M. 1984, ApJ, 277, L57

\bibitem{}
Lewin, W.H.G., van Paradijs, J., Taam, R.E. 1993, SSRv, 62, 223

\bibitem{}
Linares, M., Watts, A.L., Wijnands, R., et al. 2009, MNRAS, 392, L11

\bibitem{}
Lund, N. 1985, in: X-Ray Instrumentation in Astronomy, ed.\ J.L.\ Culhane, Porc.\ SPIE 597, p.~95

\bibitem{}
Lund, N., Brandt, S., Budtz-J\oe rgensen, C., et al. 2003, 411, L231

\bibitem{}
Maraschi, L., Cavaliere, A. 1977, in Highlights in Astronomy, ed.\ E.A.\ M\"uller, Vol.~4, 127

\bibitem{}
Markert, T.H., Laird, F.N., Clark, G.W., et al. 1979, ApJS, 39, 573

\bibitem{}
Mason, K.O., Charles, P.A., White, N.E., Culhane, J.L., Sanford, P.W., Strong, K.T. 1976,
MNRAS 177, 513

\bibitem{}
M\'endez, M., Cottam, J., Paerels, F. 2002, arXiv:astro-ph/0207277

\bibitem{}
Mereghetti, S., G\"otz, D., Borkowski, J., Walter, R., Pedersen, H. 2003, A\&A, 411, L291

\bibitem{}
Migliari, S., Fender, R. 2006, MNRAS, 366, 79

\bibitem{}
Migliari, S., Tomsick, J.A., Maccarone, T.J., Gallo, E., Fender, R.P., Nelemans, G., 
Russell, D.M. 2006, ApJ, 643, L41

\bibitem{}
Migliari, S., Tomsick, J.A., Miller-Jones, J.C.A., et al. 2009, ApJ, submitted

\bibitem{}
Molkov, S.V., Grebenev, S.A., Lutovinov, A.A. 2000, 357, L41

\bibitem{}
Murdin, P., Penston, M.J., Penston, M.V., et al. 1974, MNRAS, 169 25

\bibitem{}
Nelemans, G., Jonker, P.G., Marsh, T.R., van der Klis, M. 2003, MNRAS, 348, L7

\bibitem{}
Nelemans, G., Jonker, P.G., Steeghs, D. 2006, MNRAS, 370, 255

\bibitem{}
Nelemans, G., Yungelson, L.R., van der Sluys, M.V., Tout, C.A. 2009, MNRAS, submitted

\bibitem{} 
Parsignault, D.R., Grindlay, J.E. 1978, ApJ, 225, 970

\bibitem{} 
Piraino, S., Santangelo, A., Ford, E.C., Kaaret, P. 1999, A\&A, 349, L77

\bibitem{} 
Ricker, G.R., Atteia, J.-L., Crew, G.B., et al. 2003, in 
Gamma-Ray Burst and Afterglow Astronomy 2001: A
Workshop Celebrating the First Year of the HETE Mission, 
G.R.\ Ricker \&\ R.K.\ Vanderspek (eds.), AIP Conf.\ Ser., 662, 3

\bibitem{} 
Rothschild, R.E., Blanco, P.R., Gruber, D.E., et al. 1998, ApJ, 496, 538

\bibitem{} 
Schatz, H., Bildsten, L., Cumming, A. 2003, ApJ, 583, L87

\bibitem{} 
Shahbaz, T., Watson, C.A., Zurita, C., Villaver, E., Hernandez-Peralta, H. 2008, PASP, 120, 848

\bibitem{} 
Shirasaki, Y., Kawai, N., Yoshida, A., et al. 2003, PASJ, 55, 1033

\bibitem{} 
Smith, D.A., Levine, A.M., Bradt, H.V., et al. 1999, ApJ, 526, 683

\bibitem{}
Strohmayer, T., Bildsten, L. 2006, in {\it Compact stellar X-ray sources},
W.H.G.~Lewin \&\ M.~van der Klis (eds.), Cambridge Astrophysics Series 39, p.~113

\bibitem{} 
Strohmayer, T.E., Brown, E.F. 2002, ApJ, 566, 1045

\bibitem{} 
Strohmayer, T.E., Markwardt, C.B. 2002, ApJ, 577, 337

\bibitem{} 
Strohmayer, T.E., Markwardt, C.B., Kuulkers, E. 2008, ApJ, 672, L37

\bibitem{} 
Suzuki, M., Sakamoto, T., Takahashi, D., et al. 2004, in: Third Rome Workshop on Gamma-Ray Bursts
in the Afterglow Era, eds.\ M.~Feroci, F.~Frontera, N.~Masetti \&\ L.~Piro, ASP Conf.\ Ser.\ 312,
p.~138

\bibitem{} 
Swank, J.H., Becker, R.H., Boldt, E.A., Holt, S.S., Serlemitsos, P.J. 1978, MNRAS, 182, 349

\bibitem{} 
Tawara, Y., Kii, T., Hayakawa, S., et al. 1984, ApJ, 276, L41

\bibitem{} 
Ubertini, P., Lebrun, F., Di Cocco, G., et al. 2003, A\&A, 411, L131

\bibitem{}
van der Klis, M., 1995, in: The Lives of the Neutron Stars,
eds.\ M.A.~Alpar, \"U.~K\i z\i lo\u glu \&\ J.~van Paradijs, NATO ASI Ser.\ C,
Volume 450, Kluwer Academic Publishers, p.~301

\bibitem{} 
van Paradijs, J., Lewin, W.H.G. 1986, A\&A, 157, L10

\bibitem{} 
van Straaten, S., Ford, E.C., van der Klis, M., M\'endez, M., Kaaret, P. 2000, ApJ, 540, 1049

\bibitem{} 
van Straaten, S., van der Klis, M., M\'endez, M. 2003, ApJ, 596, 1155

\bibitem{} 
Verner, D.A., Ferland, G.J., Korista, K.T., Yakovlev, D.G., 1996, ApJ, 465, 487

\bibitem{} 
Villasenor, J.N., Dill, R., Doty, J.P., et al. 2003, in
Gamma-Ray Burst and Afterglow Astronomy 2001: A
Workshop Celebrating the First Year of the HETE Mission,
G.R.\ Ricker \&\ R.K.\ Vanderspek (eds.), AIP Conf.\ Ser., 662, 33

\bibitem{} 
Warwick, R.S., Marshall, N., Fraser, G.W., et al. 1981, MNRAS, 197, 865

\bibitem{} 
Werner, K., Nagel, T., Rauch, T., Hammer, N.J., Dreizler, S. 2006, A\&A, 450, 725

\bibitem{} 
Westergaard, N.J., Kretschmar, P., Oxborrow, C.A., et al. 2003, A\&A, 411, L257

\bibitem{} 
Wijnands, R. 2001, ApJ, 554, L59

\bibitem{} 
Wilms, J., Allen, A., McGray, R. 2000, ApJ, 542, 914

\bibitem{} 
Winkler, C., Courvoisier, T. J.-L., Di Cocco, G., et al. 2003, A\&A, 411, L1

\bibitem{} 
Woosley, S.E., Heger, A., Cumming, A., et al. 2004, ApJS, 151, 75

\bibitem{} 
Woosley, S.E., Taam, R.E. 1976, Nat, 263, 101

\end{thebibliography}
\end{document}